\documentclass[prd,aps,a4paper,twocolumn,eqsecnum,nofootinbib,floatfix]{revtex4-1} 

\usepackage{amsmath}
\usepackage{amsfonts}
\usepackage{amssymb}	
\usepackage{graphicx}
\usepackage{bm}
\usepackage{color}
\usepackage{commath}
\usepackage{hyperref}
\usepackage{comment}
\allowdisplaybreaks

\def\TEOBResumS{\texttt{TEOBResumS}}

\usepackage[T1]{fontenc}

\newcommand{\pinf}{p_{\infty}}
\newcommand{\be}{\begin{equation}}
\newcommand{\ee}{\end{equation}}
\newcommand{\bea}{\begin{eqnarray}}
\newcommand{\eea}{\end{eqnarray}}
\newcommand{\bel}{\begin{align}}
\newcommand{\eel}{\end{align}}

\def\GMc2{G M_{\odot} c^{-2}}

\def\F{{\cal F}}
\def\lm{{\ell m}}

\def\lm{{\ell m}}

\def\lm{{\ell m}}
\def\g{{\gamma}}

\def\F{{\cal F}}

\def\TEOBResumS{\texttt{TEOBResumS}}

\def\TEOBd{{\texttt{TEOBResumS-Dal\'i}}}
\def\LEOBa{{\tt LEOB-PM$_{a_0}$}}
\def\LEOBss{{\tt LEOB-PM$_{\tt SS}$}}
\def\LEOBfact{{\tt LEOB-PM$_{\tt SSfact}$}}

\def\G12{\Tilde{G}_{ap}}

\begin{document}
\title{A novel Lagrange-multiplier approach to the effective-one-body dynamics \\ of binary systems in post-Minkowskian gravity}
\author{Thibault \surname{Damour}${}^{1}$}
\author{Alessandro \surname{Nagar}${}^{1,2}$}
\author{Andrea \surname{Placidi}${}^{3}$}
\author{Piero \surname{Rettegno}${}^{3}$}

\affiliation{${}^1$Institut des Hautes Etudes Scientifiques, 91440 Bures-sur-Yvette, France}
\affiliation{${}^2$INFN Sezione di Torino, Via P. Giuria 1, 10125 Torino, Italy}
\affiliation{${}^3$Dipartimento di Fisica e Geologia, Università di Perugia,
INFN Sezione di Perugia, Via Pascoli, I-06123 Perugia, Italy}
	
\begin{abstract}
We present a new approach to the conservative dynamics of binary systems, within the effective one-body (EOB) 
framework, based on the use of a Lagrange multiplier to impose the mass-shell constraint. When applied to the 
post-Minkowskian (PM) description of the two-body problem in Einsteinian gravity, this Lagrange-EOB (LEOB) approach 
allows for a new  formulation of the conservative dynamics that avoids the drawbacks of the recursive definition of EOB-PM Hamiltonians.
Using state-of-the-art values of the waveform and radiation reaction, based on factorized and resummed 
post-Newtonian (PN) expressions, we apply our new formalism to the construction of 
an aligned-spin, quasi-circular, inspiraling EOB waveform model, called {\tt LEOB-PM}, 
that incorporates analytical information up to the 4PM level, completed by 4PN contributions up to the sixth order 
in eccentricity, in the orbital sector, and  by 4.5PN contributions, in the spin-orbit sector. 
In the nonspinning case, we find that an uncalibrated {\tt LEOB-PM} model delivers 
maximum EOB/NR unfaithfulness ${\bar{\cal F}}_{\rm EOBNR}$ 
(with the Advanced LIGO noise in the total mass range $10-200M_\odot$) varying between $0.2\%$ and $1\%$ 
over all the nonspinning dataset of the Simulating eXtreme Spacetime (SXS) Numerical Relativity (NR) 
catalog up to mass ratio $q=15$.  It also delivers excellent phasing agreement with the $q=32$ configuration 
of the RIT catalog. We also found consistency between binding energies within a few percent at the NR merger location. 
Then, when NR-informing the dynamics of the model (both orbital and spinning sectors) by using  
only 17 SXS dataset, we  find that the EOB/NR unfaithfulness (compared to 530 spin-aligned SXS waveforms) 
has a median value of $5.39\times 10^{-4}$, or $6.13\times 10^{-4}$ (depending on the chosen description of spin-spin interactions),
reaching at most $\sim 1\%$ in some of the high-spin corners. Although this global performance is less good 
(by  a factor 3.67) than some of the state-of-the-art PN-Hamiltonian-based models, our finding shows that the 
LEOB approach is a promising route toward robustly incorporating both PM and PN information 
in the EOB description of the dynamics. 
\end{abstract}
	
\maketitle

\section{Introduction}

Upcoming observations of binary black hole coalescences by present and future gravitational wave (GW) detectors require
ever-improved waveform models. In turn, this requires an ever more accurate  analytical knowledge of the dynamics
of gravitationally interacting binary systems. One of the prime approaches to the description of the dynamics of (and GW 
emission from) binary systems, is the effective one-body (EOB) formalism~\cite{Buonanno:1998gg,Buonanno:2000ef,Damour:2001tu}. This formalism was historically developed within the post-Newtonian (PN) approach to Einsteinian gravity (see~\cite{Blanchet:2013haa} for a review). 
References~\cite{Damour:2016gwp,Damour:2017zjx} showed how to extend
the validity of the EOB approach to fast motions by reformulating it
 within the post-Minkowskian (PM) approach to Einsteinian gravity. The latter works, together with
Ref.~\cite{Cheung:2018wkq}, triggered an intense activity in PM gravity using either 
quantum scattering amplitudes (notably~\cite{Bern:2019crd,Bern:2019nnu,Bern:2021dqo,Bern:2021yeh,Damgaard:2023ttc,Bern:2024adl}) or worldline
 approaches (notably \cite{Dlapa:2021vgp,Jakobsen:2023ndj,Dlapa:2024cje,Driesse:2024xad}). 
The usefulness of  incorporating PM results in the EOB formalism has been notably explored in Refs.~\cite{Bini:2018ywr,Damour:2019lcq,Antonelli:2019ytb,Damgaard:2021rnk,Khalil:2022ylj,Damour:2022ybd,Rettegno:2023ghr,Ceresole:2023wxg,Buonanno:2024vkx,Buonanno:2024byg} (for other approaches to PM gravity, see Refs.~\cite{Cristofoli:2019neg,Kalin:2019rwq,Kalin:2019inp}).
The EOB dynamics comprise a conservative sector, traditionally described by a Hamiltonian, together with 
an additional dissipative sector, described by a radiation-reaction force. The  Hamiltonian-based EOB 
approach (HEOB), initially introduced in PN gravity, was generalized to  PM gravity in Refs.~\cite{Damour:2016gwp,Damour:2017zjx,Bini:2018ywr,Damour:2019lcq,Antonelli:2019ytb,Damgaard:2021rnk,Khalil:2022ylj,Damour:2022ybd,Rettegno:2023ghr,Ceresole:2023wxg,Buonanno:2024vkx,Buonanno:2024byg}. 
In particular, Ref.~\cite{Buonanno:2024byg} constructed for the first time a complete 
inspiral-merger-ringdown waveform model, {\tt SEOBNR-PM}, for nonprecessing spinning 
black holes using a PM-version of the HEOB approach. The latter model has shown good
agreement with NR data for the binding energy without using any NR calibration.
Its  waveform also showed good consistency with NR data after the introduction
of a NR-tuned parameter. 
 
The aim of the present work is to introduce a new way, called Lagrange-EOB (LEOB), 
of incorporating PM (and PN as well) results in the conservative sector of the EOB formalism. 
The name Lagrange-EOB comes from a crucial use of a Lagrange multiplier~\cite{Lagrange:1788}
in the EOB action. Our main motivation for this new way of formulating the EOB dynamics 
is to avoid some drawbacks (illustrated in Fig.~\ref{Fig:ps_khalil} below)
of the HEOB approach to PM gravity, linked to the {\it recursive definition} 
of an explicit EOB-PM Hamiltonian, $H_{\rm PM}^{\rm EOB}(q,p)$ (first suggested
in Ref.~\cite{Damour:2017zjx}, and extended to higher PM orders in Refs.~\cite{Damour:2019lcq,Bini:2020nsb,Khalil:2022ylj,Buonanno:2024vkx,Buonanno:2024byg}). 
The LEOB approach is fruitfully applied here to the construction 
of PM-informed waveform models, with very promising performances (see notably  
Figs.~\ref{fig:barF_full} and \ref{fig:barF_full_4PM_SS}).

The outline of the paper is as follows. In Sec.~\ref{sec:start} we summarize the fundamentals of 
the EOB approach, we briefly review the HEOB approach to PM gravity pointing some issues 
that come with it, and we introduce the basic elements of our approach and our
gauge fixing choice. 
In Sec.~\ref{sec:PMstuff} we specifically deal with PM information, recasting it consistently within the LEOB approach. 
We include known (local) information up to the physical 5PM level for both orbital and spinning sectors\footnote{When dealing with spinning black holes, each power of spin adds a physical PM order.}.
Since 5PM orbital terms are currently not fully known and 5PM spin-orbit terms contain nonlocal (hyperbolic) information, we
replace them by  PN results.
We hence complete the local PM dynamics with local and (elliptic) nonlocal 4PN-accurate information in the nonspinning sector and 
4.5PN-accurate terms in the spin-orbit sector. 
The reader interested in the main results can skip the technical sections described above and directly jump
to Sec.~\ref{sec:leobWF}  to have a sense of the structure of the PM-based LEOB 
waveform model and to the following Sec.~\ref{sec:performance} where the performance 
of the model is thoroughly evaluated. Concluding remarks are collected in Sec.~\ref{sec:end}.
The paper is completed by a few Appendixes. 
In particular, 
Appendix \ref{app:ss} is devoted to assessing the performance of an alternative LEOB model using a different treatment of the spins up to the physical 5PM order;
Appendix \ref{app:data} concerns the procedure used to calibrate the model spinning sector using NR data;
Appendix \ref{app:gauge_freedom} discusses the gauge freedom of the LEOB approach in the nonspinning case; 
in Appendix \ref{app:4PM_others} we review the techniques adopted to deal with the orbital 4PM information in other HEOB-based works;
in Appendix~\ref{app:complete_Alocal_6PN}
we recast in LEOB form the currently known Tutti Frutti dynamics up to 7PM-6PN accuracy, and we follow up on that with the derivation of the 6PN-accurate nonlocal terms up to the eighth order in eccentricity, in Appendix \ref{app:nonlocal}.

Throughout the paper, we use units with $c=1$. We also often set $G=1$, except when it is useful to keep track of the PM order.

\section{Motivation and fundamentals}
\label{sec:start}

\subsection{PM dynamics within the HEOB approach}
For simplicity, we restrict our study to binary systems with aligned  spins. We use the following notation (with 
$m_1 \geq m_2$ and $M \equiv m_1+m_2$)
\begin{subequations}
\bea
 S_1 & = m_1 a_1 = Gm_1^2 \chi_1 = G M m_1 \tilde a_1\, , \\
S_2 &=  m_2 a_2 = Gm_2^2 \chi_2 = G M m_2 \tilde a_2 \,.
\eea
\end{subequations}

In this situation the spins 
enter the dynamics as constant parameters without adding new degrees of freedom.
The basic building blocks of the EOB dynamics of (aligned) spinning binary black holes are:
\begin{itemize}
	\item[(i)] The EOB energy map\footnote{This map was found to arise at successive PN orders in Refs.~\cite{Buonanno:1998gg,Damour:2000we,Damour:2015isa}, and was proven to be exactly valid in Ref.~\cite{Damour:2016gwp}.} relating the real total energy, $E$,  of the binary system in the center of mass (cm) frame
to the effective energy, $E_{\rm eff}= - P_0$, of the 	effective particle of mass $\mu= \frac{m_1 m_2}{m_1+m_2}$ describing the cm-frame
relative motion:
\bea \label{energymap}
E &= M \sqrt{1 + 2 \nu \left(\frac{E_{\rm eff}}{\mu}-1  \right)}  \nonumber \\
 & \equiv  M \sqrt{1 + 2 \nu \left(\g-1  \right)} \,.
\eea
Here, $M= m_1+m_2$, $\nu = \mu/M=m_1 m_2/M^2$, and we defined the $\mu$-rescaled effective energy as\footnote{$\g$ is $>1$ for scattering motions and $<1$ for bound motions.}
\be
\g \equiv \frac{E_{\rm eff}}{\mu} \equiv - \frac{P_0}{\mu}\,.
\ee
\item[(ii)] A relativistic mass-shell condition for the effective particle which was originally written in the form
    \begin{equation}
    	\label{eq:mass_shell_condition}
    g^{\mu \nu}(X^\lambda) P_\mu P_\nu + \mu^2 + Q(X^\mu,P_\mu)=0\,,
    \end{equation}
  with an (energy-independent) effective metric  $g_{\mu\nu}(X^\lambda)$, and an additional term 
  $Q(X^\mu,P_\mu)$ gathering contributions that are higher-than-quadratic in $P_\mu$; 
  here (see below) we shall instead use a formally purely geodesic mass-shell condition involving an
  energy-dependent effective metric:
  \be \label{eq:mass_shell_condition2}
   g^{\mu \nu}(X^\lambda, \g) P_\mu P_\nu + \mu^2 =0\, ;
  \ee
\item[(iii)] a radiation-reaction force ${\mathcal F}_\mu$; 
   \item[(iv)] a waveform decomposed in multipoles:
\be
h_+-ih_\times = \dfrac{1}{D_L}\sum_{\ell=2}^{\ell_{\rm max}}\sum_{m=-\ell}^{\ell}h_{\lm}\, {}_{-2}Y_\lm \ ,
\ee
where $D_L$ is the luminosity distance to the source and ${}_{-2}Y_\lm$ are 
the $s=-2$ spin-weighted spherical harmonics.
 \end{itemize}
 All the above EOB building blocks depend on $(m_1, m_2, a_1, a_2)$.
  
The mass-shell condition Eq.~\eqref{eq:mass_shell_condition}, or  Eq.~\eqref{eq:mass_shell_condition2}, 
is determined by working with a 
gauge-invariant description of the conservative dynamics, such as the 
Delaunay Hamiltonian~\cite{Buonanno:1998gg}, or the scattering  angle~\cite{Damour:2016gwp}.
  
When considering nonspinning bodies the effective metric can be taken to be spherically symmetric, i.e. of the general
form
\begin{align}
\label{geff}
g_{\mu\nu}(X^\lambda)dX^\mu dX^\nu&=-A(R)c^2 dT^2+B(R)dR ^2\nonumber\\
&+R^2 C(R)( d \theta^2 + \sin^2 \theta d\phi^2)\,.
\end{align}
When working in PM gravity (i.e. when expanding in powers of Newton's constant $G$) the effective metric was
shown~\cite{Damour:2016gwp} to coincide with the Schwarzschild metric at the first post-Minkowskian (1PM) order, i.e. at $O(G^1)$.
One way to incorporate higher PM contributions \cite{Damour:2017zjx} is to add  a non-geodesic term $Q(X^\mu,P_\mu)$ 
having a PM expansion of the form
\begin{align} 
\label{QPM}
\frac{Q(X^\mu,P_\mu)}{\mu^2}&= q_2(\g, \nu) \left(\frac{G M}{R}  \right)^2 +  q_3(\g, \nu) \left(\frac{G M}{R}  \right)^3 \nonumber\\
&+ q_4(\g, \nu) \left(\frac{G M}{R}  \right)^4 + \cdots
\end{align}
Here, we used an ``energy gauge" of the E-type \cite{Damour:2019lcq,Bini:2020nsb} where the dependence of $Q(X^\mu,P_\mu)$
on $P_\mu$ is restricted to be a dependence only on $P_0= - \mu \g$. 

In this approach to EOB-PM, one can, without loss of generality,  choose the effective metric $g_{\mu\nu}(X^\lambda)$ to be the
Schwarzschild metric (of mass $M=m_1+m_2$), i.e.  $A_S(R)= 1- 2\frac{GM}{R}$, $B_S(R)=1/A_S(R)$ and $C_S(R)=1$ 
(in Schwarzschild coordinates). In this ``post-Schwarzschild" approach, $Q(X^\mu,P_\mu)$ [and all the PM-expansion coefficients 
$q_n(\g, \nu)$] vanishes as the symmetric mass ratio $\nu$ tends to zero. 

The explicit map between the expansion coefficients $q_n(\g, \nu)$ entering the EOB mass-shell condition Eq.~\eqref{eq:mass_shell_condition} and the
PM-expansion coefficients $\chi_n(\gamma,\nu)$ of the scattering angle,
\begin{equation}
	\label{eq:chi_exp}
	\chi \left(j,\gamma\right) = \sum_{n} 2 \frac{\chi_n(\gamma,\nu)}{j^n}\,,
\end{equation}
[where $j \equiv J/(M \mu)$ is the rescaled angular momentum of the system in center of mass frame] has been determined in
Refs.~\cite{Damour:2017zjx,Bini:2020nsb}. 

When starting from an E-type mass-shell condition,  Eq.~\eqref{eq:mass_shell_condition}, there are several
ways to absorb the PM-expanded $Q$ term, $Q = \sum_{n\geq 2} q_n(\g, \nu) (GM/R)^n$, into an apparently purely geodesic 
mass-shell condition of the form \eqref{eq:mass_shell_condition2}, say
\be \label{massshell2}
 g_*^{\mu \nu}(X^\mu, \g) P_\mu P_\nu + \mu^2 =0\ ,
\ee
with an energy-dependent effective metric $ g_*^{\mu \nu}(X^\lambda, \g)$.
One way is to incorporate $Q(R, \g)$ in the $A$ contribution, $ - P_0^2/A(R)= - \mu^2 \g^2/A(R)$.
This leads to 
\be
 \label{eq:QtoA}
 \frac1{A_{S^*}(R, \g,\nu)}= \frac1{A_S(R)}- \frac{Q(R, \g,\nu)}{\g^2}\,,
 \ee
or equivalently,
 \be 
 \label{AS*}
 A_{S^*}(R, \g,\nu)= \frac{A_S(R)}{1- \frac1{\g^2} A_S(R)Q(R, \g,\nu) }\ .
 \ee
The mass-shell condition, Eq.~\eqref{massshell2}, with the latter  $A=A_{S^*}(R, \g,\nu)$, together with Schwarzschild
values of the other metric coefficients, $B=B_S(R)=1/(1-2/R)$, and $C=C_S(R)=1$, is then strictly equivalent to the E-type
version of Eq.~\eqref{eq:mass_shell_condition}.
Such a procedure was introduced  in Ref.~\cite{Damour:2017zjx} which showed 
that the high-energy limit ($\g \to \infty$) of the mass-shell condition Eqs.~\eqref{massshell2}-\eqref{AS*}, 
was defining (rather remarkably) a 4PM-accurate {\it energy-independent} high-energy effective 
metric $\lim_{\g\to \infty}   g_*^{\mu \nu}(X^\mu, \g)$ equivalent to the Amati-Ciafaloni-Veneziano~\cite{Amati:1990xe}
high-energy scattering angle.

References~\cite{Antonelli:2019ytb,Khalil:2022ylj,Buonanno:2024vkx} advocated defining PM Hamiltonians
based on using the following PM-reexpansion of $A_{S^*}(R, \g,\nu)$, Eq.~\eqref{AS*},
 \begin{equation}
	A_{S^*}^{\rm PM}\left(u,\gamma, \nu\right) = 1-2u + a^*_2(\gamma,\nu)u^2+a^*_3(\gamma,\nu)u^3+... \, ,
\end{equation}
where $u \equiv \frac{GM}{R}$, together with a recursive replacement of $\gamma$ in terms of the Hamiltonian
of a test-mass in a Schwarzschild metric
\be
\hat{H}_S\equiv \sqrt{(1-2u)\left(\mu^2 + (1-2u)P_R^2 + \frac{P_\varphi^2}{R^2}\right)} \ .
\ee
Appendix B.2 of Ref.~\cite{Khalil:2022ylj} gives the 4PM-accurate link between
the PM-expansion coefficients $ a^*_n(\gamma,\nu)$ and the scattering coefficients $ \chi_n(\gamma,\nu)$.
This link also follows from inserting the $q_n[ \chi_n]$ relations of Refs.~\cite{Damour:2017zjx,Bini:2020nsb} 
in Eq.~\eqref{AS*}. As a check (and since it is needed in the other gauge we shall use below) we derived it
by directly computing the scattering angle predicted by a general PM metric $ g_*^{\mu \nu}(X^\mu, \g)$.

When considering (aligned) {\it spinning} bodies, various forms of the EOB mass-shell 
condition have been proposed both in the PN approach, e.g.~\cite{Damour:2001tu,Barausse:2009xi,Damour:2014sva}, 
and in the PM one~\cite{Khalil:2022ylj,Buonanno:2024vkx}. All the approaches involve 
an effective metric which is a deformed version of the Kerr metric, and spin-orbit 
coupling terms involving two different combinations of the two spins $S_1$ and $S_2$, e.g., in our convention (recalling $a_i \equiv S_i/m_i$), 

\begin{subequations}
\begin{align}
S &\equiv S_1+S_2= m_1 a_1+ m_2 a_2 \ ,\\
S_*&\equiv \frac{m_2}{m_1} S_1 + \frac{m_1}{m_2} S_2= m_2 a_1+ m_1 a_2 \ .
\end{align}
\end{subequations}
 Here,  we shall consider a mass-shell condition involving 
an energy-dependent deformation of a Kerr metric (in Boyer-Lindquist coordinates), namely
\be
\label{eq:mass_shell_full}
- \frac{(P_0 + {\mathcal G} P_\varphi )^2}{A}+ \frac{P_R^2}{B}+ \frac{ P_\varphi ^2}{R_c^2}+ \mu^2=0\,,
\ee
where all metric coefficients $A, B, R_c, {\mathcal G}$ depend on $R, \g, a_1, a_2, \nu$.
Here the spin-orbit (LS) coupling term $  {\mathcal G} P_\varphi$ (proportional to the orbital
angular momentum $P_\varphi= L$) gathers all the terms  which are odd in spins, with a radial coefficient of the general form
\be
\label{eq:GSandGSstar}
 {\mathcal G}(R, \g, a_1, a_2, \nu)= G_S(R, \g,  a_i,\nu)  S +  G_{S_*}(R, \g,  a_i,\nu)  S_*\,,
\ee
where $G_S$ and $ G_{S_*}$ are even functions of the spin variables. 
The other metric coefficients, $A(R, \g,  a_i,\nu)$, 
$B(R, \g,  a_i,\nu)$, $R_c^2(R, \g,  a_i,\nu)$,
  are even in spins, and are energy-dependent deformations of their Kerr analogs. Let us recall
 the values of $A$, $B$, $R_c^2$, and  ${\mathcal G}$ in the probe limit of a test black hole (with vanishing spin because
 of the scaling $S_2= G m_2^2 \chi_2 $ with $|\chi_2| < 1$) moving around a Kerr black hole (in  Boyer-Lindquist coordinates),
  as functions of two Kerr mass and spin parameters, $M_K$, $S_K = M_K a_K$:
 \be
 \label{eq:rcK}
 R_{c \,\rm K}^2= R^2 + a_K^2 \left(1 + \frac{2 GM_K}{R}\right) \ , 
 \ee
 \be
 A_K(R,a_K)= \frac{1 + \frac{2 GM_K}{R_c}}{1 + \frac{2 GM_K}{R}} \left(1 - \frac{2 GM_K}{R_{c \,\rm K}} \right) \ ,
 \ee
 \be
 B_K(R,a_K)= \frac{R^2}{R_{c \,\rm K}^2} \frac{1}{A_K(R,a_K)} \ ,
 \ee
 and 
 \be \label{calGprobe}
 {\mathcal G}(R, a_K, L)= \frac{2 G L S_K}{R R_{c \,\rm K}^2}\,.
 \ee
 The precise definitions of the coupling functions, $ G_S(R, \g,  a_i)$, $  G_{S_*}(R, \g,  a_i)$, 
 $A(R, \g,  a_i)$, $B(R, \g,  a_i)$, and  $R_c(R, \g, a_i)$, that we shall use will be given below. 
 
When using the EOB dynamics to compute accurate waveforms it has been traditional to use an explicit Hamiltonian
$H_{\rm eff}^{\rm EOB}({\bf Q}, {\bf P})$ obtained by solving the mass-shell condition with respect to $P_0= - \mu \g$.
[The real Hamiltonian $H^{\rm EOB}({\bf Q}, {\bf P})$ is then derived from $H_{\rm eff}^{\rm EOB}({\bf Q}, {\bf P})$ 
by using the energy map \eqref{energymap}.]  Explicit expressions for $H_{\rm eff}^{\rm EOB}({\bf Q}, {\bf P})$  are
obtained by recursively solving the mass-shell condition. This was initiated in  Sec. VII of Ref.~\cite{Bini:2020nsb} using
standard PM expansions. A slightly different formulation of the 4PM-level solution was later given in Ref.~\cite{Khalil:2022ylj}.

The main motivation for the novel approach to PM gravity proposed here is that the just-defined HEOB 
formalism based on such PM-based Hamiltonians obtained by recursively solving the EOB 
mass-shell condition, Eqs.~\eqref{eq:mass_shell_condition} or Eq.~\eqref{massshell2},
have several unsatisfactory features. First, the combination of the intricate $\gamma$-dependence 
of the PM coefficients with the recursive steps required to get an explicit  Hamiltonian at a given PM 
accuracy  makes the final expression of the Hamiltonian as a function of the phase space variables, 
$H_{{\rm eff}, n{\rm PM}}(r,p_r,p_\varphi)$, quite involved and difficult to handle 
when solving (for $n \geq 4$) the associated EOB equations of motion, especially when considering radiation-reacted dynamics.
Second, the Hamiltonians defined by recursively solving 
the mass-shell condition  involve (at high PM orders) both positive {\it and negative} powers 
of the (rescaled, squared) Schwarzschild Hamiltonian [using $p_\lambda \equiv P_\lambda/\mu$, $r \equiv R/(GM)$]
\begin{equation}
\hat{H}_S^2   \equiv\frac{H^2_{{\rm eff},S}}{\mu^2}	= (1-2u) \left[1 + (1-2u) p_r^2 + p_\varphi^2u^2\right] \ .
\end{equation}
Focusing for simplicity on the orbital dynamics, this implies that the 4PM squared radial potential 
$V^2_{\rm eff}(r,j) \equiv H^2_{{\rm eff},4 {\rm PM}}(r,p_r=0,p_\varphi=j)$ used in Ref.~\cite{Buonanno:2024byg}
 is a $\nu$-deformation of the test-particle squared  radial potential 
\be \label{VS2}
V^2_S(r,j) =  \left(1-\frac2{r} \right) \left[1 + \frac{j^2}{r^2}\right]\,,
\ee
with two unsatisfactory features: (i) it is anchored on the Schwarzschild horizon $ r=2$ (i.e. $R= 2GM$) and (ii)
instead of vanishing at $r=2$ it is singular there, namely 
\begin{equation} \label{VSEOBPM2}
	V^2_{\rm eff, SEOB-PM}(r,j) \overset{r \to 2}{\approx} -\frac{5 \nu\bigg(1+j^2/r^2\bigg)}{256(1-2\nu)^{3/2}\hat{H}_S^9}\,.
\end{equation}
The presence of this singularity was already pointed out in Appendix B of the Supplemental Material of Ref.~\cite{Buonanno:2024byg}, 
where the authors mention that it might limit the flexibility that can be gained from incorporating high order calibration parameters
in the EOB Hamiltonian.

While it is possible that this divergent behavior around $r=2$ could be alleviated by defining some \textit{ad hoc} resummation of the 
PM Hamiltonian, any resummation is likely to produce a rather intricate final expression for the Hamiltonian. The LEOB formalism 
presented in the next section will allow us to circumvent both problematic issues by removing entirely the need to solve the 
mass-shell condition for defining a Hamiltonian.  

\begin{figure}[t]
	\center	
	\includegraphics[width=0.475\textwidth]{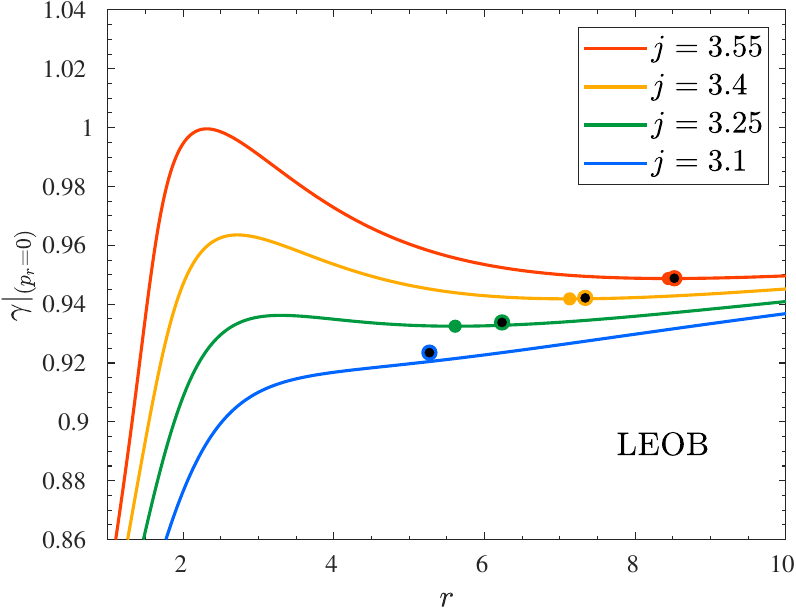}\\	
	\includegraphics[width=0.475\textwidth]{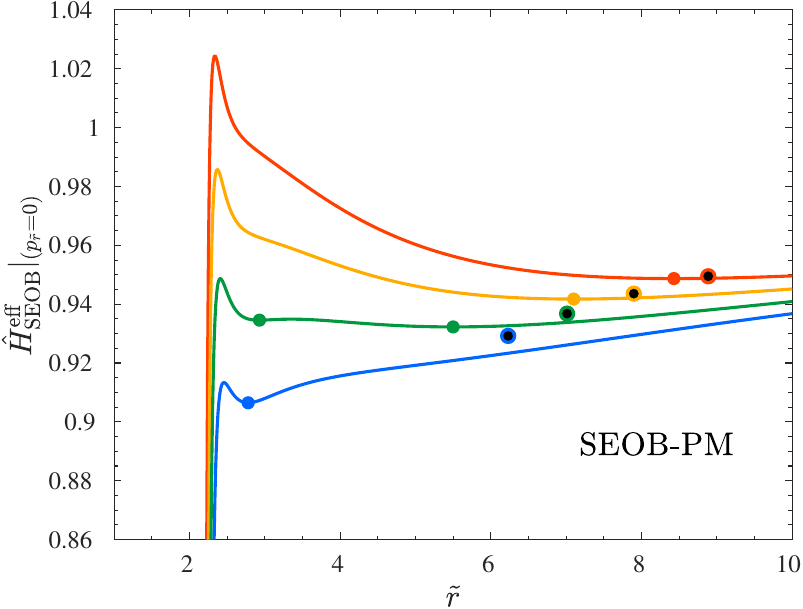}
	\caption{\label{Fig:ps_khalil}Radial potentials for inspiraling orbits against the associated radial coordinate for the PM-informed orbital dynamics determined by our new 
	LEOB approach (upper panel) and the SEOB-PM model of Ref.~\cite{Buonanno:2024byg} (bottom panel). 
	For each curve, obtained at different fixed values of the angular momentum $j$, the colored dots indicate the location of the 
	stable circular orbits.
	We also show, as black dots, the actual radial position of the system as it moves along 
	the full radiation-reacted dynamics. See text for more details.}
\end{figure}

To further motivate our change of paradigm for incorporating PM information into the EOB dynamics, we compare in Fig.~\ref{Fig:ps_khalil} 
the radial potentials~\footnote{As above, each radial potential is defined as the orbital effective energy  
at zero radial momentum versus the radial coordinate, for a given value of $j$. In other words, they define the set of turning points in the $(r, \g)$ plane at fixed $j$.}
 computed with our new LEOB approach (top panel) and with the latest HEOB take on PM gravity~\cite{Buonanno:2024byg} (bottom panel).
 The radial potentials are plotted against the corresponding radial coordinate, for different fixed values of the angular momentum $j$
  (indicated in the inset). For each value of $j$ we also show, with black dots, the actual position of the effective body subjected 
  to the full (i.e.~with $p_r\neq0$) radiation-reacted dynamics, determined by using, for simplicity, the Newtonian radiation reaction forces
 \begin{subequations}
 \begin{align}
 	&\hat{{\cal F}}_\varphi^N =\frac{8 \nu  p_r^2 p_\varphi}{5 r^3}-\frac{16 \nu  p_\varphi^3}{5 r^5}-\frac{16 \nu  p_\varphi}{5 r^4}\,, \\
 	&\hat{{\cal F}}_r^N =\frac{56 \nu  p_r p_\varphi^2}{5 r^5}-\frac{8 \nu  p_r}{15 r^4}\,.
 \end{align}
 \end{subequations}
 Finally, the colored dots in each curve highlight the positions of the respective local minima, corresponding to stable circular orbits. 
 The LEOB radial potentials are very similar to the radial potentials defined by the usual PN-based EOB Hamiltonians. They have 
 a single maximum (unstable circular orbit), and a single minimum (stable circular orbit), and are
 qualitatively similar to the probe radial potential $V_S(r,j)$, defined by Eq.~\eqref{VS2}. By contrast, the {\tt SEOB-PM} radial potentials
 are rather dissimilar to $V_S(r,j)$. In particular, they exhibit, for some intermediate values of $j$ (see the green curve),
 two maxima and two minima, and have a strongly singular behaviour around $r_{\rm SEOB}=2$, in keeping with Eq.~\eqref{VSEOBPM2}.

\subsection{LEOB approach}
\label{sec:leob}
The basic idea of the Lagrange-EOB approach is to replace the EOB Hamiltonian action
\begin{align}
	&S[X^\mu,P_\mu] = \int [P_\mu d X^\mu]^{\rm on-shell}\cr&\qquad= \int P_i d X^i - H_{\rm eff}(X^i, P_i) d T_{\rm eff} \ ,
\end{align}
where the superscript ``on-shell" means that one has solved the mass-shell condition for $P_0$ 
[to get $P_0^{\rm on-shell}= - H_{\rm eff}(X^i, P_i)$], by the action
\begin{equation}  \label{Laction}
	S[X^\mu,P_\mu,e_{\rm L} ] = \int P_\mu d X^\mu -   e_{\rm L}\, \mathcal{C}\left(X^\mu,P_\mu\right) d \tau \ .
\end{equation}
Here, $\mathcal{C}\left(X^\mu,P_\mu\right)$ denotes the EOB mass-shell condition~\eqref{eq:mass_shell_condition}
(or any equivalent condition), while $e_{\rm L}$ is a Lagrange multiplier. The choice of normalization of $e_{\rm L}$
determines a corresponding normalization for the evolution parameter $\tau$. For instance, the choice 
$e_{\rm L}= \frac12$ (imposed after variation) corresponds to using the proper time when considering
a geodesic constraint  $\mathcal{C}\left(X^\mu,P_\mu\right)=  g^{\mu \nu}(X^\mu, P_\mu) P_\mu P_\nu + \mu^2$.

The (Euler-Lagrange) variational equations obtained from the action \eqref{Laction} read
\begin{subequations}
\begin{align}
	\label{ELX}
	\frac{d X^\mu}{d\tau} &= e_{\rm L}\, \frac{\partial \mathcal{C}}{\partial P_\mu}\,, \\
	\label{ELP}
	\frac{d P_\mu}{d\tau} &= -e_{\rm L}\, \frac{\partial \mathcal{C}}{\partial X^\mu}\,, \\
	\label{ELC}
	\mathcal{C} &= 0\,.	
\end{align}
\end{subequations}
When going beyond the conservative dynamics,  these equations of motion are generalized 
by adding a radiation-reaction force $\mathcal{F}_\mu$ to Eq.~\eqref{ELP}, i.e.~by replacing it by
\begin{equation}
\label{ELPrr}
\frac{d P_\mu}{d\tau} = -e_{\rm L}\, \frac{\partial \mathcal{C}}{\partial X^\mu} + \mathcal{F}_\mu\,,
\end{equation}	
with the condition
\begin{equation}
	\label{Fcond}
	\frac{d X^\mu}{d \tau} \mathcal{F}_\mu=0 \ ,
\end{equation}
ensuring the preservation of the constraint $\mathcal{C} = 0$ along the radiation-reacted evolution.

In practice,  it is usual to work with a dimensionless rescaled action $\hat S  = \frac{S}{G m_1 m_2}$,
corresponding to dimensionless rescaled dynamical variables
 $ x^\mu = \frac{X^\mu}{G M}$ and  $ p_\mu = \frac{P_\mu}{\mu}$.  As already mentioned, we have
  $p_0  \equiv - \g$, where the so-defined energy variable $\g$ generalizes the Lorentz factor of conservative
 scattering motions in two ways. First, $\g$ is no longer constant under the evolution, but varies because
 of the inclusion of the radiation-reaction force $\mathcal{F}_\mu$, and, second, the $\g$ dependence
 of the scattering mass-shell condition must be analytically continued from the $\g >1$ domain, to
 the bound-state domain $ \g <1$. Note also again the link between
 the rescaled effective energy  $\g = -p_0 $ and the $M$-rescaled real cm energy  
 \be  \label{h2}
 h(\g,\nu) \equiv \frac{E}{M} = \sqrt{1+ 2 \nu (\g-1)} \ .
 \ee
 
 For notational simplicity, we
 will drop the hats on $S$, $e_{\rm L}$, $\mathcal{C}$, $\tau$, and $\mathcal{F}_\mu$, and write
\begin{equation}
	S[x^\mu,p_\mu,e_{\rm L} ] = \int \left[p_\mu \frac{d x^\mu}{d \tau} - e_{\rm L}\, \mathcal{C}\left(x^\mu,p_\mu\right)\right]d\tau \ ,
\end{equation}

\begin{subequations}
\begin{align}
	\label{eq:Lagrange_eqs_x}
	\frac{d x^\mu}{d\tau} &= e_{\rm L}\, \frac{\partial \mathcal{C}}{\partial p_\mu}\,, \\
	\label{eq:Lagrange_eqs_p}
	\frac{d p_\mu}{d\tau} &= -e_{\rm L}\, \frac{\partial \mathcal{C}}{\partial x^\mu}+ \mathcal{F}_\mu\,, \\
	\label{eq:Lagrange_eq_C}
	\mathcal{C} &= 0 \ ,	
\end{align}
\end{subequations}
with 
\begin{equation}
		\label{eq:constraint_gpp}
		\mathcal{C} = g^{\mu \nu}(x^\lambda) p_\mu p_\nu + 1 + Q(x^\mu,p_\mu)\,,
\end{equation}
or
\begin{equation}
		\label{eq:constraint_gppp}
		\mathcal{C} = g_*^{\mu \nu}(x^\lambda,p_0) p_\mu p_\nu + 1 \,,
\end{equation}
and
\begin{equation}
	\label{eq:fluxes_eq}
	\frac{d x^\mu}{d \tau} \mathcal{F}_\mu = 0\,.
\end{equation}
		
It is also convenient, in practice, to fix the Lagrange multiplier  by requiring the parameter $\tau$ to be the 
(rescaled) ``real'' time $t_{\rm real}=  \frac{T_{\rm real}}{GM}$  associated with the dynamical evolution of the binary system in the cm frame, rather than the effective time $ x^0= t_{\rm eff}$ entering the EOB dynamics. 
Using $\tau = t_{\rm real}$ in Eq.~\eqref{eq:Lagrange_eqs_x} for $\mu=0$, together with
$ x^0= t_{\rm eff}$, yields
\be \label{Lx0}
\frac{d x^0}{d\tau} =\frac{d t_{\rm eff}}{dt_{\rm real}} = e_{\rm L}\, \frac{\partial \mathcal{C}}{\partial p_0}= -  e_{\rm L}\, \frac{\partial \mathcal{C}}{\partial \g} \ .
\ee
 On the other hand, $ t_{\rm real}= \frac{T_{\rm real}}{GM}$ is related to  $t_{\rm eff}= \frac{T_{\rm eff}}{GM}$  via $ dT_{\rm eff} \wedge  dE_{\rm eff}= dT_{\rm real} \wedge  dE_{\rm real}$,
which implies, using Eq.~\eqref{h2},
\be
\frac{d t_{\rm eff}}{d t_{\rm real}}= \frac{d E_{\rm real}}{d E_{\rm eff}}= \frac1{\nu} \frac{dh}{d\g}=\frac{1}{h}.
\ee
Comparing the latter equation with the zeroth equation of motion \eqref{Lx0} yields the following
condition on the Lagrange multiplier
\begin{equation}
		e_{\rm L} = -\frac{1}{h} \left(\frac{\partial \mathcal{C}}{\partial \gamma}\right)^{-1}\,.
\end{equation}
We now have everything we need to solve the equations of motion above within the EOB approach.
In practice it is not necessary to solve all of them. 
The constraint equation \eqref{eq:Lagrange_eq_C} is only necessary to determine the initial conditions, since it will remain satisfied
along the full evolution because of the condition \eqref{eq:fluxes_eq}.\footnote{It is, however, useful to control the accuracy of our
numerical integration by checking that $\mathcal{C}$ stays zero during the evolution.}
The latter condition determines $\mathcal{F}_0$
in terms of $\mathcal{F}_i$:
\be \label{F0}
\mathcal{F}_0= - \frac{d x^i}{d t_{\rm eff}}\mathcal{F}_i =  - h\frac{d x^i}{d t_{\rm real}}\mathcal{F}_i \ .
\ee
We then end up with only seven first-order evolution equations for the seven variables, 
$x^i, p_i$ and $p_0 = - \g$, namely
\begin{subequations}
	\begin{align}
	\label{eq:eob_el_xi}
\frac{d x^i}{d t_{\rm real}} &= -\frac{1}{h} \left(\frac{\partial \mathcal{C}}{\partial \gamma}\right)^{-1} \frac{\partial \mathcal{C}}{\partial p_i}\,, \\
\label{eq:eob_el_pi}
\frac{d p_i}{d t_{\rm real}} &= \frac{1}{h} \left(\frac{\partial \mathcal{C}}{\partial \gamma}\right)^{-1} \frac{\partial \mathcal{C}}{\partial x^i} + \mathcal{F}_i\, ,\\
\label{eq:eob_el_gamma}
\frac{d \gamma}{d t_{\rm real}} &=-\mathcal{F}_0\,,	
\end{align}	
\end{subequations}
where $\mathcal{F}_0$ must be replaced from Eq. \eqref{F0},
and where we used the fact that the mass-shell constraint $\mathcal{C}$ has no explicit time dependence.

Given some explicit mass-shell constraint $\mathcal{C}(x^i, p_i, \g)$,
and some explicit spatial radiation-reaction force $\mathcal{F}_i(x^i, p_i, \g)$,
these equations provide a complete description, within the EOB framework, of the dynamical evolution of the binary system 
as described by the seven variables $(x^i, p_i, \g)$.

The traditional Hamiltonian EOB dynamics is regained when using a mass-shell constraint 
 in the explicitly solved form
\begin{equation}
\label{eq:cH}
	\mathcal{C}_H  \equiv \hat{H}_{\rm eff}(x^i,p_i) - \gamma = 0\,,
\end{equation}
which implies 
\begin{subequations}
\begin{align}
	&\frac{\partial \mathcal{C}}{\partial \gamma} = -1\ ,\\
	 &\frac{\partial \mathcal{C}}{\partial p_i} = \frac{\partial \hat{H}_{\rm eff}}{\partial p_i} \ , \\
	 &\frac{\partial \mathcal{C}}{\partial x^i} = \frac{\partial \hat{H}_{\rm eff}}{\partial x^i} \ .
\end{align}
\end{subequations}
Inserting these expressions in Eqs.~\eqref{eq:eob_el_xi}-\eqref{eq:eob_el_gamma} yields the 
usual EOB Hamilton equations plus an explicit evolution equation for  $\hat{H}_{\rm eff}$.

Summing up, our  Lagrange-EOB approach avoids the need to solve the mass-shell constraint for $p_0 = -\g$
(and the ensuing problematic issues discussed above) at the small cost of having one 
additional evolution equation \eqref{eq:eob_el_gamma} for  $\gamma$. 
As will be discussed below, the simplification and flexibility brought about by this novel approach have a notable impact on 
the development of PM-based EOB models built upon the analytical information that is currently available. 
We expect even more benefits when  higher-order PM-perturbative results will become available.
In addition, though our approach is motivated by the intricate structure of PM results, it can be directly used
when completing PM results by PN-based ones (as we shall do below), and also when using only PN-based results.

\subsection{Gauge flexibility}
\label{sec:gauge}

As already mentioned, the LEOB approach makes it easy to work with  a geodesic-like EOB mass-shell condition involving
an energy-dependent metric. For non-spinning bodies, this means (in rescaled form),
 \begin{equation}
    	\label{massshell3}
   \mathcal{C} = g^{\mu \nu}(x, \g) p_\mu p_\nu + 1 \,,
    \end{equation}
while the aligned-spin mass-shell condition, with dimensionless rescaled variables,
\be \label{eq:rescaled_var}
r_c\equiv \frac{R_c}{GM}, \, \, \tilde{a}_i = \frac{a_i}{GM},\, \, \hat{S} = \frac{S}{GM^2}, \, \, \hat{S}_* = \frac{S_*}{GM^2},
\ee
(where one should distinguish $\tilde{a}_i = \frac{a_i}{GM}$ from $\chi_i= \frac{a_i}{G m_i}$)
 would take the general form
\begin{equation}
	\label{eq:mass_shell_cond_m_rescaled}
	-\frac{[\gamma- \mathcal{G}(r,\gamma,\tilde{a}_i)p_\varphi]^2}{A(r,\gamma,\tilde{a}_i)}+\frac{p_r^2}{B(r,\gamma,\tilde{a}_i)}+\frac{p_\varphi^2}{r_c(r,\gamma,\tilde{a}_i)^2}+1=0\,,
\end{equation}
where
\be
\label{eq:GSresc}
{\mathcal G}= G_S(r, \g, \tilde{a}_i) \hat{S} +  G_{S_*}(r, \g, \tilde{a}_i) \hat{S}_*\,.
\ee

The rescalings in Eq.~\eqref{eq:rescaled_var} incorporate the fact that black hole spins $S_i$ are proportional to $G$. 
Throughout this paper, we shall count the PM order of any given quantity by taking into account the scaling $S_i \propto G$, referring to it as the \emph{physical} PM order. There exists also a different way of counting the PM order for spinning bodies, which we shall
call the \emph{formal} PM order, where the
proportionality $S_i \propto G$ is ignored.\footnote{The formal PM order is useful to make contact with works that count loop orders (e.g. Refs.~\cite{Jakobsen:2022fcj,Jakobsen:2023hig}), since a formal $n$PM contribution corresponds to a $(n-1)$-loop correction.} 
For instance, a term in the energy  $\sim G^3 M^2 S^2/R^5$ is of formal 3PM order and physical 5PM order.

In PM gravity, the functions  $(A,B,\mathcal{G},r_c)$ admit general $G$-expansions.
In the even-in-spin sector, they take the form
\begin{subequations}
\begin{align}
	\label{eq:A_pm}
	&A^{\rm PM}(r, \g, \tilde{a}_i,\nu)=1+\sum_{n \geq1} \frac{a_n(\g,\tilde{a}_i,\nu)}{r^n},\\
	&B^{\rm PM}(r, \g, \tilde{a}_i,\nu)=1+\sum_{n \geq1} \frac{b_n(\g,\tilde{a}_i,\nu)}{r^n},\\
	&\left[r_c^{\rm PM}(r, \g, \tilde{a}_i,\nu)\right]^2=r^2 \left[1 +\sum_{n \geq2} \frac{r_{c,n}(\g,\tilde{a}_i,\nu)}{r^n} \right],
\end{align}
\end{subequations}
where the  coefficients $X_n(\g,\nu,\tilde{a}_i)$ are even functions of the individual spin variables $\tilde a_1$ and $\tilde a_2$.

The index $n$ in the above coefficients $X_n(\g,\nu,\tilde{a}_i)$ labels the physical PM order. 
Remembering that $\frac1r= \frac{GM}{R}=O(G)$, and that each black hole spin parameter $a_i = G m_i \chi_i$ (which contains a factor $G$)
 has the dimension of a length,  and therefore enters in
the dimensionless combination $\frac{a_i}{R}= \frac{G m_i \chi_i}{R}$ which is both $O(G^1)$ and $O(\frac1r)$, we see that
the index $n$ counts at the same time all the powers of $\frac1r$ and all the powers of $G$ including those present in the spins. 
By contrast, the formal PM order would be obtained by subtracting the exponent of $a_i$ from the exponent of $\dfrac1r$. 
For instance, the probe black-hole value of $\left[r_c^{\rm PM}\right]^2$ reads
\begin{align}
	\left[r_c^{\rm PM}\right]^2_{\rm probe BH}&= r^2 + \tilde{a}_2^2 \left(1+ \frac2{r}\right)\cr&=r^2\left[ 1+ \left(\frac{ \tilde{a}_2}{r}\right)^2\left(1+ \frac2{r} \right)   \right],
\end{align}
where the term $ \big(\frac{ \tilde{a}_2}{r}\big)^2$ is a physical 2PM correction, and the term 
$ \big(\frac{ \tilde{a}_2}{r}\big)^2   \frac2r$ is a physical 3PM correction.

In the odd-in-spin sector, the $G$-expanded gyro-gravitomagnetic functions read
\begin{subequations}
\begin{align}
	 G_S^{\rm PM}(r, \g, \tilde{a}_i,\nu) &=\frac{1}{r \, r_c^2(a_K)} g_S\left(u,\gamma,\tilde{a}_i\right)  \ ,\\
	 G_{S_*}^{\rm PM}(r, \g, \tilde{a}_i,\nu) &=\frac{1}{r\, r_c^2(a_K)}g_{S_*}\left(u,\gamma,\tilde{a}_i\right)\,, 
\end{align}
\end{subequations}
where the effective Kerr parameter $a_K$ will be chosen below and where the coefficients $g_S$ and $g_{S_*}$ 
are expanded as a series in $u \equiv \frac1r$, with coefficients that are polynomials
in $\tilde{a}_i $, namely,
\begin{subequations}
	\begin{align}
	\label{eq:gS}
		&g_S\left(u,\gamma,\tilde{a}_i\right) =  g_0 (\g, \tilde{a}_i,\nu) + g_1 (\g, \tilde{a}_i,\nu) u \nonumber \\
		&\quad+ g_2 \left(\gamma,\tilde{a}_i,\nu\right) u^2 + g_3 \left(\gamma,\tilde{a}_i,\nu\right) u^3\,, \\
        \label{eq:gSstar}
	&g_{S_*}\left(u,\gamma,\tilde{a}_i\right)= g_{*0}(\g, \tilde{a}_i,\nu) + g_{*1} (\g, \tilde{a}_i,\nu) u \nonumber \\
	&\quad+ g_{*2} \left(\gamma,\tilde{a}_i,\nu\right) u^2 + g_{*3} \left(\gamma,\tilde{a}_i,\nu\right) u^3\,,
	\end{align}
\end{subequations}
where the $g_n(\g, \tilde{a}_i,\nu)$ and $g_{*n}(\g, \tilde{a}_i,\nu)$ parameters are even functions of the spins.
In this case the $n$ subscript indicates the relative PM order with respect to the leading-order spin-orbit couplings $g_0(\g, \tilde{a}_i,\nu)$ and $g_{*0}(\g, \tilde{a}_i,\nu)$.
As for the absolute PM order, the probe black-hole value of the spin-orbit coupling is given by  ${\mathcal G}(R, a_2, L)= \frac{2 G L S_2}{R R_c^2}$,
Eq. \eqref{calGprobe}, which starts at the formal 1PM order, but [because $S_2=O(G)$] at the physical 2PM order.
Consequently, each $g_n$ and $g_{*n}$ term will be at physical $(n+2)$PM order.

As usual \cite{Damour:2016gwp,Damour:2017zjx}, these PM coefficients are constrained by enforcing that the PM-expansion of the real cm scattering angle (with $j \equiv p_\varphi$), 
\begin{equation}
	\label{chiPM}
	\chi^{\rm PM}\left(j,\gamma,\tilde{a}_i\right) = \sum_{n} 2 \frac{\chi_n(\gamma,\tilde{a}_i,\nu)}{j^n}\,,
\end{equation}
coincides with the PM-expansion of the EOB scattering angle, obtained by PM expanding the integral
\begin{equation}
	\label{eq:chi_equation}
	\chi^{\rm EOB} = -\pi-2\int^{+\infty}_{r_{\rm min}}  dr \, \dfrac{\partial }{\partial p_\varphi}p_r(\gamma,p_\varphi,r,\tilde{a}_i)\,,
\end{equation}
where 
\begin{equation} \label{eq:pr_inverted}
p_r(\gamma,p_\varphi,r,\tilde{a}_i)= \sqrt{B \left(-1+\frac{\gamma_{\rm orb}^2}{A}-\frac{p_\varphi^2}{r_c^2}\right)}\,,
\end{equation}
with 
\be
\g_{\rm orb} \equiv \g - p_\varphi \mathcal{G}\,.
\ee
In Eq.~\eqref{eq:chi_equation}, $r_{\rm min}$ denotes the largest real root of $p_r(\gamma,p_\varphi,r,\tilde{a}_i)=0$. 
In this computation, one assumes that $\g >1$. 

Let us separate even and odd-in spin contributions in the PM coefficients of the scattering angle, i.e.
\begin{equation}
	\chi_n(\gamma,\tilde{a}_i,\nu) = \chi_n^{\rm even}(\gamma,\tilde{a}_i,\nu) + \chi_n^{\rm odd}(\gamma,\tilde{a}_i,\nu)\,.
\end{equation}
In general, considering that all the PM coefficients of the EOB functions appearing in Eq.~\eqref{eq:pr_inverted} are by construction even in the spins, the matching procedure described above yields, for any given physical PM order, say the $n$th, a relation connecting the even-in-spin scattering angle coefficient $\chi^{\rm even}_n(\gamma,\nu,\tilde{a}_i)$
to the PM coefficients $a_n(\g, \tilde{a}_i,\nu)$,  $b_n(\g, \tilde{a}_i,\nu)$, $r_{c,n}(\g,\tilde{a}_i,\nu)$ and another connecting its odd-in-spin counterpart, $\chi^{\rm odd}_n(\gamma,\nu,\tilde{a}_i)$, to $g_{(n-2)}(\g, \tilde{a}_i,\nu)$, $g_{*(n-2)}(\g, \tilde{a}_i,\nu)$ (and previously determined coefficients). 

Let us show explicitly the first few iterations.
At physical 1PM order there is no spin dependence in the scattering angle. 
Introducing the usual notation (for $\g >1$)
\be
p_\infty \equiv \sqrt{\gamma^2-1}\,.
\ee
the physical 1PM relation hence reads
\begin{equation}
	\label{eq:chipotcoef_1}
	\chi_1 (\gamma,\nu) = \frac{p_\infty }{2}b_1(\gamma,\tilde{a}_i,\nu) - \frac{\gamma^2}{2p_\infty}a_1(\gamma,\tilde{a}_i,\nu)\,,
\end{equation}
to be compared with the well-known result
\be
\chi^{\rm 1PM}_1 (\gamma)= \frac{2 \g^2-1}{\sqrt{\gamma^2-1}}= \frac{2 \g^2-1}{\pinf}\,.
\ee

Starting from the physical 2PM-level, we have both even and odd-in-spin contributions to the scattering angle, $\chi_2 = \chi_2^{\rm even} + \chi_2^{\rm odd}$.

The physical 2PM even-in-spin contribution to $\chi$ is purely orbital and reads
	\begin{align}
		\label{eq:chipotcoef_2}
		&\chi_2^{\rm even} (\gamma,\nu) = -\frac{1}{8} \pi  \gamma ^2 a_1\left(\gamma,\tilde{a}_i,\nu \right) b_1\left(\gamma,\tilde{a}_i,\nu \right)\cr
		&\quad -\frac{1}{32} \pi  \pinf^2 b_1\left(\gamma,\tilde{a}_i,\nu \right)^2+\frac{1}{8} \pi  \pinf^2 b_2\left(\gamma,\tilde{a}_i,\nu \right)\cr
		&\quad +\frac{1}{8} \pi  \pinf^2 r_{c,2}\left(\gamma,\tilde{a}_i,\nu \right) +\frac{1}{4} \pi  \gamma ^2 a_1\left(\gamma,\tilde{a}_i,\nu \right)^2\cr
		&\quad-\frac{1}{4} \pi  \gamma ^2 a_2\left(\gamma,\tilde{a}_i,\nu \right),
	\end{align}
while the odd-in-spin 2PM contribution to $\chi$ is
\begin{equation}
\chi_2^{\rm odd} (\gamma,\tilde{a}_i,\nu) = -\gamma\, \pinf \left[g_0\left(\gamma,\tilde{a}_i,\nu \right)S + g_{*0}\left(\gamma,\tilde{a}_i,\nu \right) S_* \right].
\end{equation}
The full relations up to the physical 5PM level (excluding the still undetermined orbital part) can be found in the ancillary
{\tt Mathematica} file attached to this paper.

As displayed in Eqs.~\eqref{eq:chipotcoef_1}-\eqref{eq:chipotcoef_2}, there is a large freedom in the determination of the coefficients 
$a_m(\g, \tilde{a}_i,\nu)$,  $b_m(\g, \tilde{a}_i,\nu)$, $r_{c,m}(\g,\tilde{a}_i,\nu)$, for $m\leq n$, as there are more unknowns
than equations. More precisely, at each PM order the full gauge-invariant information about the (conservative) dynamics is
contained in the single function $\chi_n(\g, \tilde{a}_i,\nu)$. Therefore, one is allowed to impose two {\it a priori} constraints among
the three metric functions $a_n(\g, \tilde{a}_i,\nu)$,  $b_n(\g, \tilde{a}_i,\nu)$, $r_{c,n}(\g,\tilde{a}_i,\nu)$ so as
to finally determine the PM expansion of the full effective metric from the sole knowledge of the $\chi^{\rm even}_n(\g, \tilde{a}_i,\nu)$'s.
This freedom is a gauge freedom which goes beyond the usual coordinate freedom of general relativity (GR). 
The physical origin of the larger-than-GR  freedom of the EOB dynamics is that the EOB mass-shell condition, 
which is equivalent to a Hamiltonian, admits the larger set of canonical transformations as gauge symmetries.
We discuss our particular gauge choices in the next section, while we defer to Appendix~\ref{app:gauge_freedom} 
a brief discussion of other possible choices (restricting the analysis to the nonspinning case for simplicity), including the Kerr-Schild-like gauge recently used in Ref.~\cite{Ceresole:2023wxg}.
Let us also note that, while in the present work we will use an energy-dependent geodesic-like gauge, Eq.~\eqref{massshell3}, 
the flexibility of the LEOB approach allows for hybrid gauges where the mass-shell condition involves an additional (Finsler-type) non-geodesic $Q$ contribution, say
\begin{align}
 \mathcal{C} = g^{\mu \nu}(x, \g) p_\mu p_\nu + 1 + \hat{Q}(x^\mu,p_\mu) \ ,
\end{align}
where $\hat{Q}(x^\mu,p_\mu)$ depends on $p_\mu$ in a non-quadratic way.

\subsection{Lagrange-Just-Boyer-Lindquist gauge}
\label{gauge-fixing}

In the present work we explore one particular way of fixing the gauge freedom of our LEOB formalism. 
The gauge we choose consists in imposing two algebraic constraints on the three metric coefficients $A$, $B$ and $R_c^2$.
On the one hand, we impose (similarly to a coordinate condition  introduced long ago by Kurt Just~\cite{Just:1959}, 
see also~\cite{Coquereaux:1990qs}, in scalar-tensor gravity)
the constraint that the product $ A B$ has the same value as for a Kerr metric (of mass $M_K=M=m_1+m_2$
and spin parameter $a_K= a_0\equiv a_1+a_2$) in Boyer-Lindquist coordinates.
On the other hand, we impose the further constraint that $R_c^2$ also takes the same value as for the latter Kerr metric (in Boyer-Lindquist coordinates). Explicitly, we work in a  ``Lagrange-Just-Boyer-Lindquist" (LJBL) gauge
defined by imposing the following two conditions
 \begin{subequations}
 		\label{eq:LJBL_gauge}
	\begin{align}
		\label{eq:LJBL_gauge_AB}
		 A(r,\g,\tilde{a}_i) \, B(r,\g,\tilde{a}_i)&= \frac{R^2}{R^2_{c \,\rm K}(M,a_0)} \,, \\
		\label{LJBL_gauge_rc}
		R_c^2(r,\g,\tilde{a}_i) &= R^2_{c \,\rm K}(M,a_0)\,,
	\end{align}
\end{subequations}
where  $a_K= a_0\equiv a_1+a_2$, and [see Eq.~\eqref{eq:rcK}],
\be
R^2_{c \,\rm K}(M,a_0)= R^2 + a_0^2 \left(1+ \frac{2GM}{R}  \right)\,.
\ee
In particular, these choices will ensure that our effective EOB metric reduces, at the 1PM-level, 
to the Kerr metric in Boyer-Lindquist coordinates. In the nonspinning case, the conditions above reduce
to imposing a Just-Schwarzschild gauge where $A B=1$ and $R_c^2=R^2$.

The first condition, Eq.~\eqref{eq:LJBL_gauge_AB}, has the advantage of discarding any reference 
to the Kerr event horizon. When combined with a suitable resummation of the function $A(r,\g,\tilde{a}_i)$, the
PM information contained in the scattering function $\chi(j,\g,\tilde{a}_i)$ will then automatically determine the
radial location of some effective horizon, as was happening in the traditional, PN-based EOB
formalism, especially within the Damour-Jaranowski-Sch{\"a}fer (DJS) gauge~\cite{Damour:2000we}.

When imposing the gauge conditions \eqref{eq:LJBL_gauge}, 
Eqs.~\eqref{eq:chipotcoef_1}-\eqref{eq:chipotcoef_2} and their higher-order equivalents simplify, 
leaving as unknowns just one set of coefficients, say $a_n$.
This allows us to obtain them recursively in terms of the PM coefficients $\chi_n^{\rm even}$.
The $\chi_n^{\rm odd}$ coefficients instead completely determine the $g_n$ and $g_{*n}$ ones, 
apart from some arbitrariness (discussed below) in collecting spin contributions beyond the linear in spin level.

Following these steps, and using the known values of $\chi _n^{\rm even}(\gamma,\tilde{a}_i,\nu)$~\cite{Bern:2021yeh,Bern:2021dqo,Dlapa:2021vgp,Manohar:2022dea,Dlapa:2022lmu,Bini:2022enm,Vines:2017hyw,Vines:2018gqi,Guevara:2019fsj,Kalin:2019inp,Kosmopoulos:2021zoq,Chen:2021kxt,Aoude:2022thd,Jakobsen:2022fcj,Bern:2020buy,Bern:2022kto,FebresCordero:2022jts}, 
we can obtain the explicit form of our PM-expanded $A$ metric potential.
Recalling the definition
of  $h(\g,\nu)=E/M$ given by Eq. \eqref{h2}, the first two PM orders read
\begin{subequations}
	\begin{align}
		&a_1(\gamma,\nu) =-2\, , \\
		&a_2(\gamma,\nu) = \frac{6 \nu}{h(h+1)} \frac{(\g-1) (5 \gamma ^2-1)}{ 3 \gamma ^2-1} \,.
	\end{align}
\end{subequations}

At the physical 3PM level, spin-spin interactions start to contribute, while spin-quartic terms enter at 5PM. 
We hence write
	\begin{subequations} 
\begin{align}
&a_{3} \left(\gamma,\tilde{a}_i,\nu\right) = a_{3}^{\rm 0}\left(\gamma,\nu\right) + a_{3}^{\tilde{a}^2}\left(\gamma,\tilde{a}_i,\nu\right)\,,\\
&a_{4} \left(\gamma,\tilde{a}_i,\nu\right) = a_{4}^{\rm 0}\left(\gamma,\nu\right) + a_{4}^{\tilde{a}^2}\left(\gamma,\tilde{a}_i,\nu\right)\,, \\
&a_{5} \left(\gamma,\tilde{a}_i,\nu\right) = a_{5}^{\rm 0}\left(\gamma,\nu\right) + a_{5}^{\tilde{a}^2}\left(\gamma,\tilde{a}_i,\nu\right) + a_{5}^{\tilde{a}^4}\left(\gamma,\tilde{a}_i,\nu\right)\,.
\end{align}
	\end{subequations}
Introducing the spin combinations
\begin{align}
	\tilde{a}_0 &= \frac{a_1+a_2}{G M}=\tilde{a}_1 + \tilde{a}_2 =  \hat{S}+\hat{S}_*\ , \\
	\tilde{a}_{12} &=  \frac{a_1-a_2}{G M}=\tilde{a}_1 - \tilde{a}_2=\frac{\hat{S}-\hat{S}_*}{X_{12}} \ ,
\end{align}
where $X_{12} \equiv X_1-X_2\equiv (m_1-m_2)/M \equiv \sqrt{1-4\nu}$, we report here, for illustration, the 3PM results (obtained using the conservative contribution to $ \chi _3$)
\begin{widetext}
\begin{subequations}
\begin{align}
\label{eq:a3final_a}
&a_{3}^{\rm 0}\left(\gamma,\nu\right) = \frac{\nu }{4 \gamma ^2-1}\left\{\frac{56 \gamma  \left( \gamma ^2+\frac{25}{14}\right)}{h^2}-\frac{144 \left(5 \gamma ^2-1\right) \left( \gamma ^4- \gamma
   ^2+\frac{1}{8}\right)}{h
   (h+1)(\gamma +1) \left(3 \gamma ^2-1\right)}+\frac{96 \left[\gamma ^2 \left(\gamma ^2-3\right)-\frac{3}{4}\right] {\rm arcsinh \sqrt{\frac{\gamma-1}{2}}}}{h^2 \sqrt{\gamma ^2-1}}\right\} \,,\\
   \label{eq:a3final_b}
  &a_{3}^{\tilde{a}^2}\left(\gamma,\tilde{a}_i,\nu\right) = 3\frac{1-2 \gamma ^2}{1-4 \gamma ^2} \Bigg\{\tilde{a}_0^2 \left[1-\frac{2}{(\gamma +1) h^2}\right]
   -\frac{2 (\gamma -1)}{(\gamma +1) h (h+1)}
    \Bigg[\frac{1+2 \gamma}{1-2 \gamma ^2} \tilde{a}_0 
   			\left(\tilde{a}_0- \frac{\sqrt{1-4 \nu }\,\tilde{a}_{12}}{h}\right)\cr
   			&\quad
   			+\frac{\left(\tilde{a}_0+ \sqrt{1-4 \nu }\,\tilde{a}_{12}\right)^2}{2
   			h(h+1)}\Bigg]\Bigg\}\,.
\end{align}
\end{subequations}
\end{widetext}
The above mathematical expression of $a_{3}^{\rm 0}\left(\gamma,\nu\right)$ involves the function $f_3(\gamma)={\rm arcsinh}\sqrt{\frac{\gamma-1}{2}}/\sqrt{\gamma^2-1}$ which is originally defined for unbound motions ($\gamma>1$).  This function admits a unique analytic continuation to bound motions ($\gamma<1$), namely ${\rm arcsin}\sqrt{\frac{1-\gamma}{2}}/\sqrt{1-\gamma^2}$ [indeed $f_3(\gamma)=\frac{1}{2}-\frac{x}{12}+\frac{3 x^2}{80}+\cdots$ is analytic in $x\equiv\gamma^2-1$ around $x=0$]. 
The 4PM-level metric coefficient $a_4(\gamma,\tilde{a}_i,\nu)$ involves tail effects which introduce  a non-trivial continuation 
from unbound to bound motions. This will be discussed separately in Sec.~\ref{sec:PMstuff}.
The explicit expressions of the $a_n(\gamma,\tilde{a}_i,\nu)$'s are given in electronic form in an
ancillary file attached to this work.
The functions $a_n(\gamma,\nu)$'s for $n \geq2$ contain  pole singularities located~\footnote{For higher values of $n$ there 
appear pole singularities also at $\g^2= \frac14$, $\g^2= \frac15$, and so on.} at $\g^2= \frac13$. 
These poles will have no impact in our Lagrange-EOB approach because though $\g$ decreases during the LEOB evolution,
it will never end up as low as $\frac{1}{\sqrt{3}} \approx 0.57735$.

In the probe black-hole limit $m_2 \to 0$, implying $\nu\to0$ (or equivalently $h\to1$) and  $\tilde a_2 \to 0$
(so that $\tilde{a}_{12} = \tilde{a}_0 = \tilde a_1$), the effective metric  coincides (in agreement with the general result of Ref.~\cite{Damour:2016gwp}) with the Kerr metric with $M_K=m_1$ and $a_K=a_1$. In particular, the $A$ potential
tends to [with $u_c^2(a_K) = \frac1{r_c^2(a_K)}=\frac{u^2}{1+ a_K^2 u^2(1+2 u)}$]
\begin{align}
 A(r,\g,\tilde{a}_i) \overset{\rm probe \, BH}{=} &\frac{1 - 4 u_c^2(a_1)}{1+2 u} \nonumber \\
= &1-2u+4 a_1^2 u^4 - 4 a_1^4 u^6+\cdots
\end{align}
In particular, the 3PM spin-square coefficient $a_{3}^{\tilde{a}^2}$ in Eq.~\eqref{eq:a3final_b}
above vanishes (because of several cancellations) in the probe BH limit.

Repeating the same steps for the odd-in-spin scattering angle, $\chi _n^{\rm odd}(\gamma,\tilde{a}_i,\nu)$~\cite{Bini:2017xzy,Bini:2018ywr,Vines:2018gqi,Guevara:2019fsj,Jakobsen:2023ndj,Jakobsen:2023hig}, 
we can compute our PM-expanded gyro-gravitomagnetic function 
\begin{equation}
	\mathcal{G}  =\frac{1}{r r_c^2(\tilde{a}_0)} \left[g_S\left(u,\gamma,\tilde{a}_i\right) \hat S + g_{S_*}\left(u,\gamma,\tilde{a}_i\right) \hat S_*\right] \ ,
\end{equation}
where the coefficients $g_S$ and $g_{S_*}$ are expanded as a series in $u \equiv \frac1r$, with coefficients that are polynomials
in $\tilde{a}_i $, namely
\begin{subequations}
	\begin{align}
		\label{eq:gSfact}
		&g_S\left(u,\gamma,\tilde{a}_i\right) =   g_0 \left(\gamma,\nu\right) \big[1 + \hat{g}_1 \left(\gamma,\nu\right) u \nonumber \\
		&\quad+ \hat{g}_2 \left(\gamma,\tilde{a}_i,\nu\right) u^2 + \hat{g}_3 \left(\gamma,\tilde{a}_i,\nu\right) u^3\big]\,, \\
		\label{eq:gSstarfact}
		&g_{S_*}\left(u,\gamma,\tilde{a}_i\right)= g_{*0} \left(\gamma,\nu\right) \big[1 + \hat{g}_{*1} \left(\gamma,\nu\right) u \nonumber \\
		&\quad+ \hat{g}_{*2} \left(\gamma,\tilde{a}_i,\nu\right) u^2 + \hat{g}_{*3} \left(\gamma,\tilde{a}_i,\nu\right) u^3\big]\,,
	\end{align}
\end{subequations}
where we defined $\hat{g}_n = g_n/g_0$ and $\hat{g}_{*n} = g_{*n}/g_{*0}$ [see Eqs.\eqref{eq:gS}-\eqref{eq:gSstar}].
We remind the reader that since the leading-order spin-orbit coupling starts at the physical 2PM level, 
and given that $\hat{g}_n$ and $\hat{g}_{*n}$ have been expressed as functions of the dimensionless spin parameters $\tilde{a}_i = a_i/(G M)$,
each $\hat{g}_n$ and $\hat{g}_{*n}$ is at physical $(n+2)$PM order.

Spin-cube interactions start at 4PM, with quadratic-in-spin terms entering $\hat{g}_2$ and $\hat{g}_{*2}$. 
At this spin-cube level, there is a choice on how to dispatch the four spin combinations 
$(a_1^3,a_1^2 a_{2},a_1 a_{2}^2,a_{2}^3)$ entering $\mathcal{G}$ among the two gyro-gravitomagnetic functions. 
We have chosen to impose that each spin-squared coefficient only involves the combinations $\tilde{a}_0^2$ and $\tilde{a}_{12}^2$. Therefore, the physical 4PM and 5PM contributions to $g_S\left(u,\gamma,\tilde{a}_i\right)$ read
\begin{subequations}
	\begin{align}
		\hat{g}_{2} \left(\gamma,\tilde{a}_i,\nu\right) &= \hat{g}_{2}^{\rm 0}\left(\gamma,\nu\right) + \hat{g}_{2}^{\tilde{a}_0^2}\left(\gamma,\nu\right) \tilde{a}_0^2 + \hat{g}_{2}^{\tilde{a}_{12}^2}\left(\gamma,\nu\right) \tilde{a}_{12}^2 \ ,\\
		\label{eq:g4_spin_struct}
		\hat{g}_{3} \left(\gamma,\tilde{a}_i,\nu\right) &= \hat{g}_{3}^{\rm 0}\left(\gamma,\nu\right) + \hat{g}_{3}^{\tilde{a}_0^2}\left(\gamma,\nu\right) \tilde{a}_0^2 + \hat{g}_{3}^{\tilde{a}_{12}^2}\left(\gamma,\nu\right) \tilde{a}_{12}^2 \,,
	\end{align}
\end{subequations}
and the same decomposition holds for $\hat{g}_{*2}$ and $\hat{g}_{*3}$.

The first few spin-orbit coefficients read
\begin{widetext}
\begin{subequations}
\begin{align}
	&g_0(\gamma,\nu) = \frac{1}{(1+h)\,\g\,(1+\g)}\left[1+2\g - \frac{1-2\g-4\g^2}{h}\right], \\
	&g_{*0}(\gamma,\nu) = \frac{1+2\g}{h \, \g \, (1+\g)}\,, \\
	&\hat{g}_1(\gamma,\nu) = \frac{\nu}{(1+h)\left[1-2\g-4\g^2 - h\left(1+2\g\right)\right]}\left[2\frac{1+2\g}{1+\g}\left(5\g^2-3\right) +\frac{3-15\g-15\g^2+35\g^3}{h}\right],\\
	&\hat{g}_{*1}(\gamma,\nu) = \frac{1}{(1-\g)}\left[\frac{5\g^2-3}{1+\g}+\frac{3}{2\, h}\left(\frac{1+2\g-5\g^2}{1+2\g}\right)\right],
\end{align}
\end{subequations}
while, for instance, the first cubic-in-spin (4PM) coefficients read
 \begin{subequations}
 \begin{align}
 		&\hat{g}^{\tilde{a}_0^2}_2\left(\gamma,\nu\right) = \frac{1}{2 (1+h)^2 \left[1-2\g-4\g^2 - h\left(1+2\g\right)\right]}
 		 \Bigg[h^3 (1+2 \gamma )+h^2 \left(1+6 \gamma +4 \gamma ^2\right)\cr
 		 &\quad-h \frac{3-5 \gamma -14 \gamma
 			^2-8 \gamma ^3}{1+\gamma }-\frac{7+11 \gamma -6 \gamma ^2-4 \gamma ^3}{(1+\gamma
 			)}+3\frac{1-6 \gamma }{h}+\frac{\left(1-2\g\right)\left(3+4\g^2\right)}{h^2}\Bigg] \,,\\
 		&\hat{g}^{\tilde{a}_{12}^2}_2\left(\gamma,\nu\right) = \frac{1+\g-2h^2}{2 h (1+h)^2 \left(1+\g\right)\left[1-2\g-4\g^2 - h\left(1+2\g\right)\right]} 
 		\Bigg[-3 (1+2 \gamma )+\frac{1-6 \gamma -8 \gamma ^2}{h}\Bigg] \,, \\
 		&\hat{g}^{\tilde{a}_0^2}_{*2}\left(\gamma,\nu\right) = \frac{1}{2 (1+h)^3 (1+2\gamma)}\Bigg[-h^2 (1+2 \gamma)\left(3+h\right)-h\frac{1+9 \gamma +6 \gamma ^2}{1+\gamma
 		}\cr
 		&\quad+\frac{(1-\gamma ) (5+2 \gamma )}{(1+\gamma )}+\frac{3 (3-2 \gamma
 			)}{h}+\frac{3+2 \gamma -4 \gamma ^2}{h^2} \Bigg] \,,\\
 		&\hat{g}^{\tilde{a}_{12}^2}_{*2}\left(\gamma,\nu\right) = \frac{1+\g-2h^2}{2 (1+\gamma ) h (1+h)^3}\Bigg[3+\frac{1+6 \gamma +4 \gamma ^2}{h (1+2\gamma )} \Bigg] \,.
\end{align}
\end{subequations}
\end{widetext}
The full list of coefficients up to 5PM can be found in the ancillary file.
We also point to Sec.~\ref{sec:nonlocal_4.5PN_SO} for a discussion on the time nonlocalities that appear in $\hat{g}_3^0(\g,\nu)$ and $\hat{g}_{*3}^0(\g,\nu)$.

In the PN limit, when $\g\rightarrow1$ or, equivalently, $p_\infty\rightarrow0$, the 2PM and 3PM spin-orbit coefficients read
\begin{subequations}
	\begin{align}
	&g_0(\gamma,\nu) \sim 2 - \frac{9}{4}\nu\, \pinf^2 + O\left(\pinf^4\right), \\
	&g_{*0}(\gamma,\nu) \sim \frac{3}{2} - \left(\frac{5}{8} + \frac{3}{2}\nu\right) \pinf^2 + O\left(\pinf^4\right), \\
	&\hat{g}_1(\gamma,\nu) \sim -\frac{7}{8}\nu - \left(\frac{35}{16} - \frac{81}{128}\nu\right)\nu \, \pinf^2 + O\left(\pinf^4\right), \\
	&\hat{g}_{*1}(\gamma,\nu) \sim -\frac{7}{6} - \nu + \left(\frac{1}{72} - \frac{5}{3}\nu + \frac{3}{4}\nu^2\right) \pinf^2 + O\left(\pinf^4\right).
\end{align}
\end{subequations}

In the probe BH limit, we reproduce the Kerr spin-orbit sector, with
\be
   G_S(m_2 \to 0) = \frac{2}{r_c^2\, r}\,,
\ee
and $S_*=G m_1 m_2(\chi_1+\chi_2) \to 0$, 
while $G_{S_*}$ instead is a complicated series in $u$ with $\g$-dependent coefficients, e.g. 
\be
G_{S_*}(m_2 \to 0,\g=1) \sim \frac{3}{2} u^3\left[1 - \frac{7}{6}u + O(u^2)\right] .
\ee

\section{Local and nonlocal conservative PM contributions within LEOB}
\label{sec:PMstuff}

As was pointed out long ago \cite{Blanchet:1987wq}, tail-transported effects generate nonlocal-in-time contributions
to the two-body dynamics that arise at order $O(G^4/c^8)$ (i.e at 4PM and 4PN). As a consequence, the conservative
part of the 4PM dynamics is described by an action made of two types of contributions: a local-in-time one, $S^{\rm loc}$,
and a (time-symmetric) nonlocal-in-time one, $S^{\rm nonloc}$ \cite{Foffa:2011np,Damour:2014jta,Damour:2015isa,Galley:2015kus,Bernard:2017bvn,Foffa:2019yfl,Bini:2019nra}. The Tutti Frutti approach to binary dynamics \cite{Bini:2019nra,Bini:2020wpo,Bini:2020nsb,Bini:2020hmy} has determined, in the nonspinning case, both $S^{\rm loc}$ and  $S^{\rm nonloc}$ up to the 6PN accuracy, modulo
a few unknown numerical coefficients parametrizing $O(\nu^3)$ effects. The local action is described by a local Hamiltonian
which is applicable both to scattering motions and to bound motions. By contrast, the nonlocal action takes quite different
explicit forms when considering scattering motions or bound motions. 

In the following we address this issue by computing the local 4PM component of the spin-independent part of the 
$A$ potential in LJBL gauge (Sec.~\ref{sec:4PM_local}), by completing it with local 5PM contributions at 4PN accuracy (Sec.~\ref{sec:4PN_Alocal}), and by adding nonlocal 4PN-accurate contributions specific to elliptic-like motions (Sec.~\ref{sec:nonlocal_4PN}). 
Finally, in Sec.~\ref{sec:nonlocal_4.5PN_SO}, we also discuss the time nonlocalities present in the 4PM spin-orbit sector of the dynamics.

For a discussion of the strategies adopted for tackling this issue in previous HEOB-based works, see Appendix~\ref{app:4PM_others}.

\subsection{Local part of the (nonspinning) orbital dynamics at the 4PM level}
\label{sec:4PM_local}
Recently, Ref.~\cite{Bini:2024tft} computed the 4PM-level [$O(G^4)$] local action  to the thirtieth order in velocities (i.e.~the 15PN accuracy)  by combining results of
the Tutti Frutti approach with the results of Refs.~\cite{Bern:2021yeh,Dlapa:2021vgp} and of 
Ref.~\cite{Dlapa:2024cje}. 
Ref.~\cite{Bini:2024tft} expressed the gauge-invariant content of the local 4PM action in two different ways: (i) the
local contribution to the 4PM scattering angle, $\chi_4^{\rm loc}$, and (ii) the local 4PM contribution to the
post-Schwarzschild $Q$ potential, i.e. $q_4^{\rm loc}$.  
In the present section, we use Eq.~\eqref{eq:a4fromQ} to translate the value of 
$q_4^{\rm loc}$ derived in  
Ref.~\cite{Bini:2024tft}  into a corresponding
value for the  local contribution $a_4^{\rm loc}$ to the orbital~\footnote{In the following, the word ``orbital'' refers to
the zero-spin limit.} component of the 4PM-accurate $A$ potential in the LJBL gauge.
 
The structure of $a^{\rm loc}_4(\g,\nu)$ reads
\bea \label{a4loc}
a_4^{\rm loc}(\g,\nu)&=&\frac{\nu}{h(h+1)}a_{4,1}(\gamma)+\frac{\nu}{h^2}a_{4,2}(\gamma)\nonumber\\
&+& \frac{\nu}{h^3}\left[a_{4,3}^{\pi^2}(\gamma)  + a_{4,3}^{\rm rem}(\gamma)\right]\,,
\eea
where
\begin{widetext}
\begin{subequations}
\begin{align}
&a_{4,1}(\gamma) = \frac{107-2743 \gamma ^2+26942 \gamma ^4-128682 \gamma
   ^6+308235 \gamma ^8-343935 \gamma ^{10}+142380 \gamma ^{12}}{2 (\gamma +1) (\gamma^2 -1) \left(3 \gamma ^2-1\right)^2 (4
   \gamma^2 -1)  \left(5 \gamma
   ^2-1\right)}\,,  \\ 
&a_{4,2}(\gamma) =  \frac{3-30 \gamma ^2+35 \gamma ^4}{(\gamma^2 -1) (4 \gamma^2 -1) \left(5 \gamma ^2-1\right)} \Bigg[-\frac{9 + 191 \gamma - 126 \gamma^2 - 962 \gamma^3 + 585 \gamma^4 + 543 \gamma^5 - 900 \gamma^6 + 1908 \gamma^7}{2 \left(1-3 \gamma ^2\right)^2}  \cr
   &\quad +\frac{24 \left(3+12 \gamma ^2-4 \gamma ^4\right) {\rm arcsinh \sqrt{\frac{\gamma-1}{2}}}}{\sqrt{\gamma ^2-1}}\Bigg]\,, \\
&a_{4,3}^{\pi^2}(\gamma) = \frac{2 \pi ^2 (\gamma -1) \left(12 - 64 \gamma + 65 \gamma^2 + 5 \gamma^3 - 25 \gamma^4 - 25 \gamma^5\right)}{3 \left(5 \gamma ^2-1\right)} \cr
&\qquad +\frac{1}{\left(\gamma ^2-1\right) \left(5 \gamma ^2-1\right)} \Bigg[\left(834+2095 \gamma +1200 \gamma ^2\right){\rm K^2}\left(\frac{\gamma-1}{\gamma+1}\right) + \frac{7}{2} (\gamma +1) \left(169+380 \gamma ^2\right) {\rm E^2}\left(\frac{\gamma-1}{\gamma+1}\right)\cr
&\qquad-(1183 + 2929 \gamma + 2660 \gamma^2 + 1200 \gamma^3) {\rm K}\left(\frac{\gamma-1}{\gamma+1}\right) {\rm E}\left(\frac{\gamma-1}{\gamma+1}\right) \Bigg]\,,
\end{align}  
\end{subequations}
\end{widetext}
K and E being the complete elliptic integrals of the first and second kind, respectively, expressed as functions of the parameter $m=k^2$. 
While the expressions for the above coefficients are exact, the last contribution $ a_{4,3}^{\rm rem}(\gamma)$ is only known up to order $\pinf^{30}$ (see Ref.~\cite{Bini:2024tft}). 
The beginning of the $\pinf^{30}$-accurate value of $ a_{4,3}^{\rm rem}(\gamma)$ reads
\begin{widetext}
\be
a_{4,3}^{\rm rem}(\gamma) = 
\frac{10}{p_{\infty}^2} + \frac{346}{3} + \frac{719 p_{\infty}^2}{9} + \frac{493819 p_{\infty}^4}{16800} - \frac{31001791 p_{\infty}^6}{470400} + \frac{13674094649 p_{\infty}^8}{111767040} + \cdots + O(p_\infty^{32})\,,
\ee
\end{widetext}
The apparent $10/\pinf^2=10/(\gamma^2-1)$ pole in $a_{4,3}^{\rm rem}$ cancels against similar poles present in $a_{4,1}, a_{4,2},a_{4,3}^{\pi^2}$. 
The complete  $\pinf^{30}$-accurate value of $ a_{4,3}^{\rm rem}(\gamma)$ is given in electronic form in the ancillary {\tt Mathematica} file.


\subsection{Completing the local  orbital dynamics by 5PM contributions at 4PN accuracy}
\label{sec:4PN_Alocal}

Having obtained the purely local component of the $A$ potential at the 4PM accuracy (in our LJBL gauge),
we are in position to complete it in two different directions: 
(i) by adding the {\it local} contribution of higher PM orders up to some given PN order,
and (ii) by reintroducing the (tail-related) {\it nonlocal} contribution in a form appropriate for mildly eccentric orbits up to some given eccentricity order, and up to some given PN order.

The aim of the present section is to use the results
of the Tutti Frutti approach~\cite{Bini:2019nra,Bini:2020wpo,Bini:2020nsb,Bini:2020hmy} and add to the above-determined 4PM-level
local $A$ potential in our gauge the current
4PN-limited knowledge of the 5PM
contributions to the local orbital dynamics.

We have actually performed this calculation up to 7PM and 6PN accuracy.  However, since we are not going to use this information in the construction of our present waveform model,  we relegate the presentation of this high-order extension to Appendix~\ref{app:complete_Alocal_6PN}.
The inclusion of the nonlocal contributions to the $A$ potential will be instead discussed in the next section.
Let us denote the 4PM-accurate local orbital $A$ potential determined in the previous section as  (with $u \equiv 1/r= GM/R$)
\begin{align} 
	\label{eq:Aloc4PM}
	&A^{\rm loc}_{\leq \rm 4PM}(r,\gamma,\nu)= 1 - 2 u \nonumber\\
	&\quad+ a_2(\g,\nu) u^2 +  a_3(\g,\nu) u^3 + a_4
	^{\rm loc}(\g,\nu) u^4 \ .
\end{align}
Higher PM contributions to the local conservative dynamics are described, in our LJBL gauge, by contributions 
to $A^{\rm loc}(r,\gamma,\nu)$ proportional to higher powers of $u$. In the Tutti Frutti approach such 
contributions are purely of the power-law type, without any logarithmic dependence on $u$.\footnote{Indeed,
the definition of the local-nonlocal separation used in the Tutti Frutti approach incorporates all the 
$\ln u$ terms (including the ones associated with near-zone effects, or potential gravitons) in the nonlocal 
part of the dynamics.} As a consequence, when considering the 4PN accuracy, we must simply complete 
the 4PM-accurate $A$ potential of Eq.~\eqref{eq:Aloc4PM} by the 5PM static contributions 
proportional to $u^5 = \big(GM/R\big)^5$. In other words, the completed $A$ potential reads 
\begin{align}
	\label{eq:A4PM4PNcomp}
	A^{\rm loc}_{\rm 4PN completed}(r,\gamma,\nu)&=A^{\rm loc}_{\rm \leq 4PM}(r,\gamma,\nu)+ u^5 a_{50}^{\rm loc}(\nu) \ .
\end{align}
The value of the coefficient $a_{50}^{\rm loc}(\nu)$ can be directly obtained from the results of Ref.~\cite{Bini:2020nsb}. 
Indeed, Table XII of~\cite{Bini:2020nsb} gives the coefficients $q_n(\pinf,\nu)$ of the $u$ expansion of $\hat Q$
in the E-type energy gauge. Substituting the values of these
coefficients in our Eqs.~\eqref{anvsqn}, which give
the $a_n(\g,\nu)$'s in terms of the $q_n(\g,\nu)$'s, yields
\begin{align}
			\label{eq:4PN_loc}
			&a_{50}^{\rm loc}(\nu) = \left(-\frac{34093}{360}+\frac{3539 \pi ^2}{6144}\right) \nu
			\cr&\quad+\left(-\frac{1195}{12}+\frac{205 \pi ^2}{64}\right) \nu ^2+\frac{9 \nu ^3}{4}\,. 
		\end{align}
This value can also be obtained from the 4PN-accurate determination of the EOB Hamiltonian in Ref.~\cite{Damour:2015isa}.

\subsection{Nonlocal orbital dynamics for elliptic-like motions at 4PN accuracy}
\label{sec:nonlocal_4PN}
As already mentioned, the nonlocal part of the conservative dynamics for bound motions, first appearing at the 4PN order, has
a nonperturbative dependence on Newton's constant,
which prevents its computation within the
PM approach. However, the PN approach does give
access to explicit computations of a Delaunay-Hamiltonian description of the nonlocal dynamics
of moderately eccentric motions as an expansion
in powers of the eccentricity (which, in the DJS gauge, is equivalent
to an expansion in powers of the radial momentum $p_r$). This was first done at the 4PN level and
at the sixth order in eccentricity in Refs.~\cite{Damour:2014jta,Damour:2015isa},
and then extended to the 5PN and 6PN orders (including the purely nonlocal 5.5PN contribution),
and at the tenth order in eccentricity, in 
Refs.~\cite{Bini:2019nra,Bini:2020wpo,Bini:2020nsb,Bini:2020hmy}.

In this section we will use these results at 4PN and up to the eighth order in eccentricity, corresponding to the eighth power of $p_r$, to derive a corresponding nonlocal contribution $\delta A_{\rm nonlocal}^{\rm 4PN}(\gamma,u)$ to the local $A$ potential of the previous section. For the extension of this computation to the maximum available PN accuracy, i.e.~the 6PN, see Appendix \ref{app:nonlocal}.

We start from an orbital $A$ potential with the structure
 \begin{align}
	\label{eq:Atot_4PN}
	&A(\gamma,u)=A^{\rm loc}_{\rm 4PN completed}(\gamma,u)+\delta A^{\rm 4PN}_{\rm nonlocal}(\gamma,u)\,,
\end{align}
where $A^{\rm loc}_{\rm 4PN completed}(\gamma,u)$ is the local component, determined in Sec.~\ref{sec:4PN_Alocal} with full 4PM accuracy and with additional local 5PM contributions up to the 4PN order.

At 4PN accuracy, the DJS-gauge Tutti Frutti squared effective Hamiltonian for bound state motions,
$\hat H_{\rm eff}^2(r,p_r,p_{\varphi})$,  also contains a nonlocal contribution $H_{\rm eff, nonloc}^2(r,p_r,p_{\varphi})$ given, in terms of the DJS-gauge EOB potentials, by
\begin{align}
	\label{eq:Hnonloc_4PN}
	&\delta \hat{H}_{\rm eff, nonloc}^2(r,p_r,p_{\varphi})=[1-2(1-2u)p_r^2
	\cr&\quad+p_\varphi^2 u^2] \, \delta A_{\rm nl}(r)
	+(1-2u)^2p_r^2\, \delta\bar{D}_{\rm nl}(r) \cr&\quad+ (1-2u) \, \delta Q_{\rm nl}(r,p_r) \,.
\end{align}
The nonlocal components $[\delta A_{\rm nl}(r),\delta\bar{D}_{\rm nl}(r),\delta Q_{\rm nl}(r,p_r)]$  at 4PN accuracy (and beyond, at 5PN and 6PN) can be found in Table IV of Ref.~\cite{Bini:2020wpo} and Table VI of Ref.~\cite{Bini:2020nsb}; we report them for completeness in Tables \ref{tab:nlc_A}, \ref{tab:nlc_D}, and \ref{tab:nlc_Q}.

Note that 
$\delta Q_{\rm nl}^{\rm 4PN}(r,p_r)$
has the structure
\begin{align}
	\label{eq:Q4PNdjs}
	&\delta Q_{\rm nl,h}^{\rm 4PN}(r,p_r) =  c_4 q_{43} p_r^4  u^3 + c_6 q_{62} p_r^6  u^2  + c_8  q_{81} p_r^8  u \cr&\quad+\mathcal{O}(p_r^{10})\,.
\end{align}
The parameters $c_4,c_6,c_8$ appearing in Eq.~\eqref{eq:Q4PNdjs} are eccentricity-keying parameters. They are all equal to 1 and serve the purpose of keeping track of the order in $p_r$ (beyond $p_r^2$), which is physically equivalent to the order in eccentricity
(beyond the $e^2$ level).
  
The mapping between the DJS gauge dynamics, described by the Hamiltonian $H_{\rm eff}^2(r,p_r,p_{\varphi})$, and the (PN expansion of the) LJBL-gauge dynamics described by the $A(\gamma,u)$ potential of Eq.~\eqref{eq:Atot_4PN} is obtained by looking for a canonical transformation such that the DJS-gauge constraint 
\be
0= \hat H_{\rm eff}^2(r,p_r,p_{\varphi})- \g^2\,,
\ee
be equivalent to the corresponding LJBL constraint
\begin{equation}
	0= A(\gamma,u) [1+A(\gamma,u) \, p_r^2+p_\varphi^2 u^2] -\gamma^2 \,.
\end{equation}
The resulting $\delta A^{\rm 4PN}_{\rm nonlocal}(p_\infty,u)$ contribution to $A(\gamma,u)$, and the canonical transformation itself,  are obtained order by order in the PN expansion. This canonical transformation involves logarithms of $u$. See Refs.~\cite{Bini:2019nra,Bini:2020wpo,Bini:2020nsb,Bini:2020hmy} for similar computations.

The general structure of $\delta A^{\rm 4PN}_{\rm nonlocal}(\g,u)$ reads:
 \begin{align}
&\delta A^{ 4 \rm PN}_{\rm nonlocal}= \sum _{k=1}^{5} a_{k,5-k}^{\text{nlh,c}} \, u^k p_\infty ^{2(5-k)} \cr&\quad+ \ln u \sum_{k=4}^{5} a_{k,5-k}^{\text{nlh,ln}} \,
u^k p_\infty ^{2(5-k)} \,,
\end{align}
with an eccentricity structure
\begin{align}
	\label{eq:Anonloc_struct_4PN}
	&\delta A_{\rm nonlocal}^{\rm 4PN}(\gamma,u) = \delta A_{\rm nonlocal}^{\leq e^2}(\gamma,u)\cr&\quad+c_4\delta A_{\rm nonlocal}^{e^4}(\gamma,u)+c_6\delta A_{\rm nonlocal}^{e^6}(\gamma,u)\cr&\quad+c_8\delta A_{\rm nonlocal}^{e^8}(\gamma,u)\,,
\end{align}
 where the parameters $c_4,c_6,c_8$ are the beyond-$e^2$ eccentricity-keying parameters introduced in Eq.~\eqref{eq:Q4PNdjs}.
One can prove that $A_{\rm nonlocal}^{e^4}(\gamma,u)$, $A_{\rm nonlocal}^{e^6}(\gamma,u)$, and $A_{\rm nonlocal}^{e^8}(\gamma,u)$ all vanish along circular orbits (when taking into account the link between $\g$ and $u$ that exists for circular motions). Explicitly, we find
\begin{widetext}
\begin{align}
	\label{eq:Anonloc_4PN_expl}
	&\delta A_{\rm nonlocal}^{\rm 4PN}(\gamma,u) = u^4 \pinf^2 \left(-\frac{781}{15}+\frac{296 \gamma }{15}-\frac{1624 \ln
		2}{15}+\frac{729 \ln 3}{5}+\frac{148 \ln u}{15}\right) \nu \cr&\quad+u^5 
	\left(-\frac{781}{15}+\frac{136 \gamma }{3}-\frac{856 \ln 2}{15}+\frac{729 \ln
		3}{5}+\frac{68 \ln u}{3}\right) \nu \cr&\quad+ c_4\bigg[u^3  \pinf ^4 \left(-\frac{5608}{75}+\frac{496256 \ln 2}{225}-\frac{33048 \ln
		3}{25}\right) \nu+u^4  \pinf^2 \left(-\frac{9814}{75}+\frac{868448 \ln
		2}{225}-\frac{57834 \ln 3}{25}\right) \nu\cr&\quad+u^5 \left(-\frac{1402}{25}+\frac{124064
		\ln 2}{75}-\frac{24786 \ln 3}{25}\right) \nu\bigg] \cr&\quad+
	c_6\bigg[ u^2 p_{\infty }^6  \left(-\frac{1027}{12}-\frac{147432 \ln 2}{5}+\frac{1399437 \ln
		3}{160}+\frac{1953125 \ln 5}{288}\right) \nu \cr&\quad+ u^3
	p_{\infty }^4\left(-\frac{11297}{60}-\frac{1621752 \ln 2}{25}+\frac{15393807 \ln
		3}{800}+\frac{4296875 \ln 5}{288}\right) \nu \cr&\quad+ u^4p_{\infty }^2
	\left(-\frac{2054}{15}-\frac{1179456 \ln 2}{25}+\frac{1399437 \ln
		3}{100}+\frac{390625 \ln 5}{36}\right) \nu\cr&\quad+u^5
	\left(-\frac{1027}{30}-\frac{294864 \ln 2}{25}+\frac{1399437 \ln
		3}{400}+\frac{390625 \ln 5}{144}\right) \nu\bigg]  \cr&\quad +
	c_8 \bigg[ u \, p_{\infty }^8 \left(-\frac{35772}{175}+\frac{10834496 \ln 2}{21}+\frac{6591861 \ln
		3}{350}-\frac{27734375 \ln 5}{126}\right) \nu \cr&\quad+ u^2p_{\infty }^6
	\left(-\frac{8943}{20}+\frac{3385780 \ln 2}{3}+\frac{6591861 \ln
		3}{160}-\frac{138671875 \ln 5}{288}\right) \nu \cr&\quad+ u^3 p_{\infty }^4
	\left(-\frac{259347}{700}+\frac{19637524 \ln 2}{21}+\frac{191163969 \ln
		3}{5600}-\frac{804296875 \ln 5}{2016}\right) \nu \cr&\quad+ u^4 p_{\infty }^2
	\left(-\frac{26829}{175}+\frac{2708624 \ln 2}{7}+\frac{19775583 \ln
		3}{1400}-\frac{27734375 \ln 5}{168}\right) \nu \cr&\quad+u^5
	\left(-\frac{8943}{350}+\frac{1354312 \ln 2}{21}+\frac{6591861 \ln
		3}{2800}-\frac{27734375 \ln 5}{1008}\right) \nu\bigg].
\end{align}
\end{widetext}

Preliminary investigations suggest that the inclusion of $\delta A_{\rm nonlocal}^{e^8}(\gamma,u)$ slightly deteriorates the
 behavior of  the LEOB dynamics. 
This is probably due to the low PM order at which some of these $e^8$ terms enter, such as $O(u \, \pinf^8)$, which amounts to a 1PM correction. One possible cure for this issue might be the use of an hybrid gauge where such $O(e^8)$ terms terms are kept
in a DJS-like gauge, i.e. in the form $O(p_r^8)$.
For the time being, we have decided to use, for the orbital part of our LEOB dynamics,
a total $A(\gamma,\nu)$ potential of the form \eqref{eq:Atot_4PN}, with a 
nonlocal PN contribution, $\delta A_{\rm nonlocal}^{\rm 4PN}(\gamma,u)$, truncated at the sixth order in eccentricity, 
that is Eq.~\eqref{eq:Anonloc_4PN_expl} with $c_8=0$.

\subsection{About time-nonlocal contributions to the spin-orbit sector of the dynamics}
\label{sec:nonlocal_4.5PN_SO}

In this section we briefly address the tail-transported time nonlocalities that enter the spin-orbit part of the conservative dynamics at 5PM order.

Similar to the case of the orbital 4PM coefficient of the $A$ potential, the 5PM spin-orbit coefficients $\hat{g}_3^0(\g,\nu)$ and $\hat{g}_{*3}^0(\g,\nu)$ [see e.g.~Eq.~\eqref{eq:g4_spin_struct}] that are determined by the 5PM coefficient of the scattering angle, $\chi_5(\g,a_i,\nu)$, have also a nonlocal component, valid only for unbound motions. 
In our gauge, the nonlocal contribution
to the 5PM spin-orbit coefficients $\hat{g}_3^0(\g,\nu)$ and $\hat{g}_{*3}^0(\g,\nu)$ start at order $O(\pinf^2)$ in the low-velocity
expansion of the 5PM spin-orbit coefficients: 
\begin{subequations}
\begin{align}
	\hat{g}_3^0(\g,\nu) &= \hat{g}_{30}(\nu)+  \hat{g}_{32}(\nu)\,\pinf^2 + \cdots \,,\\
	\hat{g}_{*3}^0(\g,\nu) &= \hat{g}_{*30}(\nu)+  \hat{g}_{*32}(\nu)\,\pinf^2 + \cdots\,.
\end{align}
\end{subequations}

To the best of our knowledge there is no work where (similarly to the case of the 4PM orbital dynamics) the local parts of $\hat{g}_3^0(\g,\nu)$ and $\hat{g}_{*3}^0(\g,\nu)$ have been computed in a 5PM-exact sense. 
What has been computed so far was obtained within a PN approach, where complete results are available up to the relative 3PN order~\cite{Antonelli:2020aeb,Levi:2020kvb,Placidi:2024yld}.
Partial results were also obtained, using the Tutti Frutti strategy, in Ref.~\cite{Khalil:2021fpm}, which derived the tail-related local and nonlocal contributions to the gyro-gravitomagnetic coefficients
$g_S$ and $g_{S*}$ at the fractional 4PN order [i.e. $\frac1{c^8}$ beyond the leading-order terms $g_0(\g,\nu)$ and  $g_{*0}(\g,\nu)$] and at 
eighth order in eccentricity.

In the present work, we will circumvent the lack of 5PM-exact knowledge of the local parts of  $g_S(u,\g,\nu)$ and 
$g_{S*}(u,\g,\nu)$ by approximating their 5PM-level contributions  
$\hat{g}_3^0(\g,\nu)$ and $\hat{g}_{*3}^0(\g,\nu)$ simply by their ``static" contributions, i.e.~their limit as $\g \to 1$ (which are
immune to nonlocal effects). These correspond to the fractional 3PN terms \footnote{This is sometimes counted as corresponding to the absolute 4.5PN level in the constraint, when allowing for fast spins $\chi_i =O(1)$.} first computed in Ref.~\cite{Antonelli:2020aeb}, dubbed $g_S^{\rm N^3LO}$ and $g_{S^*}^{\rm N^3LO}$ in Eqs.~(8) and (9) therein, which in our gauge read
\begin{subequations}
	\begin{align}
		\label{eq:g3pn}
		&\hat{g}_{30}(\nu) =\left(-\frac{7771}{288}+\frac{241 \pi ^2}{384}\right) \nu +\frac{55 \nu ^2}{4}-\frac{13
			\nu ^3}{32} \,,\\
			\label{eq:g3pn*}
		&\hat{g}_{*30}(\nu) = -\frac{11}{24}-\left(\frac{581}{36}-\frac{41 \pi ^2}{48}\right) \nu +14 \nu ^2-\frac{3
			\nu ^3}{2}\,.
	\end{align}
\end{subequations}
These contributions are at the fractional 3PM level in the spin-orbit sector.
 We leave to future work the inclusion of more analytical information in the spin-orbit sector, starting with the fractional 4PN results of Ref.~\cite{Khalil:2021fpm} (
 which, however, contain an undetermined $O(\nu^2)$ coefficient in the local component).

\section{PM-based waveform model for spin-aligned binaries}
\label{sec:leobWF}
In this section we finally construct and validate two different flavors of a 
complete waveform model based on the LEOB framework and incorporating PM
information at various orders in spin (labeled as $S^n={\rm spin}^n$ with $n=0,1,2,3,4, \ldots$).

In our LJBL gauge, even-in-spin effects ($S^0$, $S^2$, $S^4$, $\ldots$) are fully incorporated 
in the $A$ potential. 
To reduce to the exact Kerr metric in the test-mass limit, we use a structure of $A$ that factors 
out a Kerr-like term (with spin parameter $\tilde a_0=\tilde a_1+\tilde a_2$). 
We hence write [with $u\equiv \frac1{r}$, and $u_c(\tilde{a}_0) \equiv \frac1{r_c(\tilde{a}_0)}$]:
\begin{align} 
	\label{Afinal}
	A\left(r,\g,\nu,\tilde{a}_i\right) &= \frac{1+2 u_c(\tilde{a}_0)}{1 + 2 u} \Bigg[A_{\rm orb}\left(u_c(\tilde{a}_0),\g,\nu\right) \cr
	&\quad+ A_{\rm SS}(u_c(\tilde{a}_0),\g,\nu,\tilde{a}_i)\Bigg]\,,
\end{align}
where $ A_{\rm orb}\left(u_c(\tilde{a}_0),\g,\nu\right)= 1- 2 u_c + O(u_c)^2$ and $A_{\rm SS}(u_c(\tilde{a}_0),\g,\nu,\tilde{a}_i)= O(\tilde{a}_i^2 u^3)$.

On the other hand, odd-in-spin effects ($S^1$, $S^3$,  $\ldots$) are described by the ${\mathcal G}$ term:
\be
\label{eq:GSresc2}
{\mathcal G}= G_S(r, \g, \tilde{a}_i) \hat{S} +  G_{S_*}(r, \g, \tilde{a}_i) \hat{S}_*\,.
\ee
with
\begin{subequations}
\begin{align}
		&G_S(r, \g, \tilde{a}_i,\nu)=\frac{g_S\left(u,\gamma,\tilde{a}_i\right)}{r r_c^2(\tilde{a}_0)}\,,\\
&G_{S_*}(r, \g, \tilde{a}_i,\nu)=\frac{g_{S_*}\left(u,\gamma,\tilde{a}_i\right)}{r r_c^2(\tilde{a}_0)}\,,
\end{align}
\end{subequations}
where $g_S\left(u,\gamma,\tilde{a}_i\right) = g_0(\g,\nu) + O(u)$
 and $g_{S_*}\left(u,\gamma,\tilde{a}_i\right) = g_{*0}(\g,\nu) + O(u)$.

Note the presence of the spin-dependent centrifugal radius $r_c(\tilde{a}_0)$ (with $\tilde{a}_0=\tilde{a}_1+\tilde{a}_2$)
as a building block in several of the metric coefficients above. We recall that it was found in Ref.~\cite{Damour:2001tu}
that, at leading PN order, the quadratic-in-spin terms in the Hamiltonian depend only on $a_0=a_1+a_2$. It was
then found in Ref.~\cite{Vines:2017hyw} that, in the aligned spin case, $a_0$ parametrizes (when accompanied by the covariant orbital angular momentum) the scattering angle at the formal 1PM order, but at all orders in spin.
These results motivate using the combination $\tilde a_0=\tilde a_1+\tilde a_2$ to absorb as many spin-dependent effects as possible.
For reasons that we shall explain in detail below, the two different flavors of LEOB models that we shall consider 
in the present exploratory study differ in the choice of the last term $A_{\rm SS}(u_c(\tilde{a}_0),\g,\nu,\tilde{a}_i)$.
A first choice approximates all spin-spin interactions by the simple use of 
 the spin-dependent centrifugal radius $r_c(\tilde{a}_0)$ as a building block in several of the metric coefficients above,
 setting then the complementary spin-spin term $A_{\rm SS}(u_c(\tilde{a}_0),\g,\nu,\tilde{a}_i)$ to zero.
 We shall refer to this simplified model (in which spin-spin interactions are treated
 in a Kerr-like way) as \LEOBa{}.
 A second choice includes the term $A_{\rm SS}(u_c(\tilde{a}_0),\g,\nu,\tilde{a}_i)$ in the $A$ potential.
 However, we found that using the full presently available (5PM-level) analytical information in $A_{\rm SS}$
 leads to a loss of accuracy in certain regions of parameter space. Therefore, we shall instead focus on a model
 where $A_{\rm SS}$ is truncated to the 4PM level.  We shall refer to this second model as \LEOBss{}.

Let us now indicate the analytical information included in our current LEOB model at various orders in spin.
If available, we include terms up to the physical 5PM level.
\begin{itemize}
\item[(i)]For the $S^0$ sector (embodied in the metric potential $A_{\rm orb}$): we omit the incompletely known orbital 5PM coefficient and include local 4PM contributions completed by (local and nonlocal) 4PN terms up to order six in eccentricity.
\item[(ii)]For the  $S^1$ sector (embodied in the radial functions $G_S$, and $G_{S_*}$):
we do not include 5PM terms as they contain hyperbolic tails and work at 4PM accuracy, completed only by the static 5PM contributions;
\item[(iii)]For the $S^2$ sector: As said above, we either describe them only through the 
centrifugal radius $r_c(\tilde{a}_0)$ (setting  $A_{\rm SS}$ to zero), or we incorporate the $A_{\rm SS}$ term (but
only up to the 4PM level).
\item[(iv)]For the $S^3$ sector: we work at the full physical 5PM accuracy by adding quadratic-in-spin terms in $G_S$ and $G_{S*}$.
\item[(v)]For the $S^4$ sector: we add the 5PM contribution through a quartic-in-spin contribution to $A_{\rm SS}$.
\end{itemize}

\subsection{Resummation of the dynamics}
\label{sec:resumA}

Since the early days of EOB theory, it was found that leaving the coefficients of the effective metric as polynomials  in $u$ entails 
an unphysical behavior in the late stages of the evolution. Therefore, we follow the usual practice of EOB-based models and resum our PM potentials in the following ways.

In the nonspinning limit, when using the factorized form \eqref{Afinal} of the $A$ potential, we resum $A_{\rm orb}$ by the following Pad\'e approximant, with $u$ as argument:
\begin{equation}
	A_{\rm orb}\left(u,\g,\nu\right) = P^1_4\left[A^{\rm 4PM\text{-}4PN}_{\rm orb}\left(u,\g,\nu\right)\right],
\end{equation}
which has the good probe limit
\begin{equation}
	A_{\rm orb} \left(r,\g,\nu=0\right) = 1 - 2 u\,.
\end{equation}

In the PM-expanded function $A^{\rm 4PM\text{-}4PN}_{\rm orb}(u)$ the logarithms $L = \ln u$ are replaced by constants 
before taking the Pad\'e approximant. They are then replaced back by their values after resummation.

When considering spinning systems, we resum the residual spin-spin contribution, 
$A_{\rm SS}(r_c,\g,\nu,\tilde{a}_i)$, entering Eq.~\eqref{Afinal} together with the orbital part, as
\begin{subequations}
\begin{align}
	\label{eq:Afact}
	&A\left(r,\g,\nu,\tilde{a}_i\right) = \frac{1+2 u_c(\tilde{a}_0)}{1 + 2 u} A_{\rm tot}\,, \\
	\label{eq:Ass}
	&A_{\rm tot} = P^1_4\left[A^{\rm 4PM\text{-}4PN}_{\rm orb}\left(u_c,\g,\nu\right) + A_{\rm SS}^{\rm 5PM}(u_c,\g,\nu,\tilde{a}_i)\right]\,,
\end{align}
\end{subequations}
where the Pad\'e approximant is intended here as a resummation in $u_c$ (with constant $\ln u_c$)
 so that in the probe limit the metric potential will reduce to $\frac{1+2 u_c}{1+2 u} (1-2 u_c)$. 
The interested reader can find the analysis of a different factorization of the spinning $A$ potential in Appendix~\ref{app:ss}.

After having defined a resummation for the $A$ potential, the corresponding  $B$ potential is given [in view of Eq.~\eqref{eq:LJBL_gauge_AB}] by 
\begin{equation}
	B = \frac{r^2}{r_c^2} \frac{1}{A}\,.
\end{equation}

The resummed version of the odd-in-spin sector, described by the gyro-gravitomagnetic 
functions $G_S$ and $G_{S_*}$, is defined  (as usually done~\cite{Damour:2014sva}) by inverse-resumming the latter  gyro-gravitomagnetic 
coefficients. Explicitly, the $G_S$ function is defined as
\begin{equation}
	G_S(r,\g,\nu,\tilde{a}_i) = \frac{g_0(\g,\nu)}{r r_c^2} P^0_3 \left[ \hat{g}^{\rm 5PM, static}_S (u,\g,\nu,\tilde{a}_i)\right]\,,
\end{equation}
where 
\begin{align} \label{hatgS}
&\hat{g}^{\rm 5PM, static}_S (u,\g,\nu,\tilde{a}_i) = 1 + \hat{g}_1\left(\g,\nu\right) u + \hat{g}_2\left(\g,\nu,\tilde{a}_i\right) u^2 \cr
&\quad +\left[\hat{g}_{30}(\nu) + \hat{g}_{3}^{\tilde{a}_0^2}(\g,\nu,\tilde{a}_i)+ \hat{g}_{3}^{\tilde{a}_{12}^2}(\g,\nu,\tilde{a}_i)\right] u^3\,,
\end{align}
which is obtained from Eq.~\eqref{eq:gSfact} while using the static 5PM-order contribution $\hat{g}_{30}(\nu)$ of Eq.~\eqref{eq:g3pn}. 
The same procedure is applied to $G_{S_*}$, starting from Eq.~\eqref{eq:gSstarfact} and using the static 5PM-order contribution of Eq.~\eqref{eq:g3pn*}.
The explicit expression for these EOB functions can be found in electronic form in the attached {\tt Mathematica} file.

\subsection{Initial data and equations of motion}
\label{sec:id}

As discussed in Sec.~\ref{sec:leob}, the Lagrange-EOB approach to the dynamics imposes 
the solution of five dynamical quantities instead of four.
The consistency of this over-determined system is ensured by the constraint equation, Eq.~\eqref{eq:mass_shell_cond_m_rescaled}, which in the LJBL gauge 
can be rewritten as
\begin{equation}
	\mathcal{C} \equiv \frac{-\g_{\rm orb}^2 + p_{r_*}^2}{A} + p_\varphi^2 u_c^2 + 1 = 0\,,
\end{equation}
where we recall that $\g_{\rm orb} \equiv \g - p_\varphi \mathcal{G}$ and we defined the radial momentum $p_{r_*}\equiv \sqrt{A/B}\, p_r$.
In the usual HEOB approach, $p_{r_*}$ is conjugate to some tortoise coordinate $r_*$, while here $p_{r_*}$ also depends on $\gamma$, so that its evolution equation must be computed as 
\begin{equation}
	\frac{d p_{r_*}}{dt} = \frac{\partial p_{r_*}}{\partial p_r} \frac{d p_r}{dt} + \frac{\partial p_{r_*}}{\partial r} \frac{dr}{dt} + \frac{\partial p_{r_*}}{\partial \g} \frac{d \g}{dt}\,.
\end{equation}

The five Euler-Lagrange equations finally read
\begin{widetext}
\begin{subequations}
	\begin{align}
		\label{eq:drdt}
		\frac{dr}{dt_{\rm real}} &=  \frac{A\, p_{r_*} r_c u}{
			h \g_{\rm orb}\left[1 - \frac{1}{2 \g_{\rm orb}}\left(\gamma_{\rm orb} ^2+p_{r_*}^2\right)\partial_\g \ln{A} - p_\varphi \partial_\g\mathcal{G}\right]
		}\,,\\
		\label{eq:dphidt}
		\frac{d\varphi}{dt_{\rm real}} &=  \frac{A\, p_{\varphi } u_c^2 + \g_{\rm orb}\, \mathcal{G}}{
			h \g_{\rm orb}\left[1 - \frac{1}{2 \g_{\rm orb}}\left(\gamma_{\rm orb} ^2+p_{r_*}^2\right)\partial_\g \ln{A} - p_\varphi \partial_\g\mathcal{G}\right]
		}\,,\\
		\frac{dp_{r_*}}{dt_{\rm real}} &=\frac{
			A^2 p_{\varphi }^2 u_c^2 u \, \partial_r r_c
			- \frac{r_c}{2 r} \left(\g_{\rm orb}^2-p_{r_*}^2\right) \partial_rA 
			- A \frac{r_c}{r} \g_{\rm orb} p_\varphi \partial_r \mathcal{G}}{
				h \g_{\rm orb}\left[1 - \frac{1}{2 \g_{\rm orb}}\left(\gamma_{\rm orb} ^2+p_{r_*}^2\right)\partial_\g \ln{A} - p_\varphi \partial_\g\mathcal{G}\right]
		} + \mathcal{F}_{r_*} \,,\\
		\frac{dp_\varphi}{dt_{\rm real}} &= \mathcal{F}_\varphi \,,\\
		\frac{d\gamma}{dt_{\rm real}} &= -\mathcal{F}_0 \,,
	\end{align}
\end{subequations}
\end{widetext}
where $\mathcal{F}_0$ is defined by Eq.~\eqref{F0}, and where
\begin{equation}
\mathcal{F}_{r_*} = A \frac{r_c}{r} \mathcal{F}_{r} + p_{r_*} \mathcal{F}_0  \, \partial_\g \ln{A} \,.
\end{equation}

As for the constraint equation, we only need to impose it for the initial data.
The constraint, Eq.~\eqref{eq:fluxes_eq}, on  $\mathcal{F}_{\mu}$, guarantees its conservation also for nonadiabatic evolutions.
We have checked numerically that this is indeed the case. For example, solving the LEOB evolution equation using the Adams-Bashforth-Moulton
standard solver implemented in {\tt MATLAB} with tolerances $\sim 10^{-10}$ we find that ${\cal C}$ increases from $\sim 10^{-15}$ during 
the inspiral to a maximum of $\sim 3\times 10^{-10}$ at merger. The corresponding fractional error on solving $\gamma$ from Eq.~\eqref{eq:cH}
increases from $10^{-15}$ during the inspiral to a maximum of $10^{-10}$ at merger.

The initial conditions are computed in the following recursive way.
For each starting radius $r_0$, we first obtain adiabatic values of
the starting circular energy and angular momentum $(\g_0,p_{\varphi, 0})$ by imposing, when setting $p_{r_*,0} = 0$, the two equations
\begin{subequations}
\begin{align}
	\label{eq:circC}
	&\mathcal{C} \left(r_0, \g_0, p_{\varphi, 0}, p_{r_*, 0} = 0\right) = 0\,, \\
	\label{eq:circdCdr}	&\frac{\partial \mathcal{C}}{\partial r} \left(r, \g_0, p_{\varphi, 0}, p_{r_*, 0} = 0\right) \Big|_{r = r_0} = 0\,.
\end{align}
\end{subequations}
These adiabatic  circular initial data are then used to compute post-adiabatic~\cite{Damour:2007yf,Damour:2012ky,Nagar:2018gnk} initial data, which allow one to reduce any initial spurious eccentricity.
The post-adiabatic initial radial momentum is computed as
\begin{equation}
	p_{r_*}^{\rm 1PA} = - \frac{r_0\, h_0}{2\, r_{c,0}} \, \partial_\g \mathcal{C}\big|_0 \mathcal{F}_{\varphi,0}\,, 
\end{equation} 
where the subscript $0$ indicates quantities computed using the adiabatic initial data described above.
Finally, we need to ensure the satisfaction of the mass-shell constraint by computing $(p_\varphi^{\rm 1PA},\g^{\rm 1PA})$, imposing again Eqs.~\eqref{eq:circC}-\eqref{eq:circdCdr} with $p_{r_*} = p_{r_*}^{\rm 1PA}$ instead of $p_{r_*} = 0$.

\subsection{Waveform and radiation reaction}
\label{subsec:RR}
Let us turn to discussing the radiation reaction and waveform implemented here.
Given the high-level of development of the factorized and resummed multipolar 
waveform, based on the PN-matched Multipolar-Post-Minkowskian formalism~\cite{Blanchet:1985sp,Blanchet:1986dk,Blanchet:1987wq,Blanchet:1989ki,Damour:1990gj,Blanchet:2023bwj,Blanchet:2023sbv},
implemented in the \TEOBResumS{} model family, we directly rely on it.
More precisely, we  employ the factorized and resummed expressions
used in the latest avatar of the \TEOBd{} model and extensively discussed in
Sec.~IIB of Ref.~\cite{Nagar:2024oyk}, with some minor simplifications due
to the fact that the model discussed here is restricted to quasi-circular binaries
only. As a reminder of the main analytical structure of the radiative sector,  each multipolar 
waveform mode is factorized as~\cite{Damour:2008gu}
\be
\label{eq:hlm}
h_\lm=h_\lm^{N}\hat{h}_\lm \hat{h}_{\lm}^{\rm NQC} \ ,
\ee
where $h_\lm^{N}$ is the Newtonian prefactor\footnote{Defined in Eqs.~(3.21)-(3.3) in Ref.~\cite{Nagar:2019wds} up to $\ell=m=5$ mode.}
and $\hat{h}_\lm$ is the factorized and resummed PN correction introduced 
in Ref.~\cite{Damour:2008gu} and 
explicitly detailed in the form we use here starting from Eq.~(12) of Ref.~\cite{Nagar:2024oyk}.
The factor  $\hat{h}_{\lm}^{\rm NQC}$ is the NR-informed next-to-quasi-circular (NQC) correction
factor that is designed to improve the behavior of the analytical waveform during
the plunge so as to smoothly connect it to the post-merger signal. For its explicit
analytic form we address the reader to Ref.~\cite{Nagar:2024oyk}, in particular Eqs.~(16) and (17) therein.

The two components of the radiation reaction force are 
[see Eq.~(21) in Ref.~\cite{Nagar:2024oyk}]
\begin{align}
\hat{{\cal F}}_\varphi &= - \dfrac{32}{5} \nu r_\Omega^4\Omega^5 \hat{f}(x;\nu) + \hat{\cal F}_\varphi^{\rm H}\ ,\\
\hat{\cal F}_r &= 0 \ ,
\end{align}
where $\hat{\cal F}_\varphi^{\rm H}$ is the angular momentum flux absorbed by
the two black holes~\cite{Damour:2014sva}. The factor $\hat{f}(x)$ is a sum of 
modes $\hat{f}_\lm$ as detailed in Eq.~(22a) and (22b) of ~\cite{Nagar:2024oyk}. 
Since here we restrict to quasi-circular binaries, we fix the noncircular correction
factors $\hat{f}_\lm^{\rm noncircular}=1$ in Eq.~(22a) of Ref.~\cite{Nagar:2024oyk}.
Similarly, we do not include the NR-informed factor $\hat{h}_\lm^{\rm NQC}$ in radiation
reaction.

The function argument $x$, that coincides with $x=\Omega^{2/3}$ in 
the circular approximation, is actually replaced in the functions above by $x=v_\varphi^2$, where $v_\varphi\equiv r_\Omega \Omega$, 
so as to suitably take into account noncircular effects during the
plunge~\cite{Damour:2006tr,Damour:2007xr,Damour:2012ky}. Generalizing the definition of the Kepler-law-preserving 
radius $r_\Omega$ of Refs.~\cite{Damour:2006tr,Damour:2014sva} to take into account the $\gamma$-derivatives introduced by the present LEOB framework, we obtain
\begin{equation}
	r_\Omega = \left\{\frac{\left(r_c^3 \psi_c\right)^{-1/2} + \mathcal{G}}{h \left(1-\frac{\g_{\rm orb}}{2} \partial_\g \ln{A}- p_\varphi \partial_\g\mathcal{G}\right)}\right\}^{-2/3},
\end{equation}
with
\begin{equation}
	\psi_c = -\frac{2}{\partial_rA} \left(\partial_r u_c + \frac{\partial_r\mathcal{G}}{u_c A} \frac{\g_{\rm orb}}{p_\varphi}\right).
\end{equation}

The LEOB dynamics introduced here is found to share with the HEOB dynamics of \TEOBResumS{} models the property 
that the {\it pure} orbital frequency~\cite{Damour:2014sva,Harms:2014dqa} $\Omega_{\rm orb}$, i.e.~the orbital frequency 
{\it without} the spin-orbit contribution, defined here as 
\begin{equation}
\label{eq:OmgOrb}
		\Omega_{\rm orb} \equiv  \frac{A\, p_{\varphi } u_c^2}{
	h \g_{\rm orb}\left[1 - \frac{1}{2 \g_{\rm orb}}\left(\gamma_{\rm orb} ^2+p_{r_*}^2\right)\partial_\g \ln{A} - p_\varphi \partial_\g\mathcal{G}\right]
}
\end{equation}
develops a local maximum during the plunge\footnote{By contrast, the PM-based EOB model of Ref.~\cite{Buonanno:2024byg} , {\tt SEOB-PM}, 
does not present the same feature and uses a time-attachment shift $\Delta t_{\rm NR}$ as additional NR-calibration parameter. 
Moreover, the NQC corrections between the two models differ, with the model of Ref.~\cite{Buonanno:2024byg} using an additional free parameter.}. 
This allows us to follow the usual procedure to add the NQC corrections and to attach the postmerger (ringdown) 
waveform~\cite{Damour:2007vq,Damour:2009kr,Nagar:2019wds}. In particular, the $\ell=m=2$ ringdown attachment 
is done as in Ref.~\cite{Nagar:2024oyk}  at the time $t_{\rm attach}=t_{\rm peak}-2GM$, where $t_{\rm peak}$ corresponds to 
the maximum of $\Omega_{\rm orb}$.

\begin{figure*}[t]
	\center	
        \includegraphics[width=0.31\textwidth]{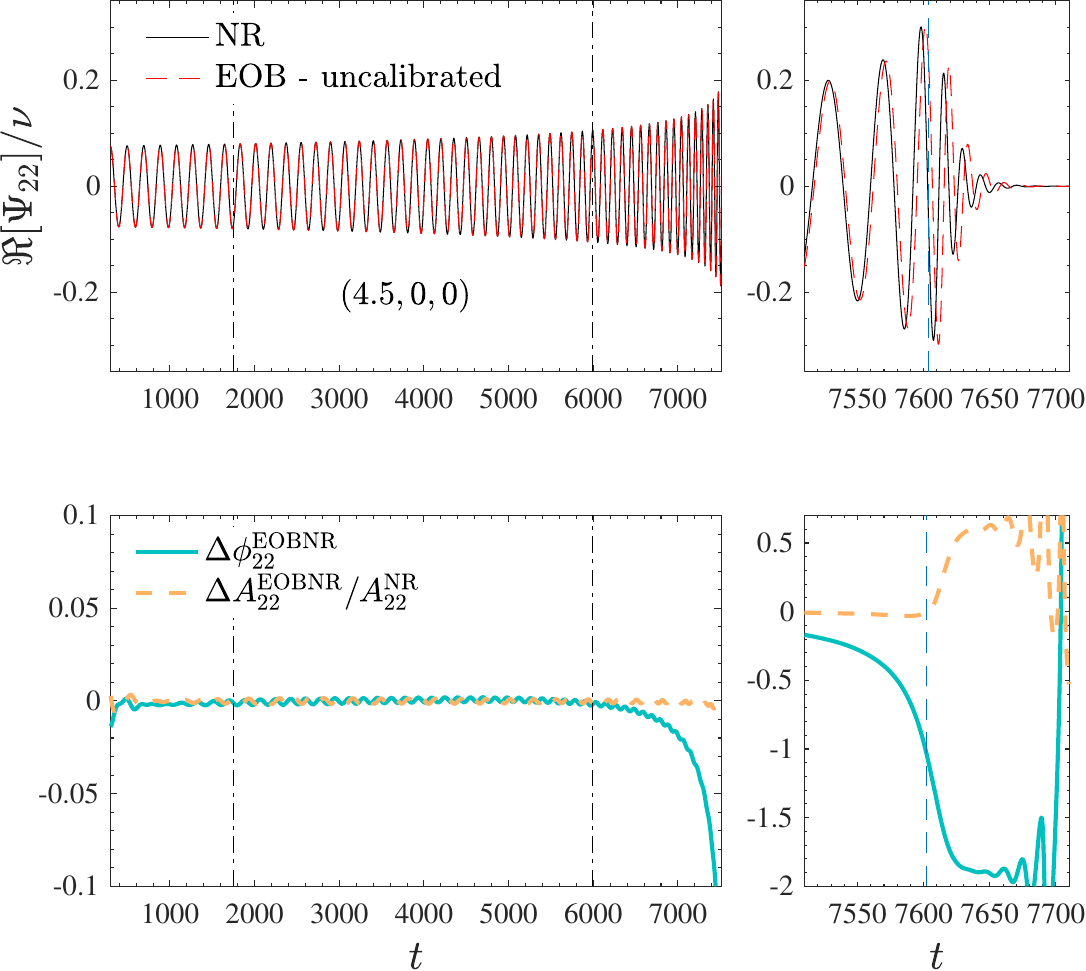}
        \hspace{2mm}
	\includegraphics[width=0.31\textwidth]{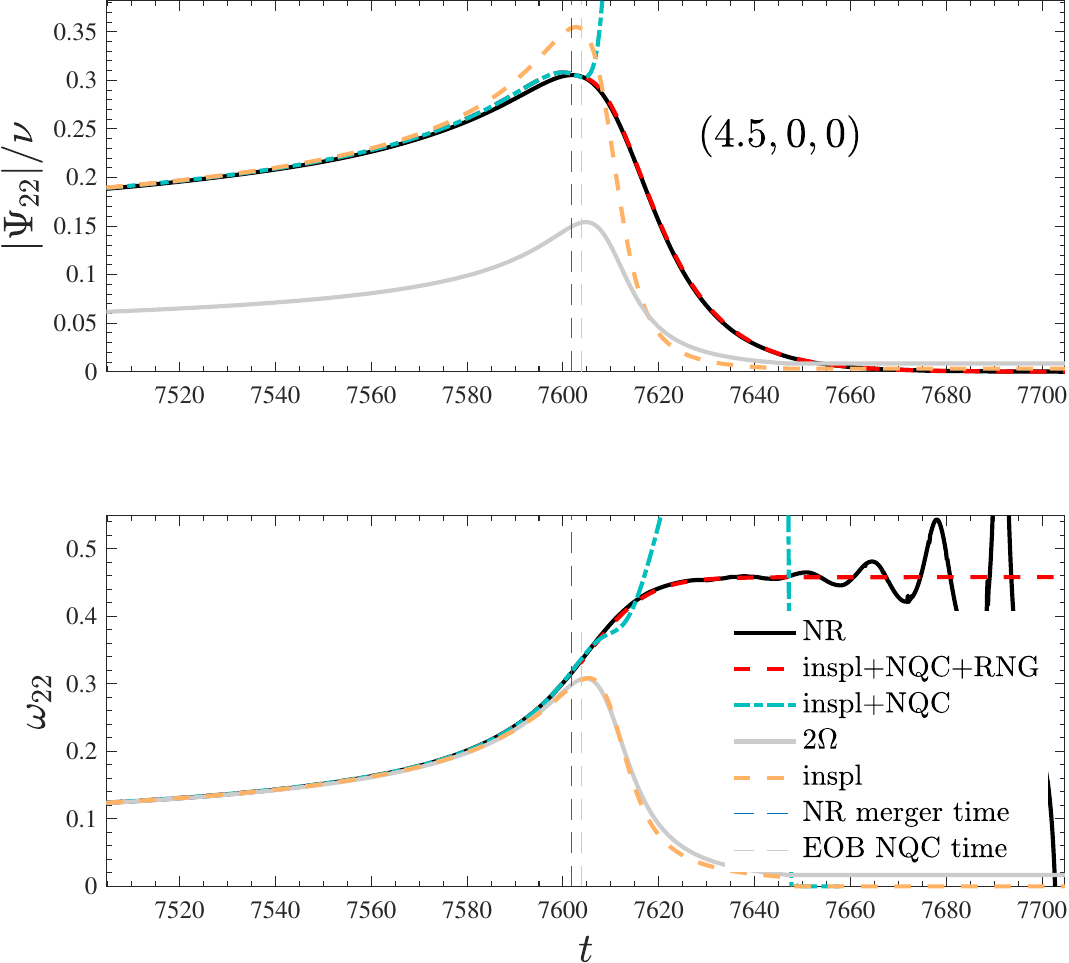}
	 \hspace{2mm}
	\includegraphics[width=0.31\textwidth]{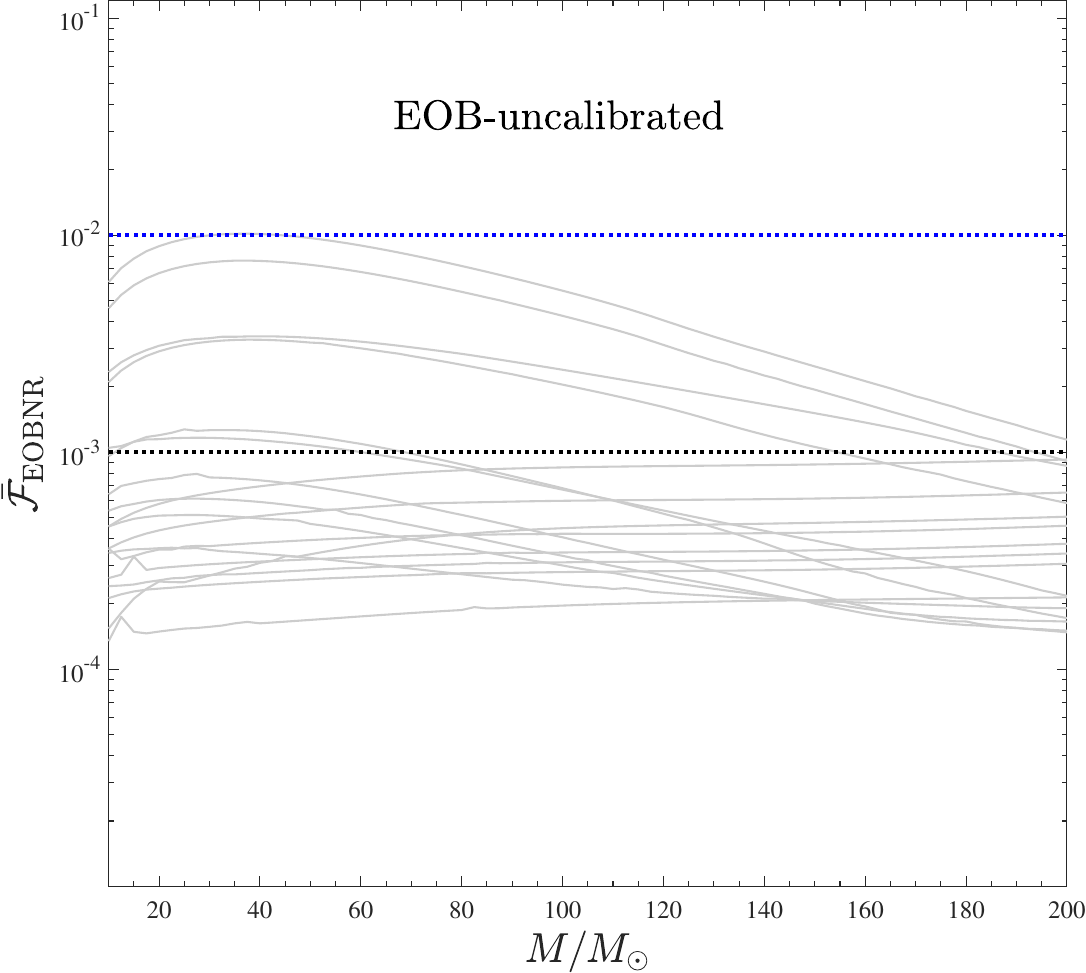}
	\caption{\label{fig:s0_leob_nocalibration}Uncalibrated nonspinning model. Performance of a 4PM-4PN, purely analytic, 
	dynamics on a sample of SXS nonspinning waveform with $1\leq q \leq 15$. The LEOB waveform is completed only 
	by NR-informed NQC corrections and ringdown, without any NR-calibrated effective parameter entering the dynamics. 
	Left panel: illustrative time-domain phasing for $q=4.5$. where the vertical, dot-dashed, lines indicate the alignment interval
	during the inspiral. The vertical dashed line marks the NR merger time, corresponding to the waveform amplitude peak.
	Middle panel: time-evolution of the waveform amplitude and frequency to illustrate the  contributions of the 
	NQC corrections and ringdown (see the text for precise description). Twice the orbital frequency $\Omega$ is represented 
	with a gray line. Right panel EOB/NR unfaithfulness with the advanced LIGO power spectral density.}
\end{figure*}

\begin{figure}[t]
	\center	
	\includegraphics[width=0.42\textwidth]{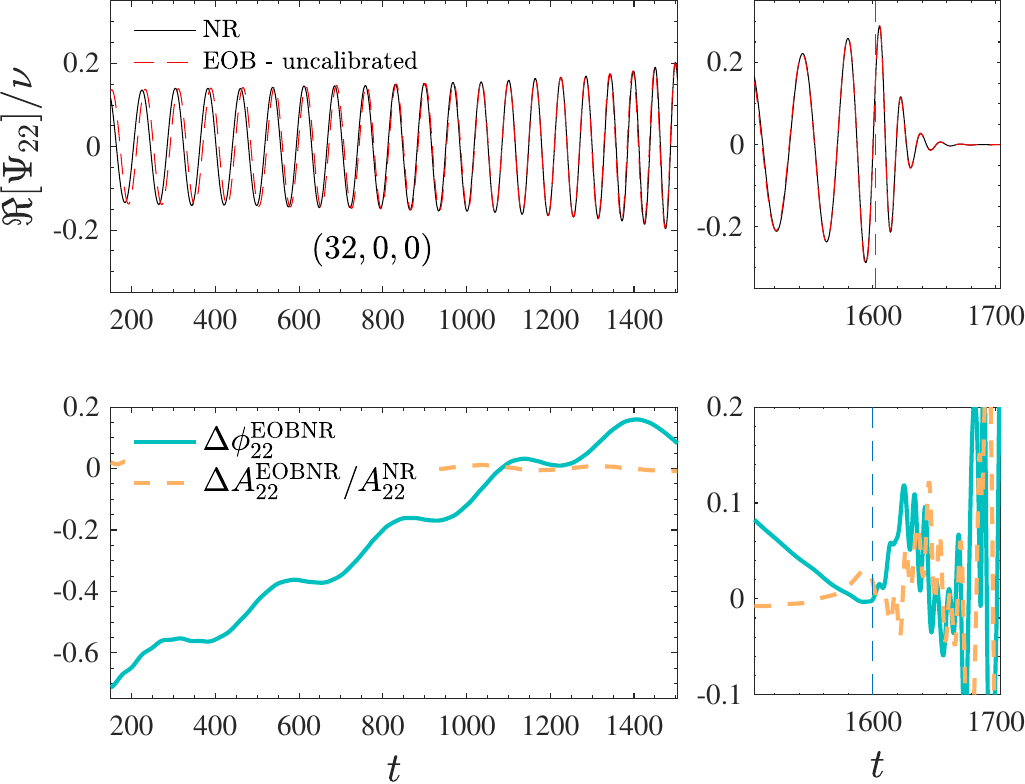}   
	\caption{\label{fig:q32}Uncalibrated, nonspinning model. Time domain phasing comparison
	for $q=32$ with the NR waveform of Ref.~\cite{Lousto:2020tnb}. The EOB and NR waveforms are
	aligned around merger time.}
\end{figure}

\begin{table*}[t]
	\caption{\label{tab:a52}Informing the orbital sector of the model using NR data to obtain
	an effective representation of the 5PM-5PN function $a_{52}^{\rm NR}(\nu)$ entering the 
	A potential at 5PM. From left to right, the columns report: an ordering index; the name of 
	the SXS simulation used; the mass ratio $q$ and the symmetric mass ratio $\nu=q/(1+q)^2$; 
	$\delta\phi^{\rm NR}_{\rm mrg}$ being an estimate of the NR phasing error at merger time, 
	equal to the $\ell=m=2$ phase difference 
	between the highest and second highest resolution available accumulated between $t=600M$ and the 
	peak amplitude of the highest resolution waveform (see also Table~II of Ref.~\cite{Nagar:2019wds});
	the first-guess values of $a_{52}^{\rm NR}$; and the quantities $\Delta\phi^{\rm EOBNR}_{22, \rm mrg}$ 
	being the phase differences $\Delta\phi^{\rm EOBNR}_{22}\equiv \phi^{\rm EOB}_{22}-\phi_{22}^{\rm NR}$ 
	computed at the NR merger (i.e., the peak of the $\ell=m=2$ waveform amplitude) for the corresponding values of $a_{52}^{\rm NR}$.}
	  \begin{center}
		\begin{ruledtabular}
   \begin{tabular}{c c c c c |c c} 
   No. & Name & $q$ & $\nu$ & $\delta\phi^{\rm NR}_{\rm mrg}$ & $a_{52}^{\rm NR}$ & $\Delta\phi^{\rm EOBNR}_{22} |_{\rm mrg}$  \\
   \hline
   \hline
   1      & SXS:BBH:0180 & 1    & $0.25$       & $-0.42$   & $66$      &  $-0.148$    \\
   2      & SXS:BBH:0169 & 2    & $2/9$  & $-0.027$  & $58$            & $-0.127$    \\
   3      & SXS:BBH:0168 & 3    & $0.1875$     & $-0.0870$ & $49$  & $-0.139$      \\
   4      & SXS:BBH:0166 & 6    & $0.1225$     & $\dots$   & $33$     &  $-0.116$      \\
   5      & SXS:BBH:0302 & 9.5 & $0.0862$     & $+0.0206$ & $22$  & $-0.040$      \\
      \end{tabular}
\end{ruledtabular}
\end{center}
\end{table*}

\begin{table}[t]
	\caption{\label{tab:nonspinning_datasets}EOB/NR maximum unfaithfulness ${\bar{\cal F}}_{\rm EOBNR}^{\rm max}$
	for the NR-uncalibrated nonspinning model, $a_{52}^{\rm NR}=0$ or for the NR-calibrated one, $a_{52}^{\rm NR}=263.55\nu-0.171$
	evaluated over a sample of nonspinning datasets up to $q=15$. The median value is $6.3\times 10^{-4}$ 
	in the uncalibrated case and $2.86\times 10^{-4}$ in the NR calibrated case.}
	\begin{center}
  \begin{ruledtabular}
	\begin{tabular}{ccc|cc}
	        & & & uncalibrated & calibrated \\
	        \hline
	  ID & $q$ & $\nu$ & ${\bar{\cal F}}_{\rm EOBNR}^{\rm max}$  & ${\bar{\cal F}}_{\rm EOBNR}^{\rm max}$\\ 	  
	  \hline
    SXS:BBH:0180   & 1   & 0.25       & 0.010161& 0.00031815 \\
	SXS:BBH:0007   & 1.5 & 0.24    &  0.0076394 & 0.00032733 \\
	SXS:BBH:0169   & 2   & 2/9 & 0.0034206 & 0.00044425 \\
	SXS:BBH:0259   & 2.5 & 0.204    & 0.0033002  & 0.00016364 \\
	SXS:BBH:0168   & 3   & 0.1875   & 0.0011738  & 0.00035578 \\
	SXS:BBH:0294   & 3.5 & 0.1728  & 0.0012745   & 0.00030085 \\
	SXS:BBH:0295   & 4.5 & 0.1488  & 0.00076691   & 0.00023531 \\
	SXS:BBH:0056   & 5   & 0.1389   & 0.00061075  & 0.00021191 \\
	SXS:BBH:0296   & 5.5 & 0.1302  & 0.00051645   & 0.0002876 \\
	SXS:BBH:0166   & 6   & 0.1224   & 0.00035937  & 0.00018594 \\
	SXS:BBH:0298   & 7   & 0.1094  & 0.00092568   & 0.0003495 \\
	SXS:BBH:0299   & 7.5 & 0.1038  &0.00065203    & 0.00028902 \\
	SXS:BBH:0063   & 8   & 0.0988  & 0.00050494   & 0.00021772 \\
	SXS:BBH:0300   & 8.5 & 0.0942 & 0.00045856    & 0.00029464 \\
	SXS:BBH:0301   & 9   & 0.09     & 0.00037875  & 0.00026429 \\
	SXS:BBH:0302   & 9.5 & 0.0862 & 0.00034149    & 0.00028545\\
	SXS:BBH:0303   & 10  & 0.0826 & 0.00030641    & 0.00022792 \\
	SXS:BBH:2477   & 15  & 0.0586  & 0.00021437   & 0.00024229 
 \end{tabular}
  \end{ruledtabular}
  \end{center}
  \end{table}

\begin{figure*}[t]
	\center	
        \includegraphics[width=0.305\textwidth]{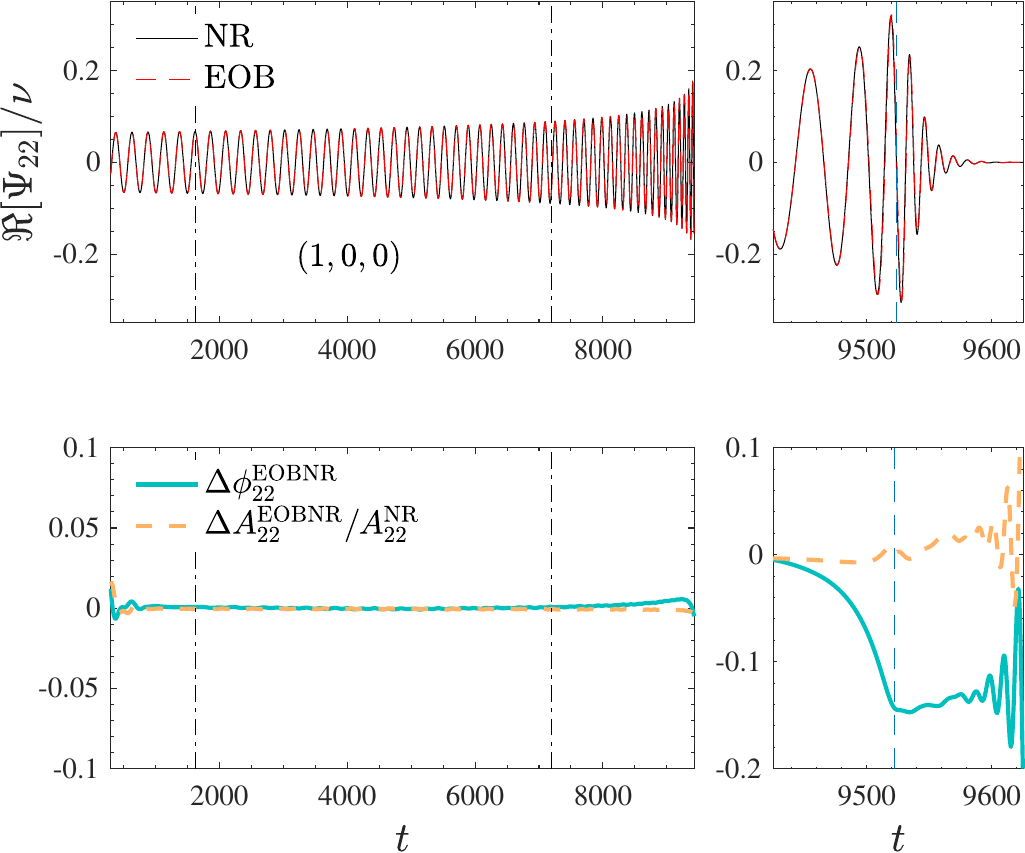}
	\includegraphics[width=0.31\textwidth]{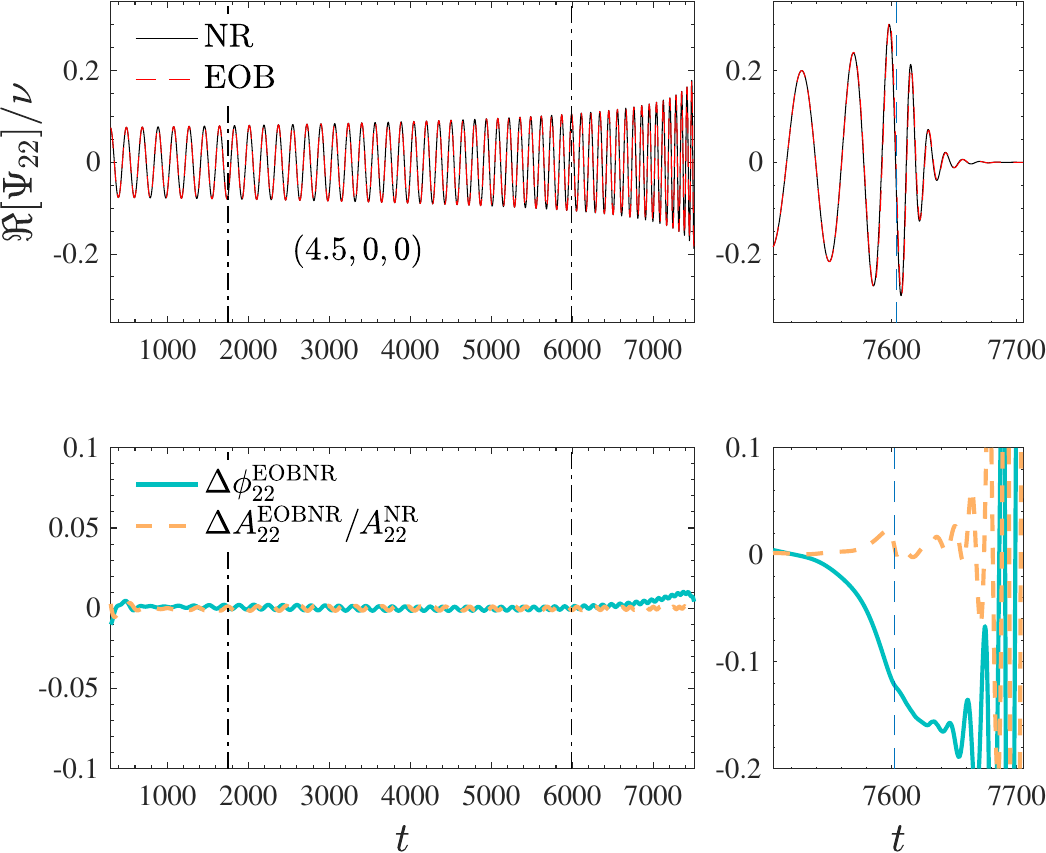}
	\includegraphics[width=0.31\textwidth]{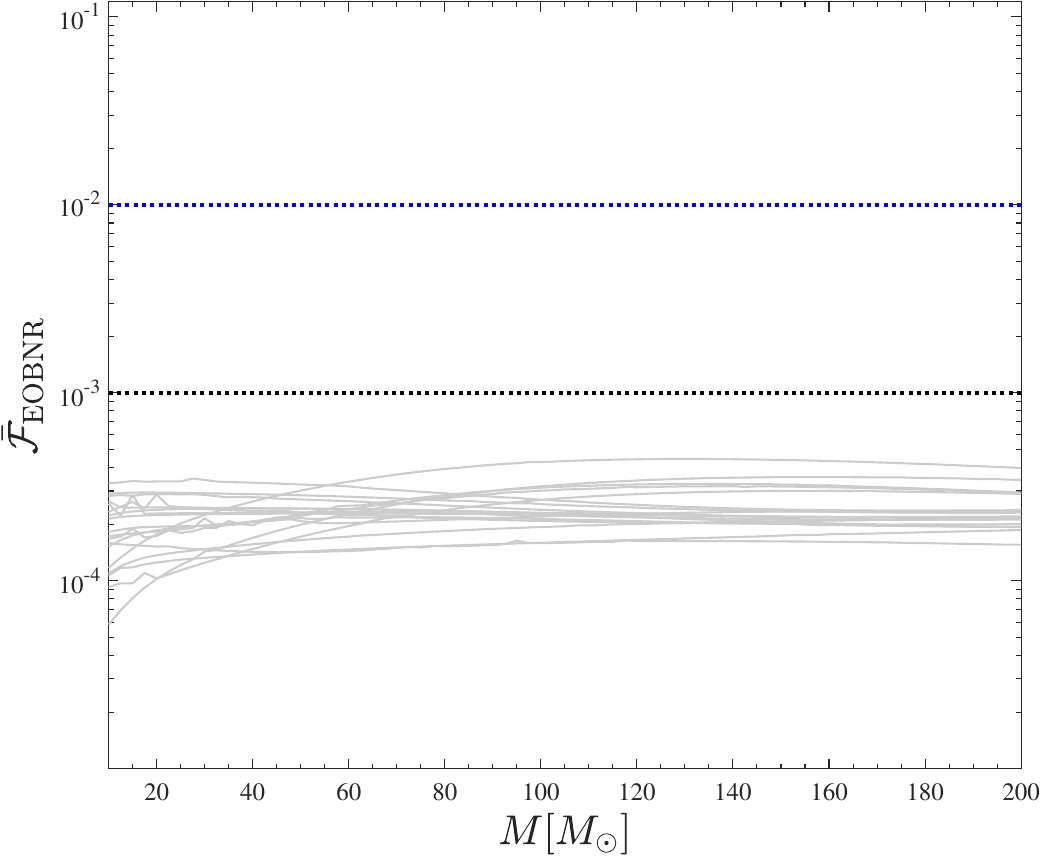}
	\caption{\label{fig:s0_leob}NR-informed, nonspinning case: performance with the 5PM-5PN parameter $a_{52}^{\rm NR}$ informed by NR simulations.
	Left and middle panels: time-domain phasings for two configurations: compare the $q=4.5$ one with Fig.~\ref{fig:s0_leob_nocalibration}. Rightmost 
	panel: $\bar{\cal F}_{\rm EOBNR}(M)$ for the same nonspinning NR datasets considered in Fig.~\ref{fig:s0_leob_nocalibration}. 
	The values of $\bar{\cal F}_{\rm EOBNR}^{\rm max}$ are listed in Table~\ref{tab:nonspinning_datasets}. For nearly equal-mass configurations the 
	NR-informed, effective, $a_{52}^{\rm NR}(\nu)$ function yields an improvement of almost two orders of magnitude.}
\end{figure*}

\begin{table*}[t]
 \caption{\label{tab:g42_coeff}Coefficients of the fits of the $\hat{g}_{32}^{\rm NR}$ data of Figs.~\ref{fig:g42_vs_a0} and~\ref{fig:g42_parspace} in Appendix~\ref{app:data} 
 in the form given by Eq.~\eqref{eq:g42fit}. The coefficients refer to either \LEOBa{} (without 3PM and 4PM residual spin-spin terms and using only equal-mass and
 equal-spin NR data) or \LEOBss{}, where these terms are considered.}
   \begin{center}
     \begin{ruledtabular}
\begin{tabular}{c |c c c c c c | c c c c c c} 
Model   & \multicolumn{12}{c}{\hspace{-28mm}$\hat{g}_{32}^{=}\equiv p_0\left(1 + n_1\tilde{a}_0 + n_2\tilde{a}_0^2 + n_3\tilde{a}_0^3 + n_4\tilde{a}_0^4 +n_5 a_0^5\right)$}\\
            &  \multicolumn{12}{c}{$\hat{g}_{32}^{\neq}\equiv \left(p_1\tilde{a}_0 + p_2\tilde{a}_0^2  + p_3\tilde{a}_0^3\right)\sqrt{1-4\nu}+ p_4\tilde{a}_0\nu \sqrt{1-4\nu} 
            +p_5\tilde{a}_0^2\nu\sqrt{1-4\nu}+p_6 \tilde{a}_0(1-4\nu)\nu + p_7 \tilde{a}_0(1-4\nu)^2\nu$}\\
            &  \multicolumn{12}{c}{$+\left(p_8\tilde{a}_{12}+ p_9\tilde{a}_{12}^2+ p_{10}\tilde{a}_{12}^3\right)\nu^2+ p_{11}\tilde{a}_0\sqrt{1-4\nu}\nu^2+p_{12}a_0^2\nu^2(1-4\nu)$}\\
            \hline
     & $p_0$ & $n_1$ & $n_2$   &  $n_3$   &  $n_4$  & $n_5$  & $p_1$ & $p_2$ & $p_3$ & $p_4$ & $p_5$ & $p_6$\\
\hline
\LEOBa{} & 84.891 & $-2.621$ & $-0.8459$  & $0.2551$ & $0.6247$ & $-0.6098$   &0  & 0   & 0  & 0  & 0 & 0   \\
                &  &  &   &  &  &    &$p_7$  & $p_8$   & $p_9$  & $p_{10}$  & $p_{11}$ & $p_{12}$   \\
                &  &  &   &  &  &    &0  & 0   & 0  & 0  & 0 & 0   \\

\hline
\LEOBss{} & 102.26 & $-0.409$ & $-0.323$  & $-0.243$ & $0.05548$ & $-0.095$   & $-1120.04$ & $-184.25$   &  $172.45$  & $-45999.59$  & $12.486$ & $15979.72$   \\
                              &  & &    &     &   &  & $p_7$ & $p_8$ & $p_9$ & $p_{10}$ & $p_{11}$ & $p_{12}$\\
                              &  & &    &     &   &  & $51250.62$ & $432.79$ & $-175.476$ & $-2827.247$ & $192213.47$ & $18952.94$
\end{tabular}
\end{ruledtabular}
\end{center}
\end{table*}

\begin{figure*}[t]
	\center	
	\includegraphics[width=0.32\textwidth]{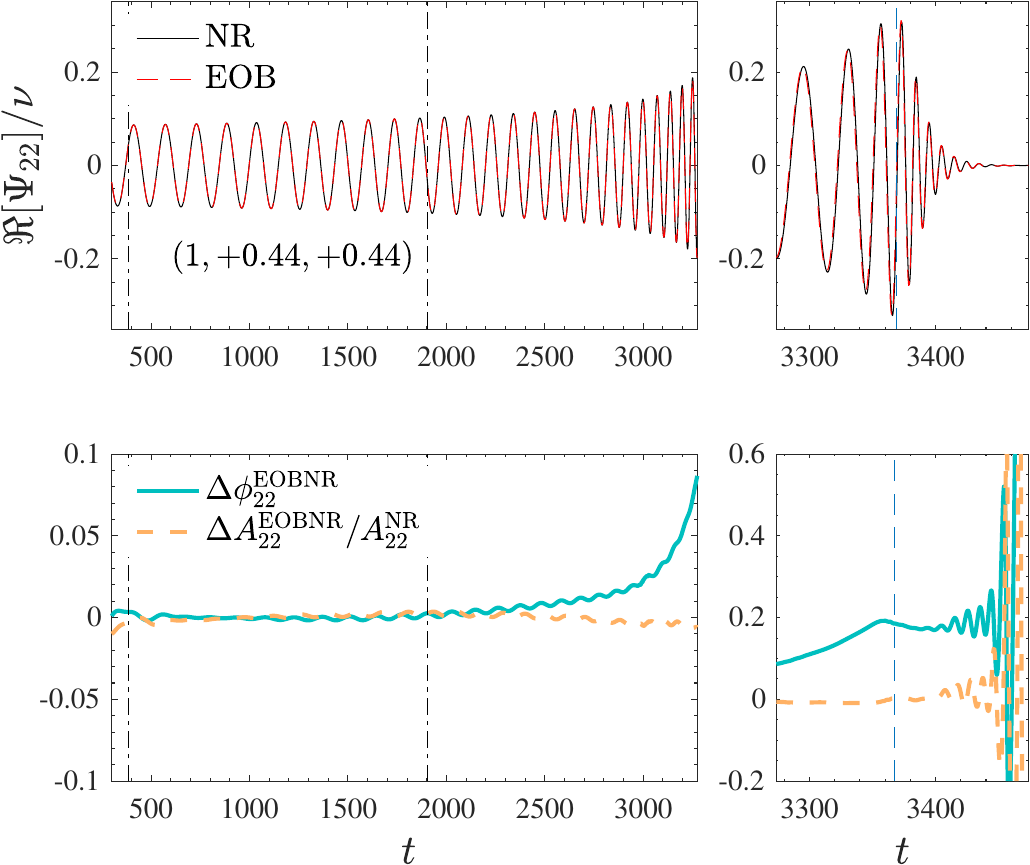}
	\includegraphics[width=0.32\textwidth]{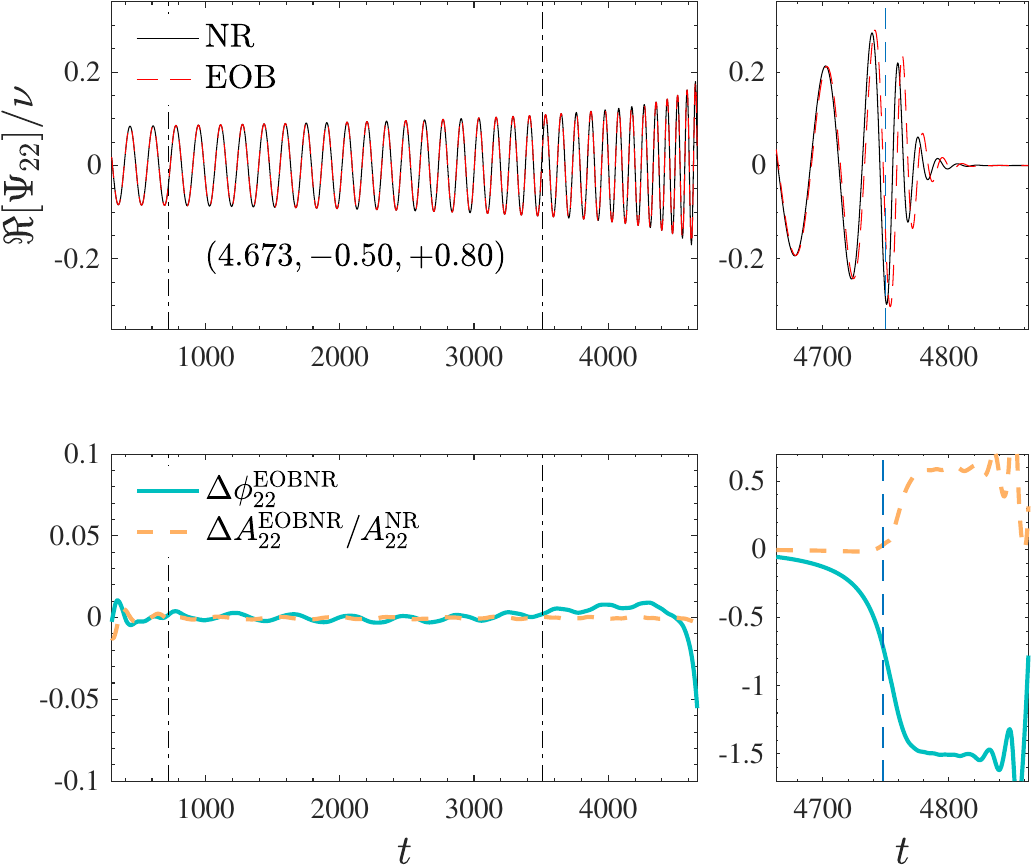} 
	\includegraphics[width=0.32\textwidth]{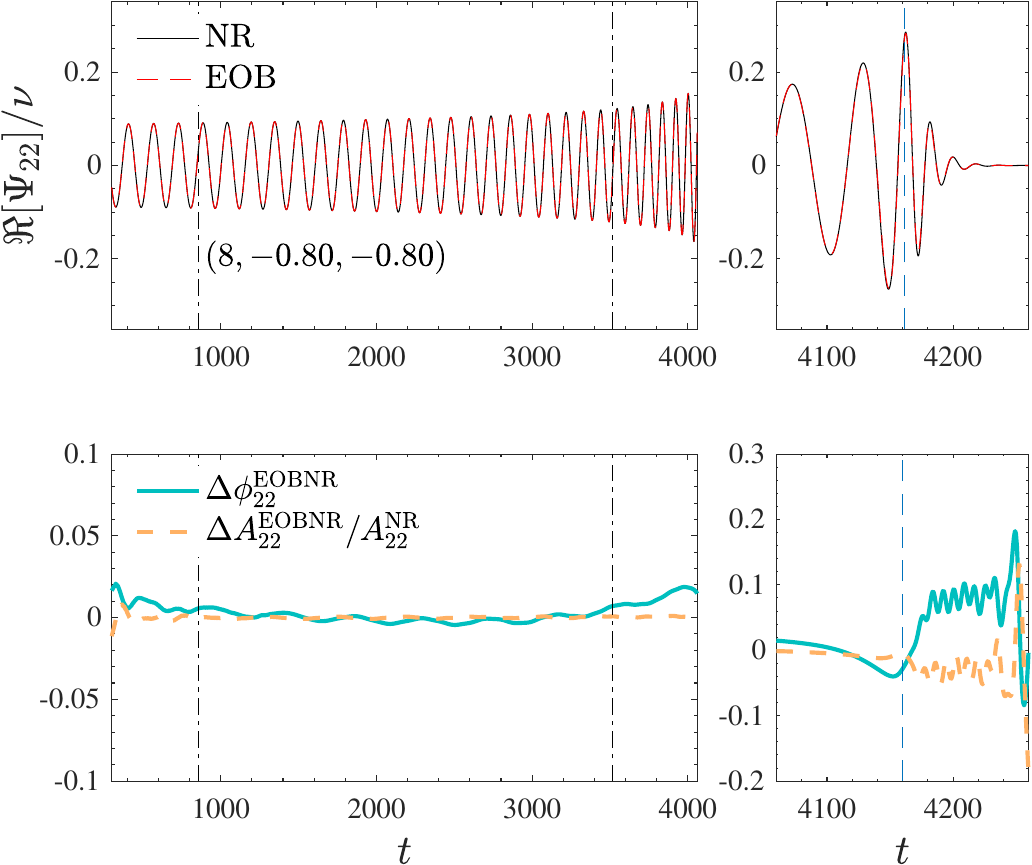} 	
	\caption{\label{fig:spin_phasing}Illustrative EOB/NR time-domain phasing comparison 
	for a selected sample of spin-aligned configurations using the \LEOBa{} model. 	
	The vertical lines in the left subpanels mark the alignment frequency interval, while the one in the right bottom panels 
	identifies the NR merger time. The rightmost panel refers to the {\tt SXS:BBH:1445} NR simulation, 
	that is also explicitly used in Ref.~\cite{Buonanno:2024byg} to show the performance of the 
	{\tt SEOBNR-PM} model developed there.
	The EOB/NR phasing agreement also in this case is rather good even if the spin-sector of the model  
	was NR-informed, through the function $\hat{g}_{32}^{\rm NR}$, by using only the equal-mass, 
	equal-spin SXS dataset listed in Table~\ref{tab:g42_eqmass} in Appendix~\ref{app:data}.}
\end{figure*}
\begin{figure*}[t]
	\center	
	\includegraphics[width=0.325\textwidth]{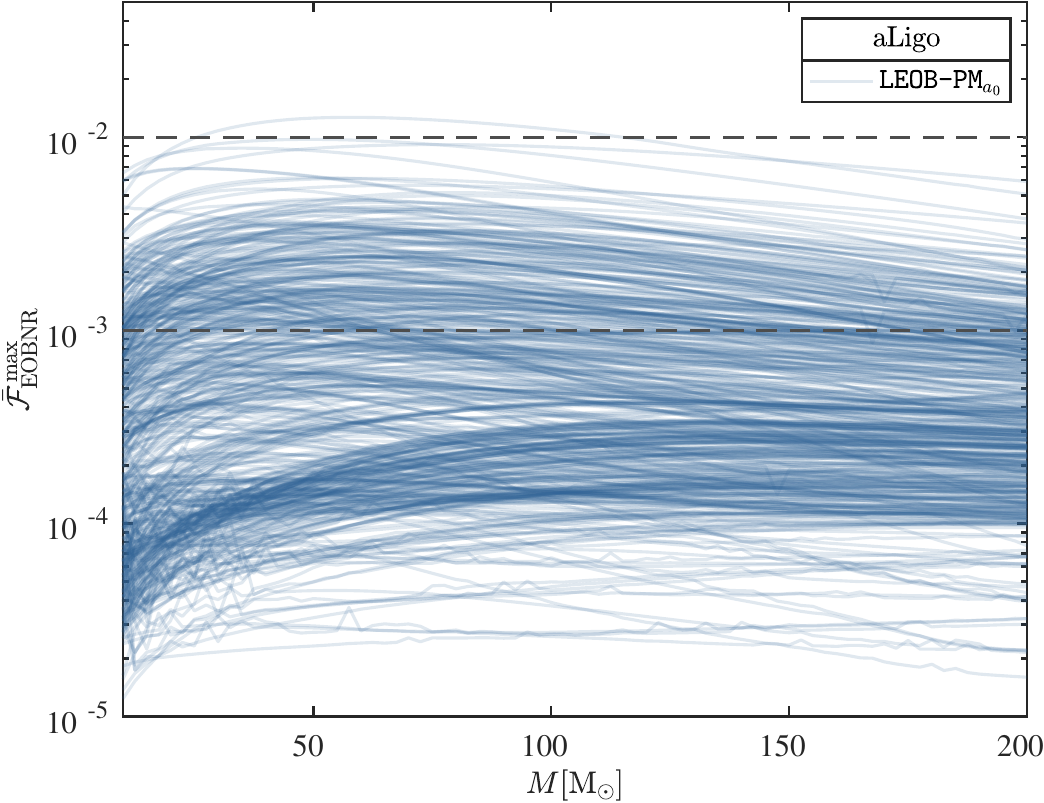} 
	\includegraphics[width=0.31\textwidth]{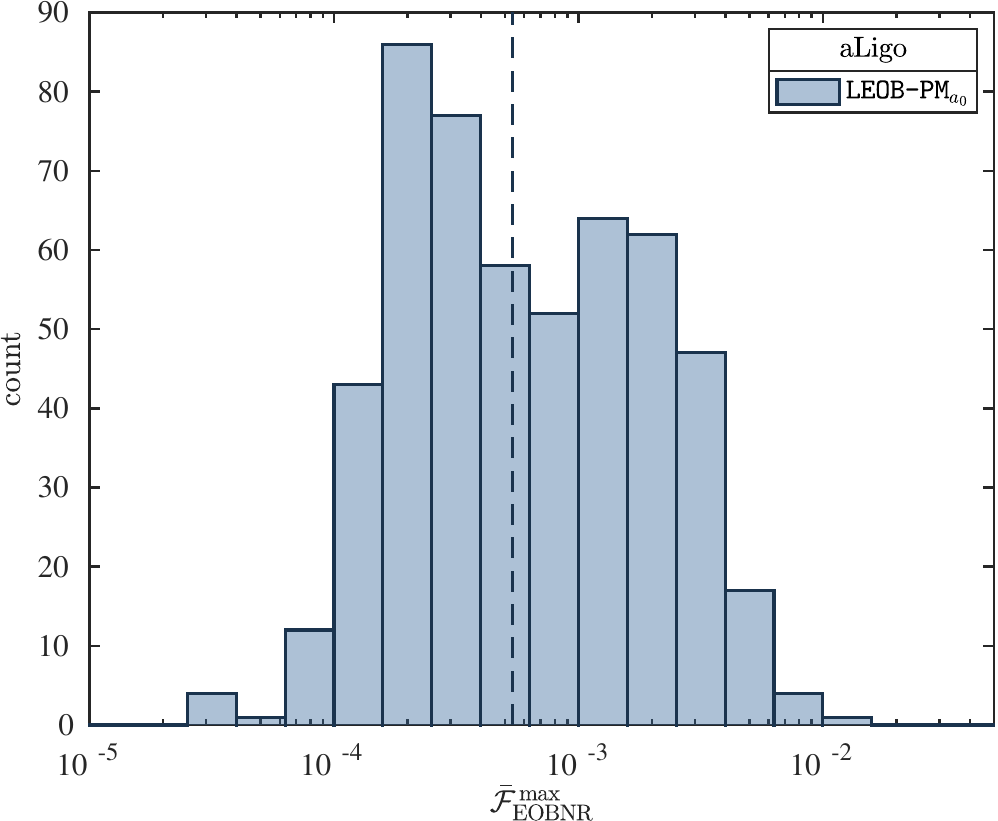}   
	\includegraphics[width=0.35\textwidth]{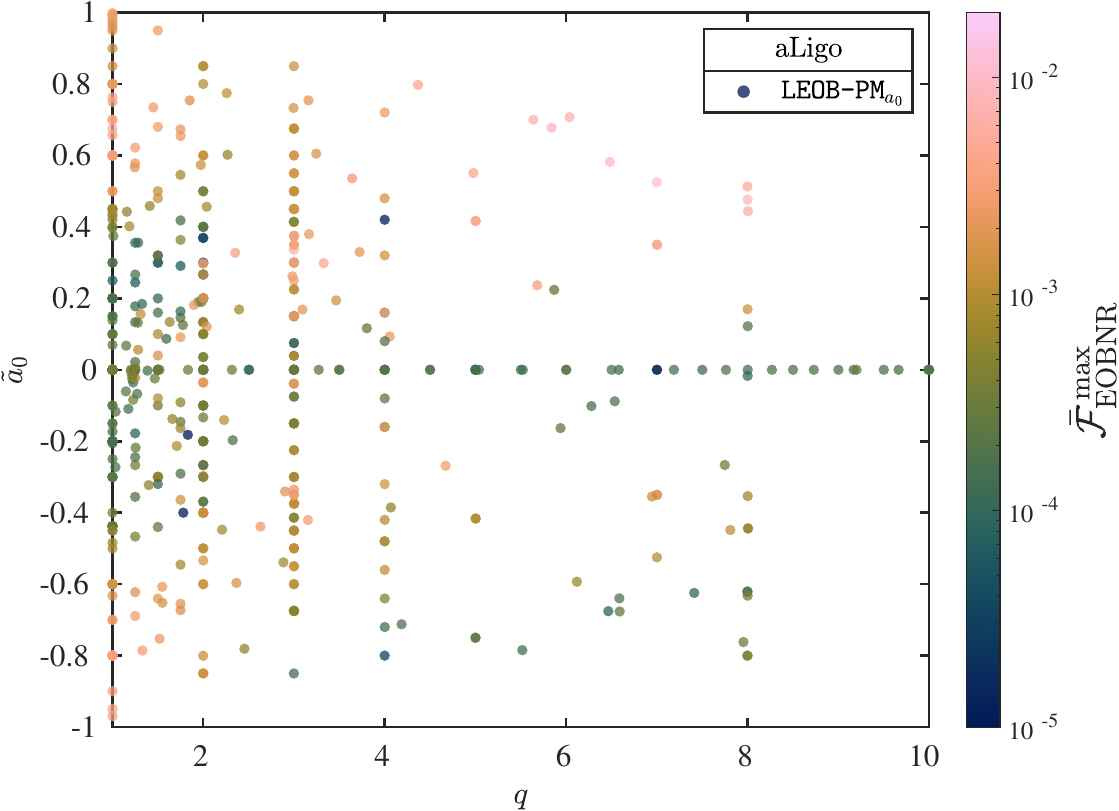}  
\caption{\label{fig:barF_full} EOB/NR unfaithfulness computation with the Advanced LIGO power spectral density
	all over a sample of 530 spin-aligned NR simulations of the public SXS waveform catalog using the \LEOBa{} model. 
	The left panel shows $\bar{\cal F}_{\rm EOBNR}(M)$, while the distribution of $\bar{\cal F}_{\rm EOBNR}^{\rm max}$
	is shown in the middle panel. The rightmost panel shows $\bar{\cal F}_{\rm EOBNR}^{\rm max}$ versus $(\tilde{a}_0,q)$.
	The performance of the model progressively degrades as the spins increase, especially when they are positive. 
	The median of the $\bar{\cal F}_{\rm EOBNR}^{\rm max}$ distribution is  $5.39\times 10^{-4}$, indicated by a vertical dashed line
	in the middle panel. Note that ${\rm Max}[\bar{\cal F}_{\rm EOBNR}^{\rm max}]\sim 0.01$ and corresponds to configurations with
	high spins and mass ratios in the range $4.5\lesssim q \lesssim 8$, as highlighted by the more pinkish points in the right panel of the figure.}
\end{figure*}

\subsection{NR-calibration parameters in the dynamics}
\label{nr:params}
Similarly to the case of \TEOBResumS{} model we will see that the performance
of the LEOB-based model can be improved by calibrating two effective parameters 
(in the dynamics) against NR simulations: one parameter enters the nonspinning 
part of the dynamics and another one the spin part.

In the orbital sector, we have chosen to NR-inform a 
parameter, dubbed $a_{52}^{\rm NR}$, that represents 
an additional (effective) 5PM-5PN term in our 
$A^{\rm 4PM\text{-}4PN}_{\rm orb}$.
The nonspinning $G^5$ term in our metric potential will then become
\be
\label{eq:a5}
a_5^0(\pinf,\nu)=a_{50}(\nu)+\nu\, a_{52}^{\rm NR}(\nu)\, p_\infty^2\,,
\ee
with $a_{50}(\nu)$ being the static 5PM contribution coming from our 4PN completion, Eqs.~\eqref{eq:4PN_loc} and \eqref{eq:Anonloc_4PN_expl}.
The $a_{52}^{\rm NR}(\nu)$ term is instead an effective function of $\nu$ to be informed from nonspinning NR simulations.

In the spin-dependent sector, we act similarly and 
include effective fractionally 4PN terms in the linear-in-spin 5PM contributions, which will then read
\begin{subequations}
\begin{align}
\hat{g}_3^0&= \hat{g}_{30}(\nu)+ \nu\, \hat{g}_{32}^{\rm NR}(\nu,\tilde{a}_i) \, p_\infty^2 \ , \\
\hat{g}_{*3}^0&= \hat{g}_{\rm * 30}(\nu) + \nu\,\hat{g}_{*32}^{\rm NR}(\nu,\tilde{a}_i) \, p_\infty^2 \ ,
\end{align}
\end{subequations}
where $[\hat{g}_{30}(\nu),\hat{g}_{30}(\nu)]$ are the analytically known static 5PM (fractionally 3PN) contributions, see Eqs.~\eqref{eq:g3pn}-\eqref{eq:g3pn*}.
The new $(\hat{g}_{32}^{\rm NR},\hat{g}_{*32}^{\rm NR})$ are effective parameters that can be informed via comparisons with NR simulations.
For simplicity, we assume in the present work that $\hat{g}_{32}^{\rm NR}=\hat{g}_{*32}^{\rm NR}$, so that in practice 
we calibrate only one parameter, $\hat{g}_{32}^{\rm NR}$ against a sample of spinning NR simulations. 
The determination of $a_{52}^{\rm NR}(\nu)$ and $\hat{g}_{32}^{\rm NR}(\nu,\tilde{a}_i)$ is presented in Sec.~\ref{nr:calibr} below. 

This NR-calibration framework differs from the usual one employed in PN-based \TEOBResumS{} models~\cite{Nagar:2014kha,Nagar:2018zoe,Nagar:2024oyk}.
In the latter, the nonspinning NR-informed coefficient (dubbed $a_6$) generally enters
the $A$ metric potential at 5PN-6PM order ($\propto u^6$).
Here, our parameters enters at the same PN order but a different PM one (proportionally to $u^5 p_\infty^2$).
The main difference however lies in the spinning sector. 
Within \TEOBResumS{}, the NR spin-sector parameter (generally called $c_{3}$) enters the gyro-gravitomagnetic functions at relative 3PN order, in place of the analytical results [$g_S^{\rm N^3LO}$ and $g_{S^*}^{\rm N^3LO}$, Eqs. (8) and (9)] of Ref.~\cite{Antonelli:2020aeb}.
The inclusion of the latter analytical 3PN terms within \TEOBResumS{} was attempted in Ref.~\cite{Nagar:2021xnh}, resulting
into a worse agreement with NR data. 
Within our current LEOB approach, we employ the corresponding analytical coefficients at fractional 3PN order [Eqs.~\eqref{eq:g3pn}-\eqref{eq:g3pn*}],
and postpone the NR information to relative 4PN order.
We discuss the performance of such model in the next Section.

\section{Model performance}
\label{sec:performance}

The evaluation of the model performance we do here closely follows  Ref.~\cite{Nagar:2024oyk}.  
We rely on: time-domain phasing analyses by computing the EOB/NR phase difference; and on the 
evaluation of the EOB/NR unfaithfulness, essentially  following the logic of 
Sec.~III of Ref.~\cite{Nagar:2024oyk}. Let us recall some relevant technical details.

Given two waveforms, $(h_1,h_2)$, the EOB/NR unfaithfulness, $\bar{\F}$, is a function of the total  mass $M$ of the binary defined as
\be
\label{eq:barF}
\bar{\cal F}(M) \equiv 1-{\cal F}=1 -\max_{t_0,\phi_0}\dfrac{\langle h_1,h_2\rangle}{||h_1||\,||h_2||},
\ee
where $(t_0,\phi_0)$ are the initial time and phase. Here $||h||\equiv \sqrt{\langle h,h\rangle}$,
with the inner product between two waveforms  defined as 
$\langle h_1,h_2\rangle\equiv 4\Re \int_{f_{\rm min}^{\rm NR}(M)}^\infty \tilde{h}_1(f)\tilde{h}_2^*(f)/S_n(f)\, df$,
where $\tilde{h}(f)$ denotes the Fourier transform of $h(t)$, $S_n(f)$ is the detector power spectral density,
and $f_{\rm min}^{\rm NR}(M)=\hat{f}^{\rm NR}_{\rm min}/M$ is the initial frequency of the
NR waveform at highest resolution. To avoid contamination from the NR junk radiation,
$\hat{f}_{\rm min}^{\rm NR}$ is taken to be 1.35 times larger than the actual initial frequency 
of $\tilde{h}(f)$ (i.e., 1.35 times the frequency where $|\tilde{h}|$ peaks; this is also the  choice 
adopted in Ref.~\cite{Pompili:2023tna}). Before taking the Fourier transform, the waveforms 
are  tapered in the time-domain to reduce high-frequency oscillations. For $S_n$, in our comparisons 
we use the zero-detuned, high-power noise spectral density of Advanced LIGO~\cite{aLIGODesign_PSD}.
For time-domain comparisons, we rely on our usual procedure described in Ref.~\cite{Baiotti:2011am},
see Eqs.~(29)-(31) therein.

Similarly to Ref.~\cite{Nagar:2024oyk}, we  first assess the performance of the model waveform in 
the nonspinning case and then we move to the two flavors of the complete spin-aligned case. 
In particular, focusing on the nonspinning case, we also quantify in detail the performance of the
{\it uncalibrated} LEOB model. Here, by uncalibrated model we mean the dynamics obtained by 
setting  $a_{52}^{\rm NR}=0$ in Eq.~\eqref{eq:a5} (and $\hat{g}_{32}^{\rm NR}=0$ if we were 
considering spinning system). Such a dynamics relies only on the analytic knowledge of the 
$A$ function together with the choice of its resummation using a $(1,4)$ Pad\'e approximant.

\subsection{Nonspinning case: uncalibrated model}
\label{sec:uncalibrated}
In the left panel of Fig.~\ref{fig:s0_leob_nocalibration}, we compare the time-domain phasing of the 
uncalibrated LEOB model, as defined above,  to a nonspinning NR dataset with $q \equiv m_1/m_2=4.5$. 
The phase difference $\Delta\phi_{22}^{\rm EOBNR}\equiv \phi_{22}^{\rm EOB}-\phi_{22}^{\rm NR}$
accumulated up to merger time is $\sim -1$~rad. The left plots use $ GM=1$ units. 
In this comparison, the dynamics is fully analytical, but the waveform is 
completed by the NR-informed  NQC corrections (to both phase and amplitude) and then by the NR-informed ringdown. 
The effect of the NQC correction factor and of the NR-informed ringdown is highlighted in the middle panel 
of Fig.~\ref{fig:s0_leob_nocalibration} for both the amplitude (top plot) and the frequency (bottom plot).
In both cases we have five curves: (i) the NR waveform, in black; (ii)(twice) the EOB orbital frequency $\Omega$; 
(iii) the {\it purely analytical} EOB resummed waveform (dashed, orange line, dubbed {\tt inspl} in the plot legend), 
without NQC correction, i.e. fixing $\hat{h}_{22}^{\rm NQC}=1$ in Eq.~\eqref{eq:hlm}; (iv) the EOB resummed 
waveform completed by NQC corrections We remark the excellent
quantitative agreement between NR and {\tt inspl} frequency up to $\sim 30$ before merger (frequencies are 
visually indistinguishable on the plot) and NR and {\tt inspl} amplitude up to $\sim 50$ before merger. Similarly,
it is also worth pointing out that $|\Psi_{22}^{\tt inspl}|$ peaks rather close (differences of order unity) to the actual
location of the NR peak; (iv) the outcome of the NR-informed NQC contributions are shown as dash-dotted light-blue
curves, dubbed {\tt inspl+NQC} in the plot. While the $\omega_{22}^{\tt inspl+NQC}$ frequency looks practically 
indistinguishable from $\omega_{22}^{\rm NR}$, the amplitude difference is noticeable. This is due to our choice of
NR-informing only two parameters around the merger amplitude, by only imposing a $C^1$ continuity condition
(see Ref.~\cite{Albanesi:2023bgi} for the discussion of this point in the test-mass limit); (v) the red curve, dubbed
{\tt inspl+NQC+RNG} shows the final result with the NR-informed ringdown attached.
The rightmost panel shows the EOB/NR unfaithfulness $\bar{\cal F}_{\rm EOBNR}$ versus $M/M_\odot$ for the same 18
nonspinning datasets considered in Ref.~\cite{Nagar:2024oyk}, which in particular include the $q=15$ nonspinning 
waveform of Ref.~\cite{Yoo:2022erv}. The information in the plot is complemented by Table~\ref{tab:nonspinning_datasets} 
which lists the value of $\bar{\cal F}^{\rm max}_{\rm EOBNR}$. We note that $\bar{\cal F}^{\rm max}_{\rm EOBNR}< 10^{-3}$ 
as $q>4$, with very small values reached for large values of $q$. 
In this respect, it is worth to evaluate the model also for large mass ratios, in particular the $q=32$ nonspinning
waveform calculated in Ref.~\cite{Lousto:2020tnb}. In Ref.~\cite{Nagar:2022icd}, Fig.~5, 
a time-domain phasing comparison between a $q=32$, resolution extrapolated, waveform and the quasi-circular 
{\tt TEOBResumS-GIOTTO} model was presented . The latter model incorporates a NR-informed parameter in the nonspinning sector,
on top of the usual NQC corrections. In our Fig.~\ref{fig:q32} we show the corresponding time-domain phasing comparison
obtained with LEOB and $a_{52}^{\rm NR}=0$. It looks consistent with, though different from, Fig.~5 of Ref.~\cite{Nagar:2022icd}.
In any case, the EOB/NR phasing agreement with $q=32$ remains remarkable considering that the model does
not use any information from this $q=32$ dataset and also the ringdown modelization was obtained putting together 
NR-information for mass ratios between $1\leq q \leq 10$ and the test-mass values of amplitude and frequency at 
merger (see Ref.~\cite{Nagar:2022icd} for additional information).

\subsection{Spinning case: NR-informed \LEOBa{} model}
\label{nr:calibr}

\begin{figure*}[t]
	\center	
	\includegraphics[width=0.32\textwidth]{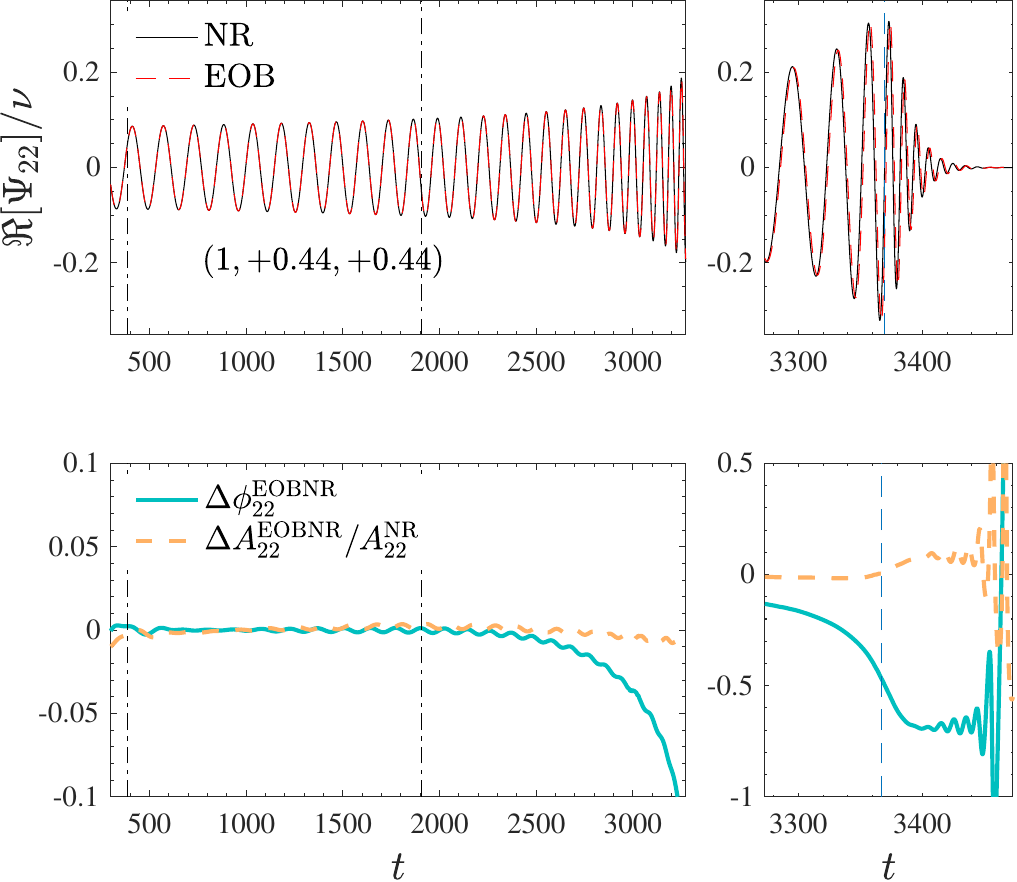} 	
	\includegraphics[width=0.32\textwidth]{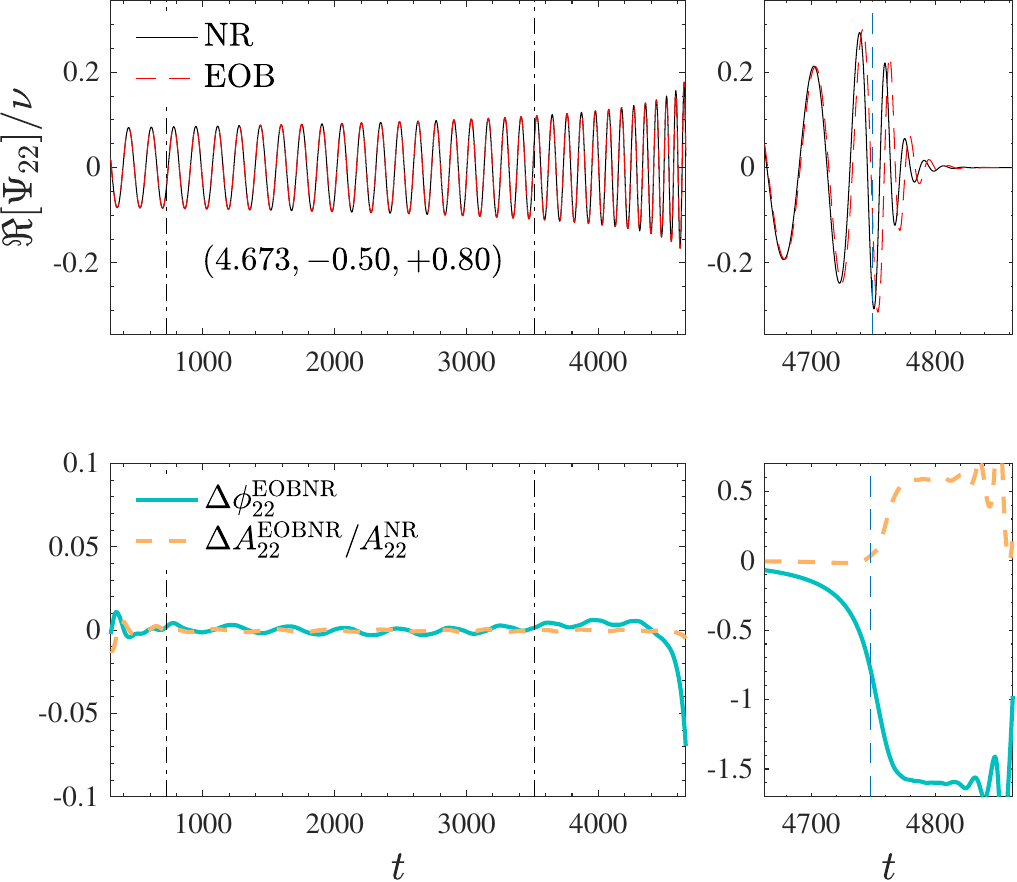} 
	\includegraphics[width=0.32\textwidth]{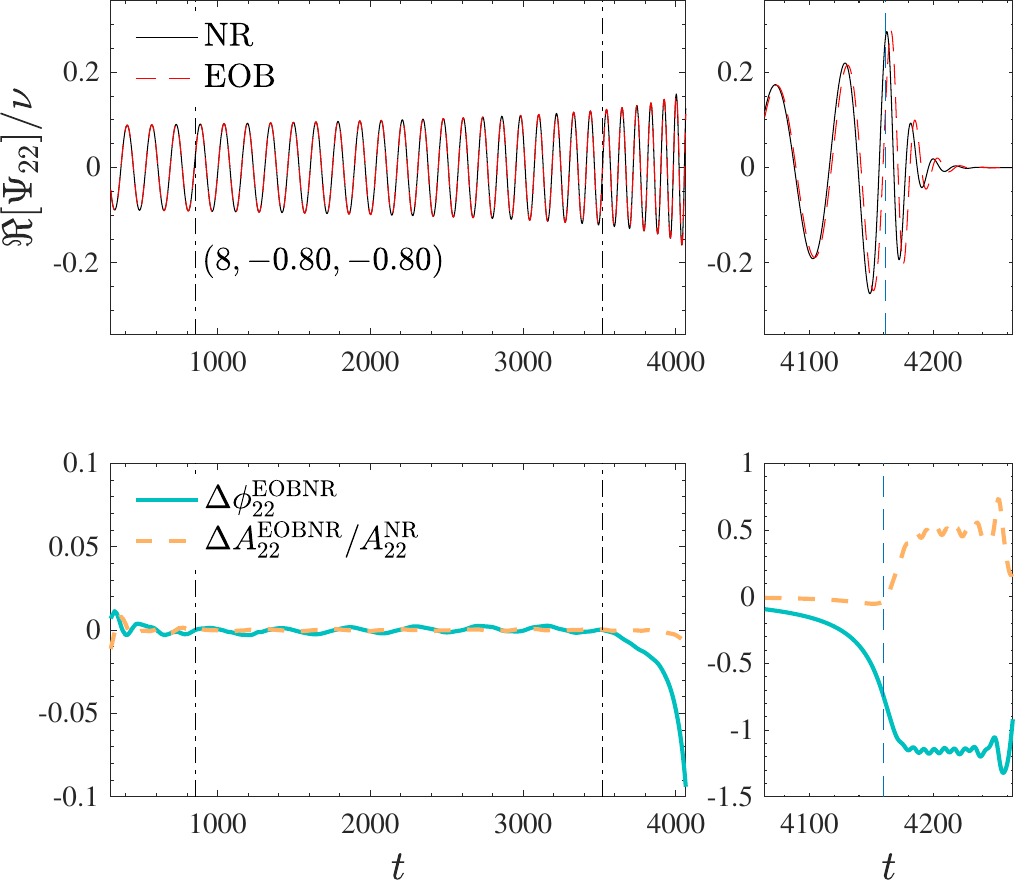} 
	\hspace{5mm}      
	\caption{\label{fig:spin_phasing_4PM}Incorporating 3PM and 4PM spin-spin effects within the $A$ function
	and related new determination of $\hat{g}^{\rm NR}_{32}$ for the \LEOBss{} model. EOB/NR time-domain phasing comparisons for the
	same configurations of Fig.~\ref{fig:spin_phasing}: note the changes in $\Delta\phi_{22}^{\rm EOBNR}$
	either in the inspiral (where it is flatter than before) and during merger ringdown. The vertical lines in the left
	subpanels mark the alignment frequency interval, while the one in the right bottom panels identifies the NR
	merger time.}
\end{figure*}
\begin{figure*}[t]
	\center	
	\includegraphics[width=0.32\textwidth]{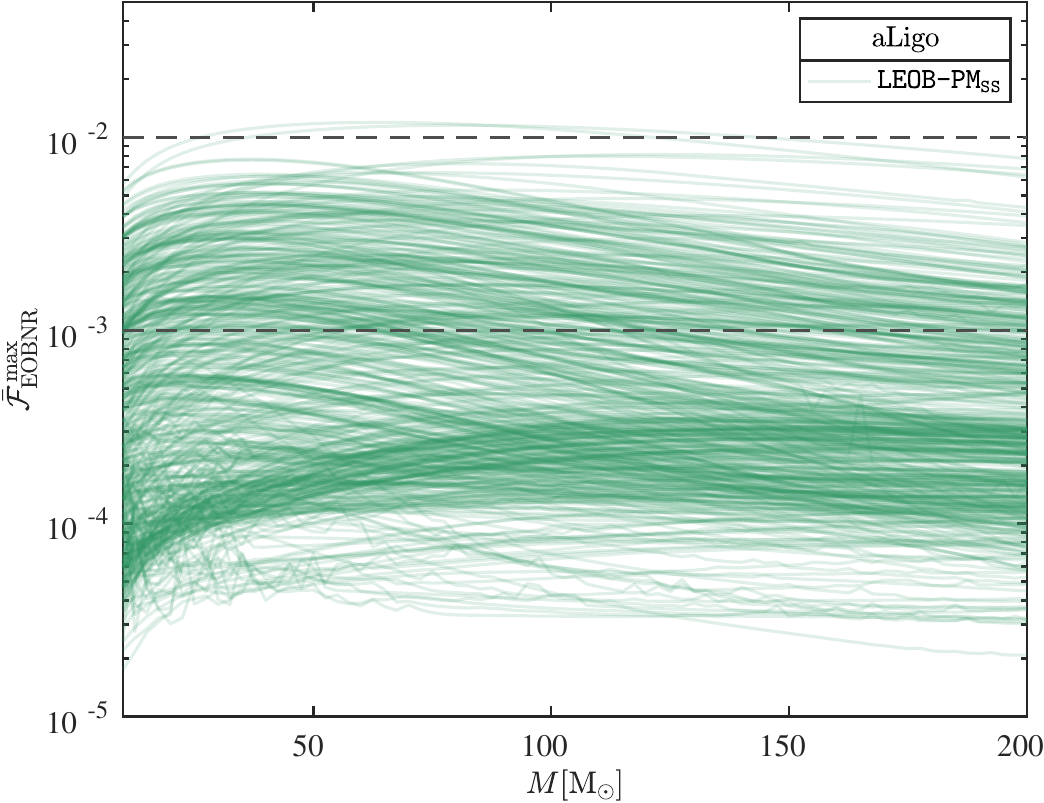} 
	\includegraphics[width=0.31\textwidth]{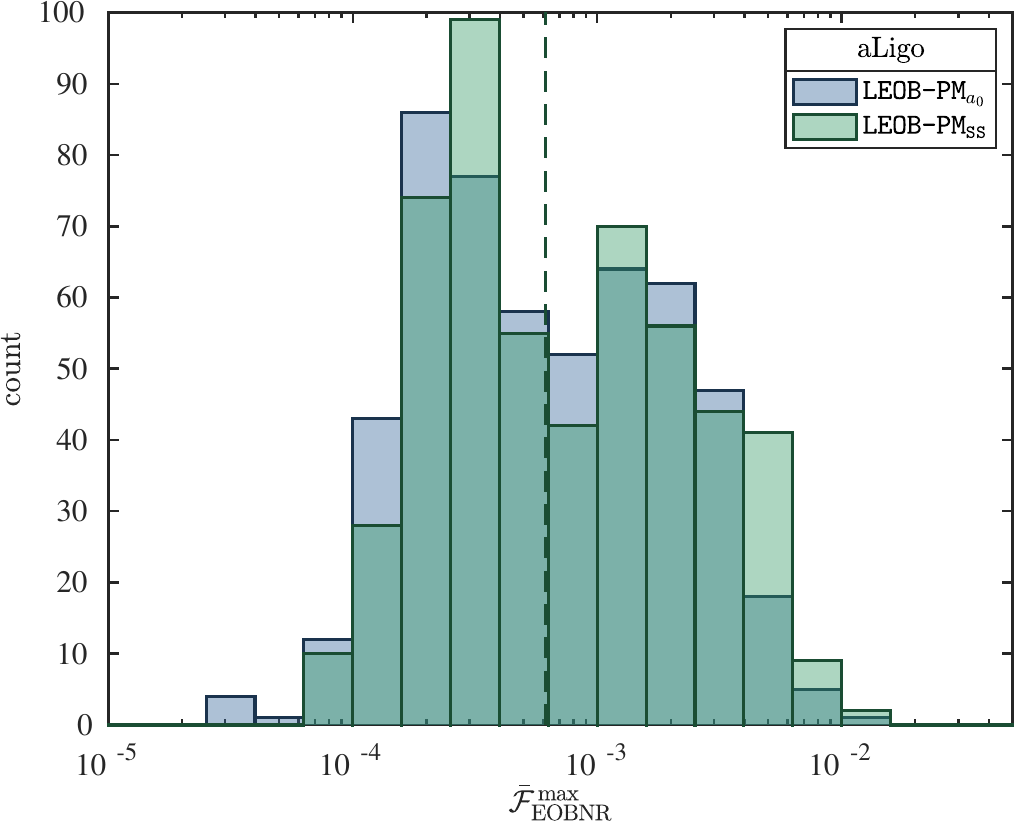}   
	\includegraphics[width=0.345\textwidth]{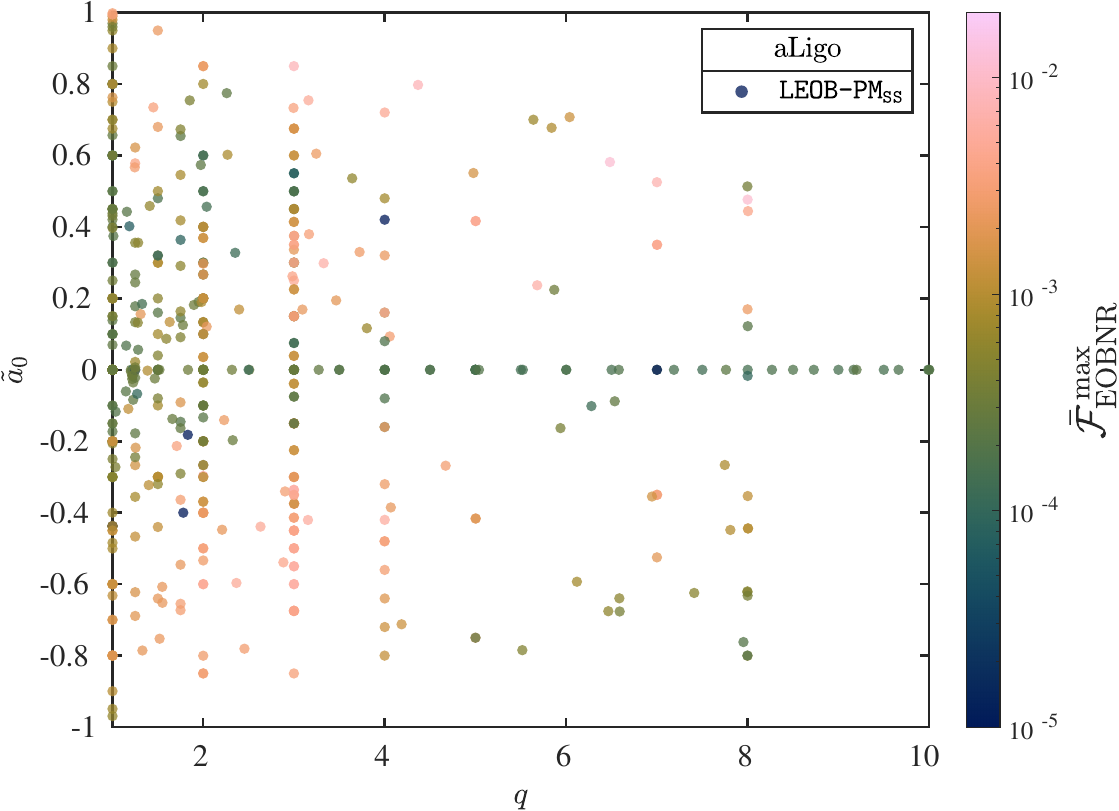}  
	\caption{\label{fig:barF_full_4PM_SS} Same as Fig.~\ref{fig:barF_full} but using the 3PN and 4PM spin-spin terms within
	the \LEOBss{} model. Note the improvement in the performance in the equal-mass, high spin corner, although globally 
	the median (dashed vertical line) has slightly worsened to $6.13\times 10^{-4}$. Note also the related different qualitative 
	behavior of $\bar{\cal F}_{\rm EOBNR}(M)$ for $M\lesssim 50 M_{\odot}$.}
\end{figure*}

\begin{figure}[t]
	\center	
        \includegraphics[width=0.4\textwidth]{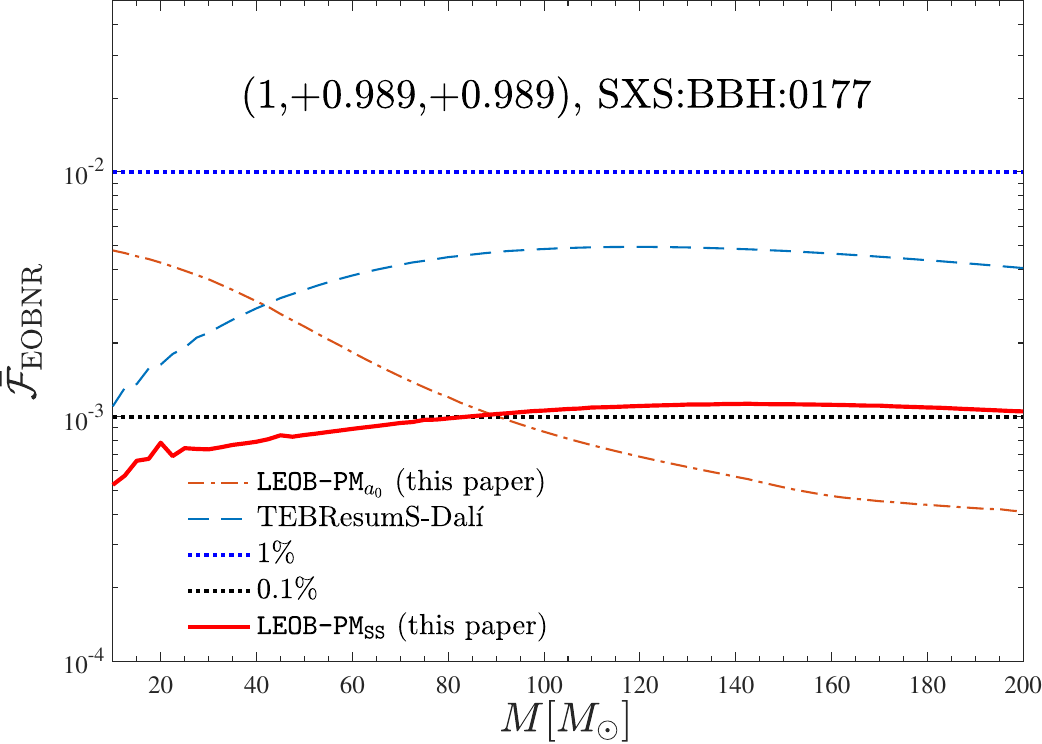}
	\caption{\label{fig:barF_099}
		Case $(1,+0.98952,+0.98952)$, {\tt SXS:BBH:0177}: comparing $\bar{\cal F}_{\rm EOBNR}$
	calculated with either \LEOBa{},  \LEOBss{} or with the \TEOBd{} model of Ref.~\cite{Nagar:2024oyk}. 
	The combined effect of the analytical spin-spin terms and the new $\hat{g}_{32}^{\rm NR}$ yields a qualitative and quantitative
	improvement of the \LEOBss{} curve with respect to either the \LEOBa{} and the  \TEOBd{} one.}
\end{figure}

Let us now turn first to discuss the performance of the \LEOBa{} model (defined by setting $A_{\rm SS}$ to zero), 
when improved by NR-informing the two parameters introduced in Sec.~\ref{nr:params} above. 
Let us focus first on the determination of $a_{52}^{\rm NR}(\nu)$ for
the nonspinning sector of the model, following precisely the procedure described in Sec.~IIIA of 
Ref.~\cite{Nagar:2024oyk}: the EOB and NR waveforms are aligned in the early inspiral and the 
parameter is chosen so that the
EOB-NR phase difference, required to be monotonically decreasing, is minimized around merger.
With this rationale in mind, we considered five datasets with mass ratios $q=(1,2,3,6,9.5)$ and determine
the corresponding values of $a_{52}^{\rm NR}$, which are reported in the sixth column of Table~\ref{tab:a52}.
One finds that they are quite accurately fitted by the following straight line
\be 
\label{a52fit}
a_{52}^{\rm NR}(\nu)=263.55\nu - 0.171 \ .
\ee
[We recall that $a_{52}^{\rm NR}(\nu)$ enters the model after multiplication by a factor $\nu$.]
The performance of such a NR-informed model for nonspinning configurations is illustrated in Fig.~\ref{fig:s0_leob}.
The leftmost panels refer to the time-domain phasing for $q=1$ and $q=4.5$, so it is possible to compare
with the corresponding phasing performance of the uncalibrated model in Fig.~\ref{fig:s0_leob_nocalibration}, with
basically a gain of $1$~rad due $a_{52}^{\rm NR}$. The plot is complemented by the values of $\bar{\cal F} _{\rm EOBNR}^{\rm max}$ 
listed in  Table~\ref{tab:nonspinning_datasets}. The latter table quantitatively highlights  the improvement reached, 
which are especially impressive near the equal-mass case.

Moving  to the spinning sector, let us tune the second effective parameter (now entering the spin sector) that we introduced 
above in Sec.~\ref{nr:params}. As a starting point, we determine $\hat{g}_{32}^{\rm NR}$ using only equal-mass, 
equal-spin datasets, both either aligned or anti-aligned with the orbital angular momentum. We do so 
using a restricted sample of spin-aligned dataset, only 12, that are reported in 
Table~\ref{tab:g42_eqmass} in Appendix~\ref{app:data}, with the corresponding best-choice of $\hat{g}_{32}^{\rm NR}$. 
The corresponding values of $\hat{g}_{32}^{\rm NR}$ are accurately fitted by a fifth-order polynomial 
in $\tilde{a}_0$ where the coefficients obtained from the fit are reported in Table~\ref{tab:g42_coeff}
with the {\tt LEOB} label.
Although the fit is done for the equal-mass, equal spin case,  in which $\tilde{a}_0=\chi_1=\chi_2$,
we, however, extend the result to the general case, using as argument $\tilde{a}_0=\tilde{a}_1 + \tilde{a}_2$.  
Figure~\ref{fig:spin_phasing} shows
the EOB/NR phasings for a few configurations, some of which were also considered in Fig.~9 of Ref.~\cite{Nagar:2024oyk}.
In particular, it is interesting to note that for the case $(8,-0.80,-0.80)$ \LEOBa{} performs {\it better} than
the  \TEOBd{} model, even if the latter received more NR-tuning for large mass ratios.
Note that this comparison might be considered merely qualitative, since the two models are very different,
different gauges are used and the spin-orbit sector of \LEOBa{} includes more analytic information, in particular
the fractional 3PN terms which were, in {\tt TEOBResumS-Dal\'i},  effective and NR-informed.
The middle panel of Fig.~\ref{fig:spin_phasing} also shows the phasing performance of {\tt SXS:BBH:1445}
that was used in Ref.~\cite{Buonanno:2024byg} to illustrate an example of time-domain phasing comparison with
{\tt SEOB-PM}, see Fig.~1 therein.  Though our phasing agreement is visually less good when looking 
at the real part of the waveform, this is quite acceptable in view of the minimal amount of calibration that we are using.
The EOB/NR performance in terms of unfaithfulness in analyzed in the three panels of Fig.~\ref{fig:barF_full}.
In the leftmost panel of Fig.~\ref{fig:barF_full} we show $\bar{\cal F}_{\rm EOBNR}$ in the range $10M_\odot\leq M\leq 200M_\odot$.
Globally, the performance of the model is rather good, with just one dataset with $\bar{\cal F}_{\rm EOBNR}^{\rm max}$ 
slightly above the $1\%$ level, as illustrated by the histogram in the middle panel of the figure. 
This corresponds to the NR simulation {\tt SXS:BBH:0202}, whose binary parameters are  
$(q,\chi_1,\chi_2) = (7.0,+0.60,0.00)$ and for which $\bar{\cal F}_{\rm EOBNR}^{\rm max} = 1.26\%$. 
As indicated by the
rightmost panel of the figure, the configurations grazing the $1\%$ are those with $\tilde{a}_0\sim 0.7$ and mass
ratio $4\leq q \leq 8$, highlighted as more pinkish points in the figure. By contrast, we also note that the equal-mass
configurations with high, positive, values of the spin have $\bar{\cal F}_{\rm EOBNR}^{\rm max}\sim 4 \times 10^{-3}$.
On the basis of our experience with the NR-calibration of {\tt TEOBResumS-Dal\'i}~\cite{Nagar:2024oyk}, the worsening 
of the performance of the model for large mass ratios is \textit{a priori} expected since we are only using equal-mass, equal-spin
data to NR-inform the function $\hat{g}_{32}^{\rm NR}$. In fact, it is remarkable that despite our very mild NR-calibration
the performance of the \LEOBa{} model remains good also for large mass ratios. 

\subsection{Spinning-case: NR-informed \LEOBss{} model}
\label{sec:ss_calibr}

Let us move on to exploring the \LEOBss{} model, in which one retains the 4PM-truncated $A_{\rm SS}$ contribution.
Here, the $A_{\rm SS}$ contribution is resummed together with the orbital contributions to the $A$ potential
by a $(1,4)$ Pad\'e approximant.
In Appendix~\ref{app:ss} we discuss another way to incorporate and resum the  $A_{\rm SS}$ contribution
[namely, by factorizing it  as $\hat{A}_{\rm SS}= 1+ \cdots$, and then resumming it with a $P^2_3$ approximant,
see Eq.~\eqref{eq:AtotFactorized}].
Appendix~\ref{app:ss} shows that this factorized approach  gives results 
substantially identical to those displayed in the previous section, but  with a worsening of 
the robustness of the model in some corners of the parameter space. 
When incorporating spin-spin corrections in the $A$ potential we need to complete the 
NR-fitted $\hat{g}_{32}^{\rm NR}$ function by adding an additional term, $\hat{g}_{32}^{\neq}$,
needed to improve the description of asymmetric (unequal-mass and unequal-spin) configurations.
The  NR-fitted $\hat{g}_{32}^{\rm NR}$ function is now separated in two
contributions
\be
\label{eq:g42fit}
\hat{g}_{32}^{\rm NR}(\nu,\tilde{a}_0,\tilde{a}_{12})=\hat{g}_{32}^=+\hat{g}_{32}^{\neq} \ ,
\ee
where $\hat{g}_{32}^=$ indicates the function obtained by fitting the values of $\hat{g}_{32}^{\rm NR}$ for
equal-mass, equal-spin binaries. We start working
with all 3PM, 4PM and 5PM contributions, determining $\hat{g}_{32}^{\rm NR}$ by inspecting the phase difference
as usual. The configurations we select are reported in Appendix~\ref{app:data}, notably Fig.~\ref{fig:g42_parspace}.
In doing so, we realized that there are  regions of the parameter space where it is necessary
[if we want to keep having a local maximum of the pure orbital frequency, Eq.~\eqref{eq:OmgOrb},
in order to proceed with the usual NQC determination and ringdown attachment]  to
complete  the term $\hat{g}_{32}^=$ obtained above by fitting equal-mass, equal-spin binaries,
by the extra term, $\hat{g}_{32}^{\neq}$, useful for describing  asymmetric configurations. 
For this reason we included several NR points in the range $3\leq q\leq 4$ and $-0.8\lesssim \tilde{a}_0\lesssim 4$, 
about 20 more than those composing the standard NR-calibration set of \TEOBd{} 
(see e.g.~\cite{Nagar:2024oyk} and references therein). These points are then fitted with a function of 
$(\nu,\tilde{a}_0,\tilde{a}_{12})$ that is modeled on the fitting function used for the effective spin-orbit 
parameter, but with more fitting coefficients. Despite the higher number of coefficients and NR anchor points,
the wave generation might remain not robust (in the sense described above) when using up to 5PM 
spin-spin, especially around the corner region of parameter space with $q=3$ and $\tilde{a}_0\sim -0.6$.

We have also explored this procedure when using the 5PM-accurate spin-spin term. However, this led
to a certain lack of robustness in the description of the dynamics. Leaving to the future an
attempt to keep the 5PM information by 
 either increasing the number of NR-anchor points or employing more sophisticated fitting procedures, we decided in the present work to work with a 4PM-accurate  $A_{\rm SS}$ contribution.
We have, however, verified that, wherever the wave generation mechanism
is robust, the 4PM and 5PM waveforms are consistent among themselves, with the 5PM ones typically
closer to the NR waveform by a few tenths of a radian around merger. This indicates that our results are
eventually more conservative than what the formalism would, in principle, allow us to get.
The result of this choice are reported in Figs.~\ref{fig:spin_phasing_4PM} (time-domain phasings) 
and~\ref{fig:barF_full_4PM_SS} (EOB/NR unfaithfulness). Concerning the time-domain phasing, it is instructive to compare
the left panel of Fig.~\ref{fig:spin_phasing_4PM} with corresponding one of Fig.~\ref{fig:spin_phasing},
corresponding to a $(8,-0.80,-0.80)$ binary. The waveform is aligned on the same frequency interval
(indicated by the two vertical lines in the left panel): the addition of high-order spin-spin terms 
yields that now $\Delta\phi^{\rm EOBNR}_{22}$ oscillates around zero during the inspiral and eventually
decreases monotonically up to merger. Similar considerations also hold for the other configurations. 
We have thus here a clear evidence of the effect of the residual, high-order, spin-spin term, an effect that cannot be
compensated by the tuning of $\hat{g}_{32}^{\rm NR}$. The calculation of the unfaithfulness is reported 
in Fig.~\ref{fig:barF_full_4PM_SS}. The results are globally consistent with the previous ones, confirming 
that the effect of these residual spin-spin information is small. We have, however, a notable improvement
of the performance for the $q=1$, high-spin cases. By contrast, the median of the distribution of maximum
unfaithfulness has slightly worsened, to  ${\rm Me}[\bar{\cal F}_{\rm EOBNR}^{\rm max}]=6.13\times 10^{-4}$.
There are two datasets corresponding to maximum unfaithfulness larger than $1\%$:
{\tt SXS:BBH:1439}, with $(q,\chi_1,\chi_2) = (6.48,+0.72,-0.32)$ and $\bar{\cal F}_{\rm EOBNR}^{\rm max} =1.19\%$;
and {\tt SXS:BBH:1441}, with $(q,\chi_1,\chi_2) = (8.00,+0.60,-0.48)$ and $\bar{\cal F}_{\rm EOBNR}^{\rm max} =1.15\%$.
Comparing the right panels of Figs.~\ref{fig:barF_full_4PM_SS} and \ref{fig:barF_full} we see that this 
result is certainly due to the behavior of the $q\simeq 3$, negative spin configurations and possibly might
be improved with a more precise determination of $\hat{g}_{32}^{\rm NR}$ (and related fitting function) in this regime.
\begin{figure*}[t]
	\center	
	\includegraphics[width=0.305\textwidth]{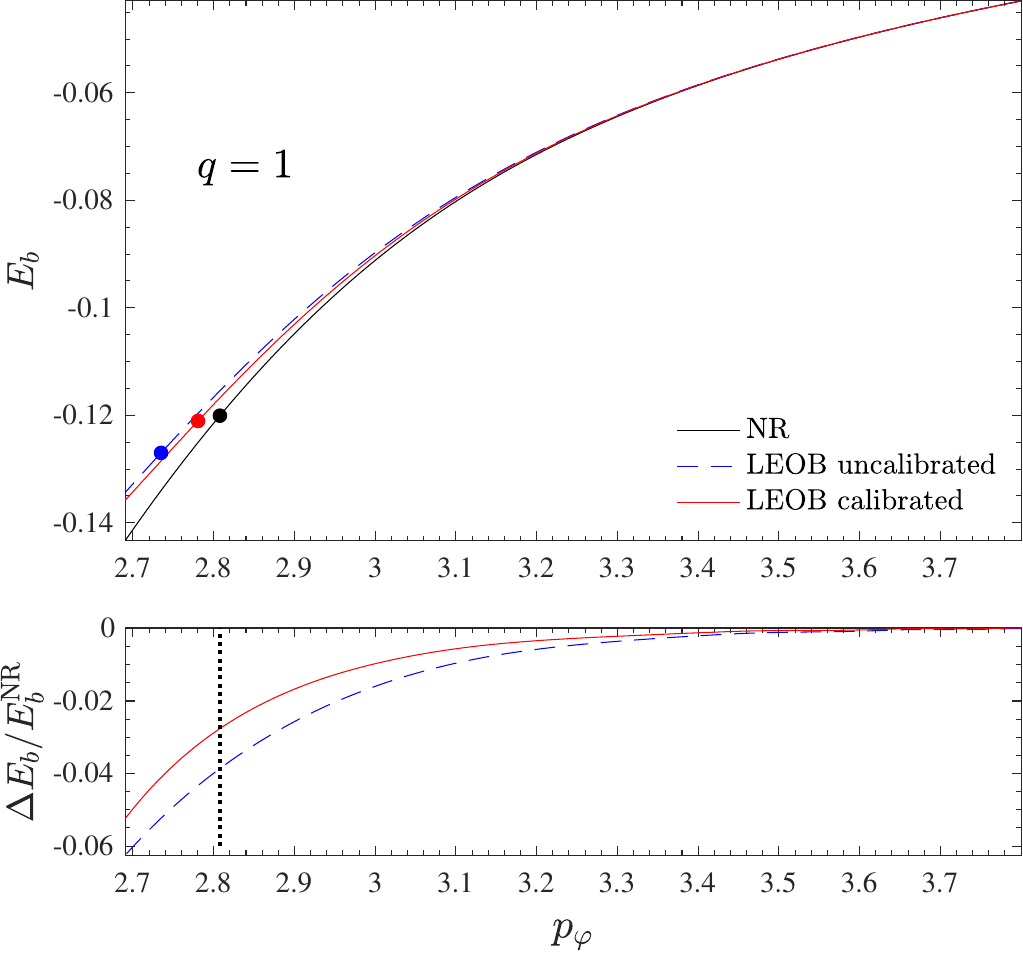} 
	\hspace{2mm}
	\includegraphics[width=0.31\textwidth]{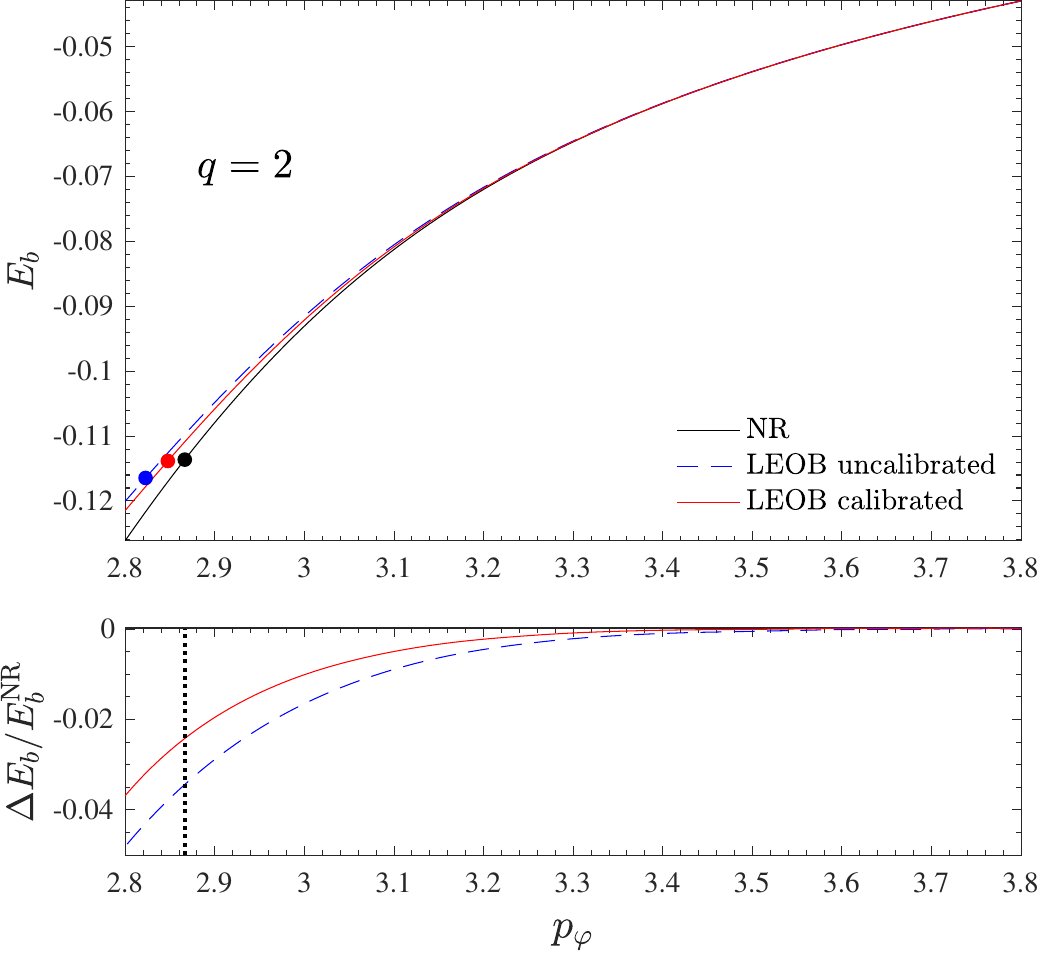}   
	\hspace{2mm}
	\includegraphics[width=0.31\textwidth]{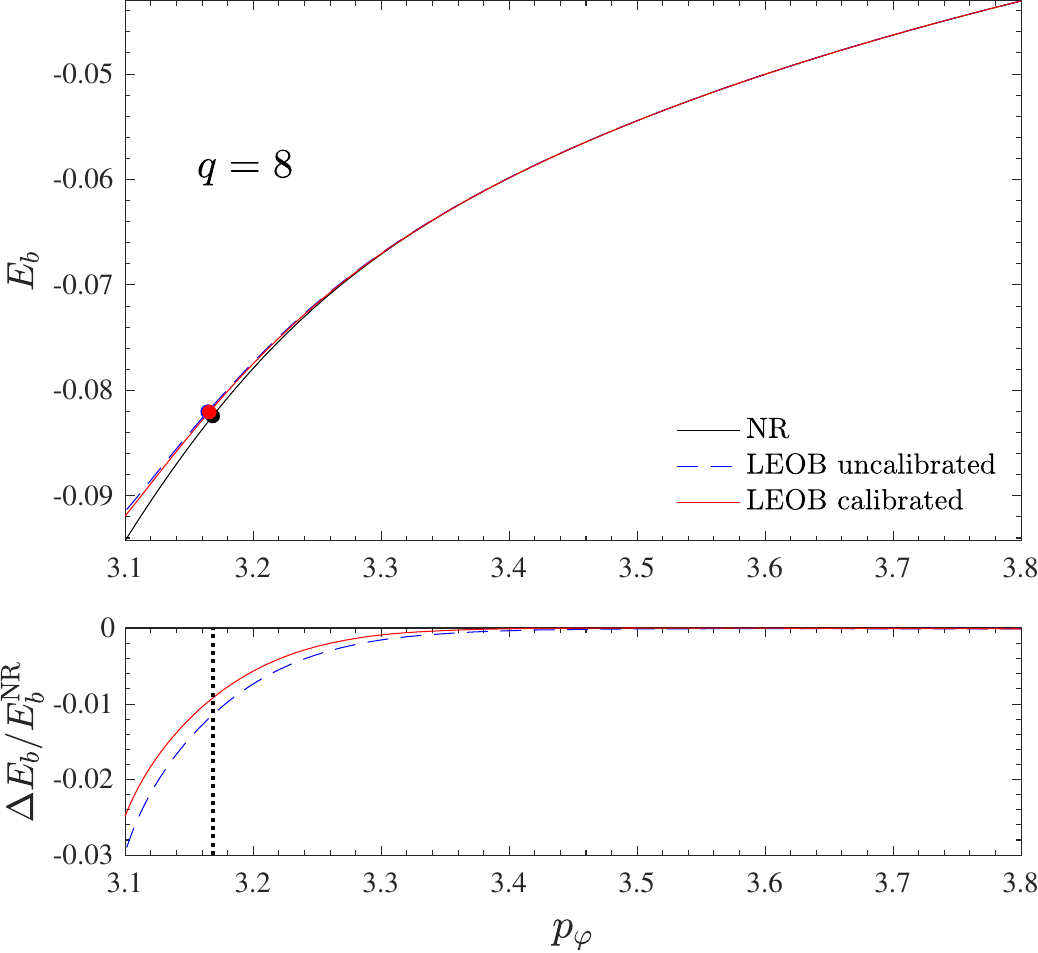}  
	\caption{\label{fig:ej}Comparison between binding energy curves versus rescaled angular momentum for three nonspinning configurations. In the bottom panel, 
	we show the relative differences $\Delta E_b/E_b=(E_b^{\rm LEOB}-E_b^{\rm NR})/E_b^{\rm NR}$. We consider two flavors of LEOB: without NR calibration in the dynamics, $a_{\rm 52}^{\rm NR}=0$,
	and with calibrated dynamics, i.e. $a_{\rm 52}^{\rm NR}$ given by Eq.~\eqref{a52fit}. Markers on the curve indicate the merger location (i.e. the location of the peak of the
	$\ell=m=2$ waveform amplitude) for each dynamics. The (absolute value of the) fractional difference at NR merger for the uncalibrated dynamics varies between $4\%$, for $q=1$ 
	to $1\%$ for $q=8$.}
\end{figure*}
As a final comment of this analysis and on the impact of spin-spin subdominant terms, 
we want to emphasize a point related to the qualitative behavior of the $\bar{\cal F}_{\rm EOBNR}(M)$ curves 
(for both \LEOBa{} and \LEOBss{}) with respect to the corresponding ones 
of  {\tt TEOBResumS-Dal\'i} in Fig.~10 of Ref.~\cite{Nagar:2024oyk}. 
For \LEOBa{} the curves typically have a local maximum between $50-100 M_\odot$ and then progressively 
decrease for large masses. This is also the case for $q=1$ datasets with high, positive, spins: the corresponding 
curves in the left panel of Fig.~\ref{fig:barF_full} are those that start at $\simeq 4\times 10^{-3}$ and then are seen 
to decrease monotonically versus $M$. On the contrary, for \TEOBd{} the curves are mostly monotonically increasing,
with a behavior qualitatively closer to the one of \LEOBss{}.
To highlight this fact, in Fig.~\ref{fig:barF_099} we display these three curves for $(1,+0.98952,+0.98952)$, 
corresponding to dataset {\tt SXS:BBH:0177}, so that the qualitative and quantitative differences are apparent. 
We remark the qualitative consistency between \TEOBd{}  and \LEOBss{} curves, although in this
second case the model actually performs much better. It is remarkable that for such a high-spin configuration
 \LEOBss{} also performs better than the quasi-circular {\tt TEOBResumS-GIOTTO} model, 
 see Refs.~\cite{Nagar:2020pcj,Riemenschneider:2021ppj}.
Since subdominant spin-spin terms are incorporated in both {\tt TEOBResumS-GIOTTO}
\TEOBd{} as corrections to the Kerr-like centrifugal radius~\cite{Damour:2014sva}, this comparison 
suggests that an approach similar to the one used here, i.e. within the corresponding PN-based $A_{\rm orb}$ 
function, might be useful to further improve the performance of these, HEOB-based, models in the 
high-spin region of the parameter space.

\subsection{Energetics in the nonspinning case}
\label{sec:energetics}
\begin{figure}[t]
	\center	
	\includegraphics[width=0.45\textwidth]{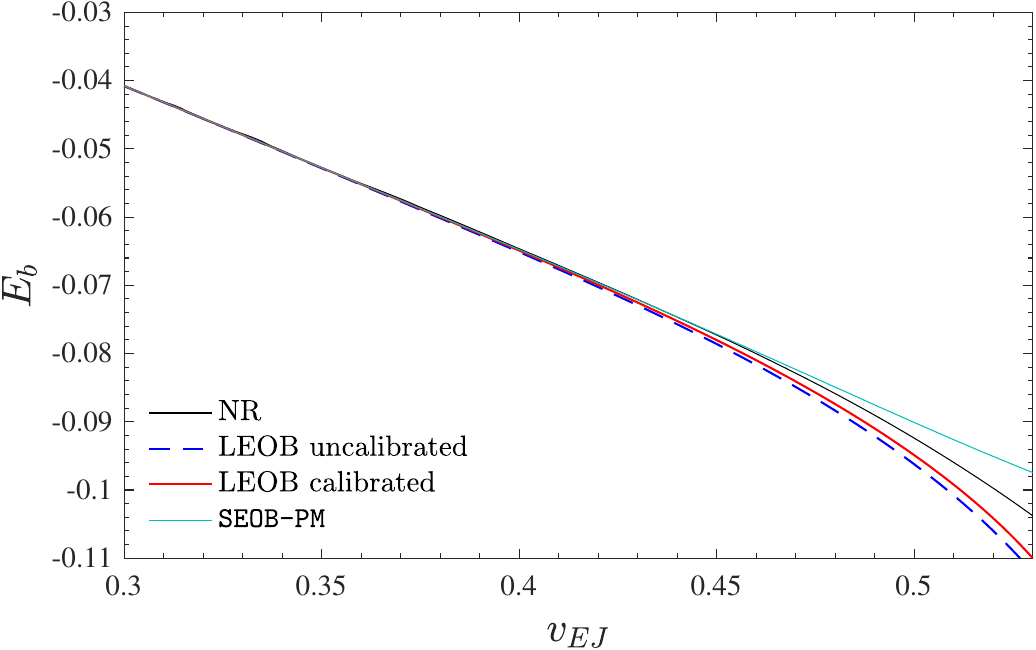} 	
	\caption{\label{fig:Eb_vs_vOmg}Complement to Fig.~\ref{fig:ej}: LEOB/NR comparison between binding energies 
	versus the velocity-like variable $v_{EJ}$ (see the text for definition) for $q=1$. 
	The figure also reports the corresponding {\tt SEOB-PM} curve taken from Fig.~2 of Ref.~\cite{Buonanno:2024byg}.
	}
\end{figure}
Let us also mention in passing the performance of our LEOB model for what concerns the dynamics,
represented in particular using the gauge-invariant relation between the energy and angular 
momentum~\cite{Damour:2011fu,Nagar:2015xqa,Ossokine:2017dge}.
In Fig.~\ref{fig:ej} we present the dimensionless binding energy $E_b=(E-M)/\mu$ as a function of $p_\varphi =J/(M \mu)$ for three illustrative 
nonspinning configurations (namely $q=1,2$ and $8$).
The NR curves were computed in Ref.~\cite{Nagar:2015xqa}, to which we address the reader for the technical details
and for their error estimate. In the plot we also show two kinds of LEOB curves: the uncalibrated ones (dashed blue lines), 
with $a_{\rm 52}^{\rm NR}=0$, or the NR-calibrated ones (solid red) with $a_{\rm 52}^{\rm NR}$ given by the 
NR-informed function of Eq.~\eqref{a52fit}.
The relative energy differences $\Delta E_b/E_b^{\rm NR}\equiv (E_b^{\rm LEOB}-E_b^{\rm NR})/E_b^{\rm NR}$ 
are shown in the bottom panels of Fig.~\ref{fig:ej}. At the NR merger location (black markers on the top panels) 
the (absolute value of the) difference for the uncalibrated model varies between $\sim 4\%$ for $q=1$ to $\sim 1\%$ for $q=8$. 
The NR-calibration improves the LEOB/NR agreement, especially for the nearly equal-mass case, consistently with 
our finding for the waveform. In this respect, it is interesting to note (see top panels of Fig.~\ref{fig:ej}) that the NR calibration 
helps make the LEOB and NR $E_b$ values at merger very close (compare the red and black dots), 
while differences remains at the angular momentum level. The LEOB/NR merger consistency is striking for 
the $q=8$ case.
As mentioned in the Introduction, Ref.~\cite{Buonanno:2024byg} developed the only other existing waveform model incorporating
PM information, i.e.~the  {\tt SEOBNR-PM}. As discussed in Sec.~\ref{sec:start}, the latter model is based on a 
recursively-defined Hamiltonian. Reference~\cite{Buonanno:2024byg} compared the {\tt SEOB-PM} 
dynamics\footnote{There is a distinguo between {\tt SEOB-PM} that is not NR-calibrated 
and {\tt SEOBNR-PM} where some NR calibration is introduced in the waveform.} 
with the NR dynamics taken from Ref.~\cite{Ossokine:2017dge}, considering several configurations 
varying the mass ratio and the individual spin. More precisely, in Ref.~\cite{Ossokine:2017dge} the NR dynamics 
is eventually represented by means of NR-informed fits, that are then used in Ref.~\cite{Buonanno:2024byg}.
Note that, instead of using, as we do, the $E_b(p_\varphi)$ relation, they rely (see Fig.~2 in Ref.~\cite{Buonanno:2024byg})
on a link $E_b(v)$, where $v$ is some velocitylike parameter extracted from the {\tt SEOB-PM} dynamics in a way that we specify below.
To give an apples-to-apples comparison between the LEOB and {\tt SEOBNR-PM}
dynamics, we present in Fig.~\ref{fig:Eb_vs_vOmg} the same kind of diagnostics for $q=1$. 
From the NR data extracted in Ref.~\cite{Nagar:2015xqa}, we compute the gauge-invariant NR orbital-frequency-like
variable  $\Omega^{\rm NR} \equiv dE/dJ= M^{-1} d E_b^{\rm NR}/dp_\varphi^{\rm NR}$ and then define
the orbital-velocity-like variable
 $v_{EJ}^{\rm NR}\equiv (M\Omega^{\rm NR})^{1/3}$.
For the two LEOB flavors (calibrated and uncalibrated) we follow the same procedure, i.e. 
compute (along the LEOB dynamics) $v_{EJ}^{\rm LEOB} \equiv (M (d E/dJ)^{\rm LEOB})^{1/3}$.
Figure~\ref{fig:Eb_vs_vOmg} compares the so computed NR curve with the LEOB ones,
either with the NR-calibration or without it. The figure also presents the corresponding {\tt SEOB-PM} 
curve displayed in Fig. 2 of Ref.~\cite{Buonanno:2024byg}\footnote{We thank the authors of Ref.~\cite{Buonanno:2024byg} 
for providing us with the {\tt SEOB-PM} data for the latter $E_b(v)$ curve, and for clarifying 
the definition they actually used for $v$.}. The figure indicates that {\tt SEOB-PM},
that is not NR-calibrated, delivers a slightly better agreement with the NR curve until 
$v_{EJ}\simeq 0.47$. After that, the two analytical descriptions are
quantitatively equivalent, although in one case the NR curve is approximated from below, 
the other from above.

\subsection{Waveform comparisons with other EOB models}
\label{sec:compare}
\begin{table}[t]
	\caption{\label{tab:model_performance}
	Median (Me) and maximum (Max) values of $\bar{\cal F}^{\rm max}_{\rm EOBNR}$,
	the unfaithfulness between EOB and SXS $\ell=m=2$ waveform data for several state-of-the-art EOB models in the spin-aligned, 
	quasi-circular case. The models in the first three rows incorporate PM information. 
	Note than only \TEOBd{} can also deal with eccentric binaries and scattering configurations. 
	The values for the {\tt SEOB} family that are extracted from the literature are computed on 441 SXS NR datasets 
	(partly private) that do not fully overlap with the 530 public ones used here. In this respect, the last row of the table reports the 
	{\tt SEOBNRv5HM} median on the public simulations computed in Ref.~\cite{Nagar:2024oyk}.}
	  \begin{center}
		\begin{ruledtabular}
   \begin{tabular}{l l l c} 
   Model  & ${\rm Me}[\bar{\cal F}^{\rm max}_{\rm EOBNR}]$ & ${\rm Max}[\bar{\cal F}^{\rm max}_{\rm EOBNR}]$ & Reference   \\
   \hline
   \LEOBa{}                 & $5.39\times 10^{-4}$   & $0.0126$ & this paper \\
   \LEOBss{}           & $6.13\times 10^{-4}$   & $0.0119$ & this paper \\
   {\tt SEOBNR-PM}            & $6.1\times 10^{-4}$   & $\sim 0.2$ & \cite{Buonanno:2024byg} \\
   \hline
   \TEOBd{}   & $3.09\times 10^{-4}$  & $6.80\times 10^{-3}$   & \cite{Nagar:2024oyk}             \\
   {\tt TEOBResumS-GIOTTO}    & $4.0\times 10^{-4}$   & $\sim 10^{-3}$         & \cite{Nagar:2023zxh}    \\
   {\tt SEOBNRv5HM}           & $1.99\times 10^{-4}$  & $\sim 2\times10^{-3}$  & \cite{Pompili:2023tna}  \\
   {\tt SEOBNRv5HM}           & $1.47\times 10^{-4}$  & $2.98 \times10^{-3}$   & \cite{Nagar:2024oyk}
   \end{tabular}
\end{ruledtabular}
\end{center}
\end{table}

We conclude the analysis of our waveform model by comparing and contrasting its phasing performance 
with the one delivered by the  {\tt SEOBNR-PM} model of Ref.~\cite{Buonanno:2024byg}.
For our purpose here, one should compare our Figs.~\ref{fig:barF_full} and~\ref{fig:barF_full_4PM_SS} 
with the left panel of Fig.~D.1 in the Supplemental Material  of Ref~\cite{Buonanno:2024byg}.
While our PM-based models, \LEOBa{}
and \LEOBss{} deliver $\bar{\cal F}_{\rm EOBNR}^{\rm max}$ values at most of the order of $0.01$ (with just 
very few configurations around this value), {\tt SEOBNR-PM}  delivers $\bar{\cal F}_{\rm EOBNR}^{\rm max}$ values
that may reach up to 0.1, with approximately $5\%$ of the configurations with $0.01<\bar{\cal F}_{\rm EOBNR}^{\rm max}<0.1$.
Both models are calibrated to NR simulations, but with several qualitative and quantitative differences. 
The {\tt LEOB} model relies on a NR-informed dynamics involving the calibration of two parameters:
one parameter, $a_{52}^{\rm NR}$, in the nonspinning sector and a second one, $\hat{g}_{32}^{\rm NR}$,
in the spinning sector. The NR-informed value of $a_{52}^{\rm NR}$ involves two
fitting coefficients, while that of $\hat{g}_{32}^{\rm NR}$ involves six fitting coefficients. These
fitting coefficients were informed by 17 NR datasets, 5 nonspinning and 12 spinning  
(notably, only  equal-mass and equal-spin ones).

The NR-calibration in {\tt SEOBNR-PM} is handled differently:
it is implemented only at the waveform level 
through the parameter  $\Delta t_{\rm NR}$ (previously used in {\tt SEOBNRv5HM}~\cite{Pompili:2023tna}) 
that defines where the merger time, $t^{22}_{\rm peak}$ is found on the EOB 
dynamics\footnote{The use of $\Delta t_{\rm NR}$ introduces a qualitative difference with respect to many EOB models, and notably 
{\tt TEOBResumS} and {\tt LEOB-PM}. While in the latter models the merger time is anchored on the EOB
dynamics, in {\tt SEOBNR-PM} and {\tt SEOBNRv5HM} it is detached from it (see Sec.~IIIC of Ref.~\cite{Nagar:2024oyk}).}. 
The parameter $\Delta t_{\rm NR}$ is fitted with a function of $\nu$ and of the spins involving 21 fitting 
coefficients\footnote{Note the presence of a sign typo in the definition of $t^{22}_{\rm peak}$ in both 
Refs.~\cite{Pompili:2023tna} and~\cite{Buonanno:2024byg}. It should 
read $t^{22}_{\rm peak}=t_{\rm ISCO}-\Delta t_{\rm NR}$  to be compatible with the 
negative $\Delta t_{\rm NR}$ obtained from the NR-informed fit.}. 
These fitting coefficients are determined using 441 NR simulations, partly private (this set only partly overlaps 
with our dataset of 530 configurations), see Eq.~(C1) of Ref.~\cite{Buonanno:2024byg}. 

Table~\ref{tab:model_performance} summarizes the comparison between all
existing spin-aligned models based on various EOB avatars, with several results taken from the literature and
in particular from Ref.~\cite{Nagar:2024oyk}. Note that the global performance of {\tt LEOB-PM} is consistent with
those of both {\tt TEOBResumS-GIOTTO} and \TEOBd{}, that share  with it most of the structure of the radiation
reaction force.

\section{Conclusions}
\label{sec:end}
We present a new way of incorporating PM (and PN results) in the conservative part of the dynamics within 
the effective one body formalism, that we call Lagrange-EOB (LEOB). This comes from the crucial use of a Lagrange
multiplier in the EOB action to impose the presence of the mass-shell constraint.
This method allows us to avoid some drawbacks of the extension of the usual Hamiltonian EOB approach based on 
a recursive definition of an explicit EOB-PM Hamiltonian, as discussed extensively in Sec.~\ref{sec:start}. 
The new LEOB approach  eventually yields one evolution equation more than the standard Hamiltonian approach.
The large flexibility of our new formalism allows us to blend together PM and PN information in various
forms and to construct reliable, robust and accurate EOB-based waveform models that incorporate PM information.
As a first, exploratory, investigation, only some special choices were here analyzed in detail. 
In particular, we showed that it is possible to construct a complete, reliable and
accurate spin-aligned waveform model building upon 4PM-4PN  analytical information in
the nonspinning sector and 3PM-3PN information in the spin-orbit sector. The model showed sufficient
flexibility and could additionally be significantly improved by  calibrating some high-order effective parameters to 
a limited sample of NR simulations.
The expressions of radiation reaction and waveform is inherited from previous PN-based EOB models of the {\tt TEOBResumS} family, 
in particular building upon Refs.~\cite{Nagar:2023zxh,Nagar:2024oyk}.
Our main findings are as follows.
\begin{itemize}
\item[(i)] In the nonspinning case, the formalism allows one to build an {\it uncalibrated} LEOB waveform model (completed by NQC
corrections and NR-informed ringdown) with unfaithfulness at most of the order of 0.01 evaluated over a meaningful sample
of nonspinning SXS datasets with mass ratio $1\le q\leq 10$ and $q=15$. Excellent EOB/NR phasing consistency is also 
found with the $q=32$ waveform of Ref.~\cite{Lousto:2020tnb}.
In addition, the energetics of the uncalibrated LEOB dynamics shows a good consistency 
with NR data down to merger, see Fig.~\ref{fig:ej}.
\item[(ii)] In the spinning case, we built a waveform model, called \LEOBa{} (incorporating
an approximate, Kerr-like description of spin-spin interactions) with minimal calibration of the dynamics whose
unfaithfulness is at most $1.26\%$ and has a median of $5.39\times 10^{-4}$ over the usual set of 530 public 
NR configurations of the SXS catalog used in previous works.
\item[(iii)] By incorporating more analytical information in the spin-spin sector (\LEOBss{} model,
as defined above) and by implementing a different NR calibration, involving more fitting coefficients and using more NR datasets, we could improve the model behavior in specific corners of the 
parameter space, e.g. the equal-mass, high spin region. Nevertheless, the global median is 
seen to increase slightly to the value $6.13\times 10^{-4}$.
\end{itemize}
The LEOB approach, together with PM information, opens a promising route toward constructing EOB-based
waveform models able to robustly exploit the analytical information available. 
However, we found that the LEOB model seems somewhat less flexible than PN-based HEOB ones. 
This might, however, be related to the factorization and resummation strategies used here, which can 
probably be improved. We leave such a study to the future. 

\begin{figure*}[t]
	\center	
	\includegraphics[width=0.32\textwidth]{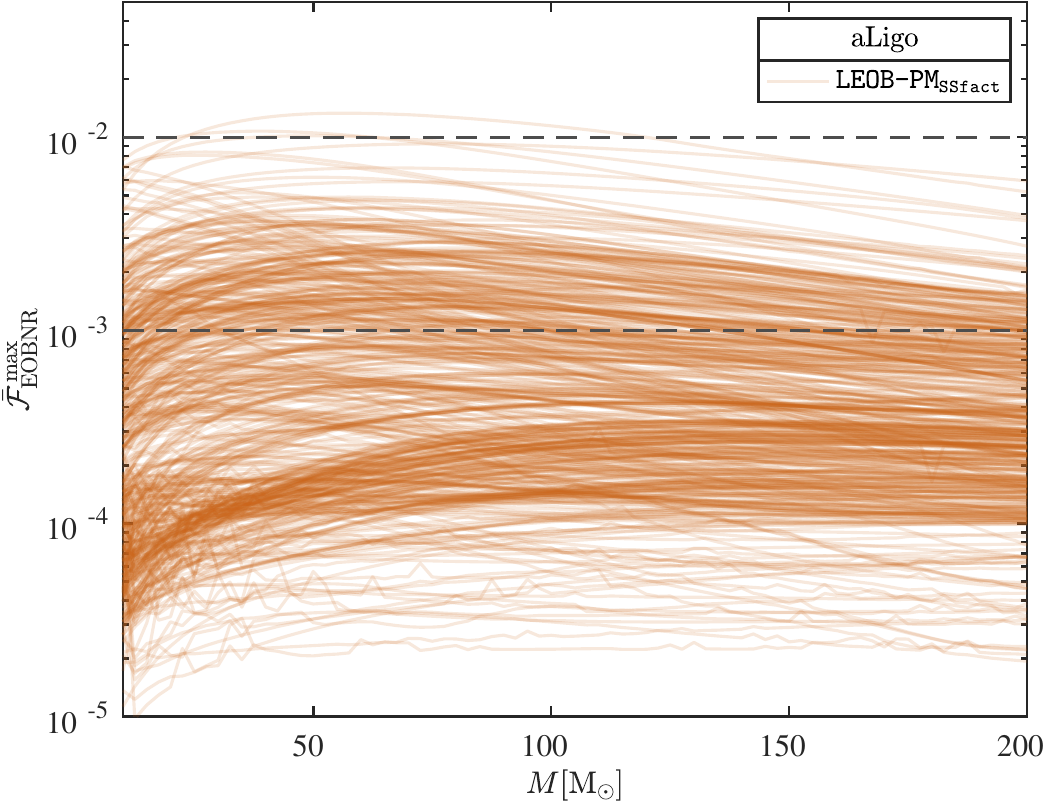} 
	\includegraphics[width=0.31\textwidth]{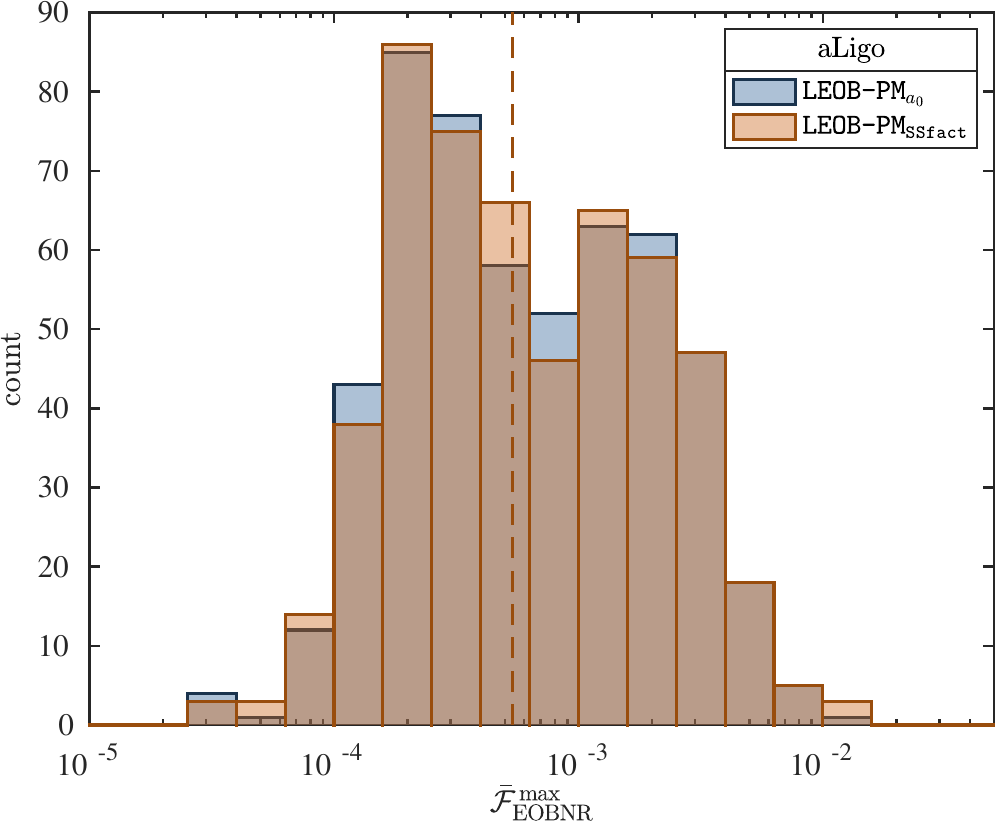}   
	\includegraphics[width=0.345\textwidth]{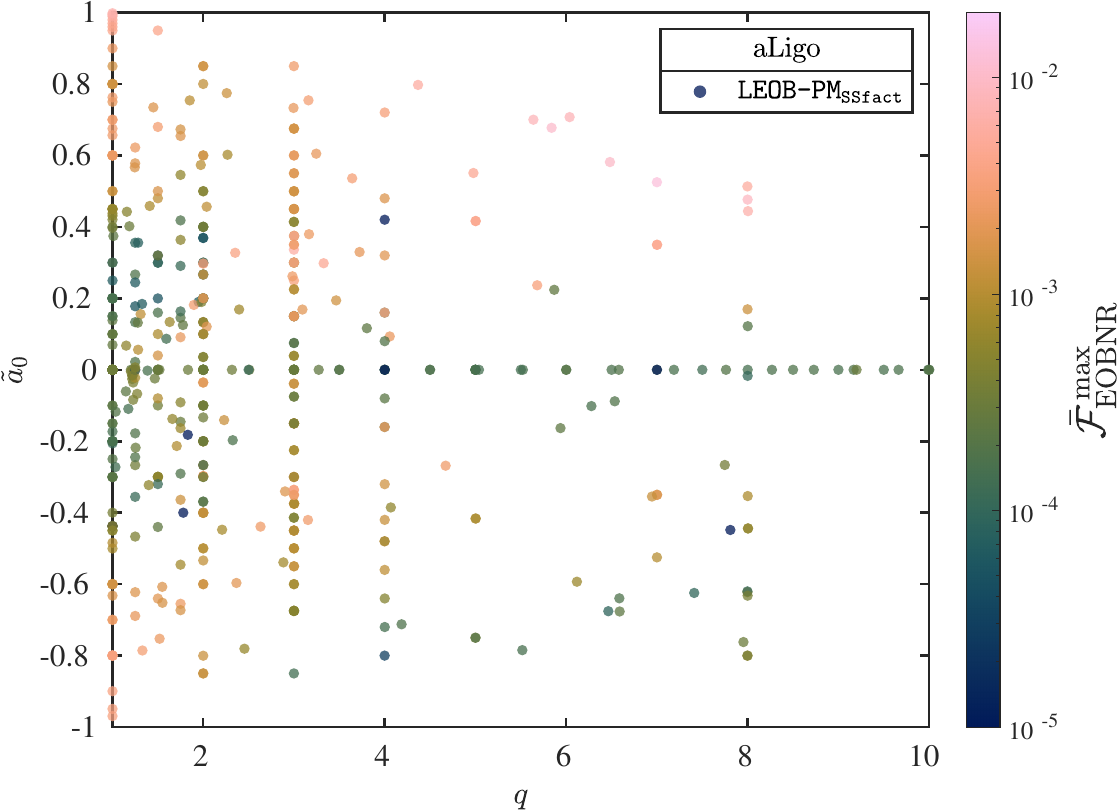}  
	\caption{\label{fig:barF_full_with_SS}EOB/NR unfaithfulness comparison and performance as in 
	Fig.~\ref{fig:barF_full} but incorporating the even-in-spin terms through the function $\hat{A}_{\rm SS}$ up to 5PM.
	Note that now we have more configurations with $\bar{\cal F}_{\rm EOBNR}^{\rm max}\gtrapprox 0.01$, in particular
	the equal-mass ones with rather high values of the spins.
	The SXS simulations {\tt SXS:BBH:1421} and {\tt SXS:BBH:1906} have been excluded, since the chosen EOB model develops a pole for the corresponding binary parameters.}
\end{figure*}

\begin{acknowledgments}
We are grateful to D.~Chiaramello for help with the EOB/NR 
unfaithfulness computations. A.~P.~and P.~R.~thank the hospitality 
and the stimulating environment of the IHES, where part of this work was carried out. 
We thank A.~Buonanno, G.~Mogull, R.~Patil and L.~Pompili for comments on the
manuscript and for the {\tt SEOB-PM} curve of Fig.~\ref{fig:ej}.
The present research was also partly supported by the ``\textit{2021 Balzan Prize for 
Gravitation: Physical and Astrophysical Aspects}'', awarded to Thibault Damour.
A.~P.~and P.~R.~acknowledge support from the Italian Ministry of University and Research (MUR) via the PRIN 2022ZHYFA2, GRavitational wavEform models for coalescing compAct binaries with eccenTricity (GREAT) and through the program
“Dipartimenti di Eccellenza 2018-2022” (Grant SUPER-C).
P.~R.~also acknowledges support from “Fondo di Ricerca d’Ateneo” of the University of Perugia. 
\end{acknowledgments}

\newpage
\appendix

\section{5PM spin-spin effects in factorized form}
\label{app:ss}

In this Appendix we report EOB/NR unfaithfulness, discussed in the main text and summarized in Fig.~\ref{fig:barF_full},
obtained with a different factorization procedure for the 5PM spin-even terms.
We rewrite the even-in-spin potential $A$, Eqs.~\eqref{eq:Afact}-\eqref{eq:Ass}, as
\begin{subequations}
	\label{eq:AtotFactorized}
\begin{align}
		&A_{\rm tot} =A_{\rm orb}\left(u_c,\g,\nu\right)\hat{A}_{\rm SS}(r,\g,\nu,\tilde{a}_i)\,, \\
		&\hat{A}_{\rm SS}(r,\g,\nu,\tilde{a}_i) = P^2_3\left[\hat{A}_{\rm SS}^{\rm 5PM}(r,\g,\nu,\tilde{a}_i)\right]\,.
\end{align}
\end{subequations}
In this formulation, we factorize the orbital term and re-expand the residual spin-spin terms as a series in $u$, as 
\begin{equation} \label{eq:hatAss}
	\hat{A}_{\rm SS}^{\rm 5PM}(r,\g,\nu,\tilde{a}_i) = {\rm Series}_u \left[1 + \frac{A_{\rm SS}^{\rm 5PM}(u_c,\g,\nu,\tilde{a}_i)}{A^{\rm 4PM\text{-}4PN}_{\rm orb}\left(u_c,\g,\nu\right)}\right]\,,
\end{equation}
which will be of the form $\hat{A}_{\rm SS}^{\rm 5PM} = 1+ O(\tilde{a_i}^2 u^3)$, 
and finally resum it through a near-diagonal Pad\'e approximant (in $u$).
The use of a $P^2_3$ approximant allows for the inclusion of 5PM analytical information in the spin-spin sector with 
only small changes to the (robust) baseline waveform model. Using the same NR calibration of the \LEOBa{} model discussed 
in the main text [see Eqs.~\eqref{a52fit} and \eqref{eq:g42fit} and leftmost column of Table~\ref{tab:g42_coeff}], 
we recompute EOB/NR mismatches against the 530 quasi-circular spin-aligned publicly-available NR 
simulations from the SXS collaboration. The full results are shown in Fig.~\ref{fig:barF_full_with_SS}: the left panel shows
$\bar{\cal F}_{\rm EOBNR}(M)$; 
the middle panel shows the distribution of $\bar{\cal F}_{\rm EOBNR}^{\rm max}$
of Fig.~\ref{fig:barF_full}, dubbed \LEOBa{}, superposed to the one obtained with 
up to 5PM spin-spin terms, dubbed \LEOBfact{}; 
the rightmost panels shows $\bar{\cal F}_{\rm EOBNR}^{\rm max}$ versus $(\tilde{a}_0,q)$. 
The overall results are very similar to the ones obtained neglecting even-in-spin PM contributions.
There are three datasets with maximum unfaithfulness larger than $1\%$:
{\tt SXS:BBH:0202}, with $(q,\chi_1,\chi_2) = (7.00,+0.60,0.00)$ and $\bar{\cal F}_{\rm EOBNR}^{\rm max} =1.33\%$;
{\tt SXS:BBH:1432}, with $(q,\chi_1,\chi_2) = (5.84,+0.66,0.79)$ and $\bar{\cal F}_{\rm EOBNR}^{\rm max} =1.07\%$;
and {\tt SXS:BBH:1439}, with $(q,\chi_1,\chi_2) = (6.48,+0.72,-0.32)$ and $\bar{\cal F}_{\rm EOBNR}^{\rm max} =1.01\%$.

However, this model could not be compared to every NR simulation, because the analytical spin-spin potential (more precisely, its $\g$-derivative) develops a pole in a small parameter-space region.
Because of this singularity, we excluded from our computations the two following NR simulations:
{\tt SXS:BBH:1421}, with $(q, \chi_1, \chi_2) = (7.81,-0.61,+0.80)$; and {\tt SXS:BBH:1906}, with 
$(q, \chi_1, \chi_2) = (4.00,+0.000058,-0.000085)$.

\section{Numerical relativity configurations}
\label{app:data}
\begin{figure}[t]
	\center	
        \includegraphics[width=0.4\textwidth]{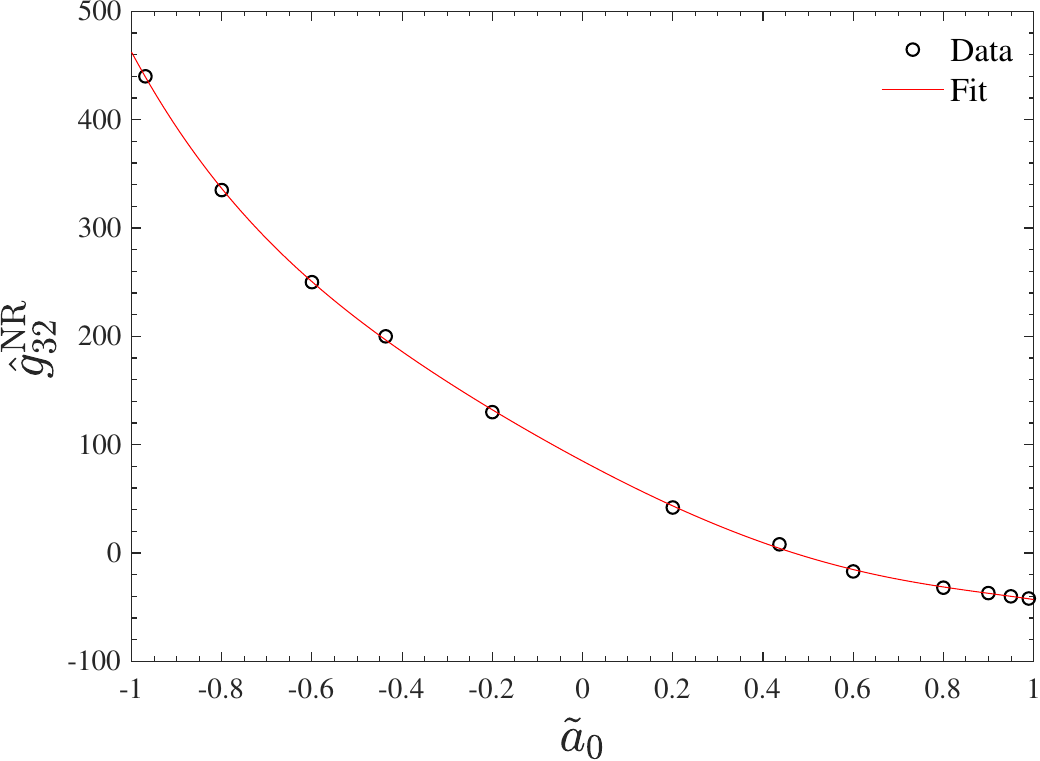}
	\caption{\label{fig:g42_vs_a0}Model \LEOBa{}: equal-mass, equal-spin values of the $\hat{g}_{32}^{\rm NR}$ reported in Table~\ref{tab:g42_eqmass} 
	and fitted with the 5th order polynomial in $\tilde{a}_0$ with the corresponding coefficients listed in the \LEOBa{} line of Table~\ref{tab:g42_coeff}.}
\end{figure}

In this section we report the details of the NR datasets that are used in the main text to determine
the two expressions of $\hat{g}_{32}^{\rm NR}$ used in the text, see Table~\ref{tab:g42_coeff}.
In particular, Fig.~\ref{fig:g42_vs_a0} shows the behavior of the best $\hat{g}_{32}^{\rm NR}$ values obtained
in absence of 3PM, 4PM and 5PM spin-spin contributions to the $A$ function. These values are also
listed for convenience in Table~\ref{tab:g42_eqmass}. 
\begin{figure*}[t]
	\center	
	 \includegraphics[width=0.45\textwidth]{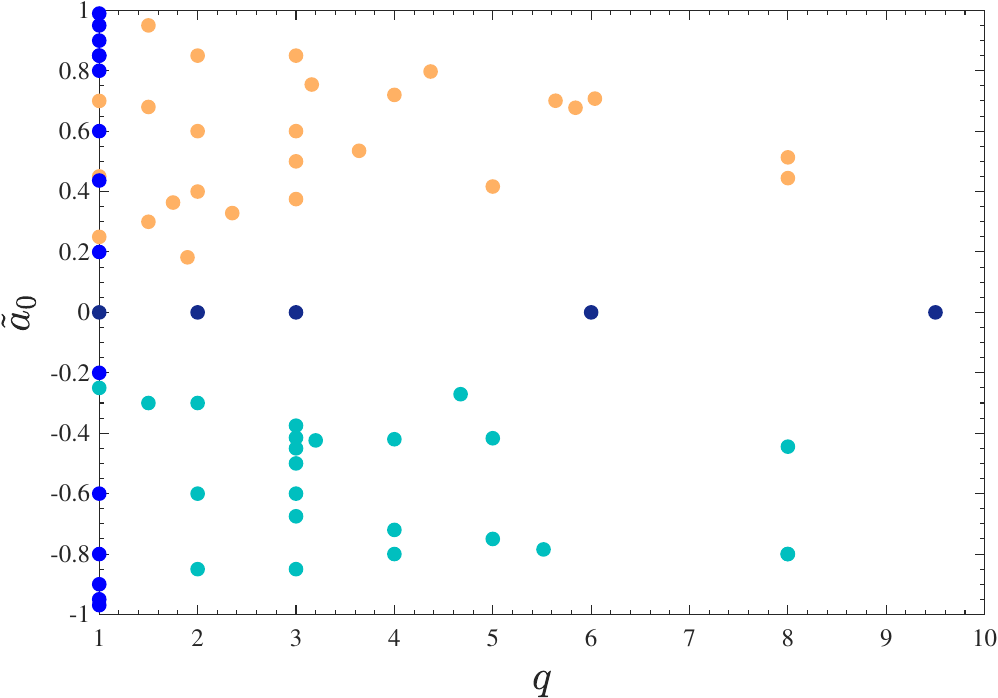}
	 \hspace{10 mm}
        \includegraphics[width=0.45\textwidth]{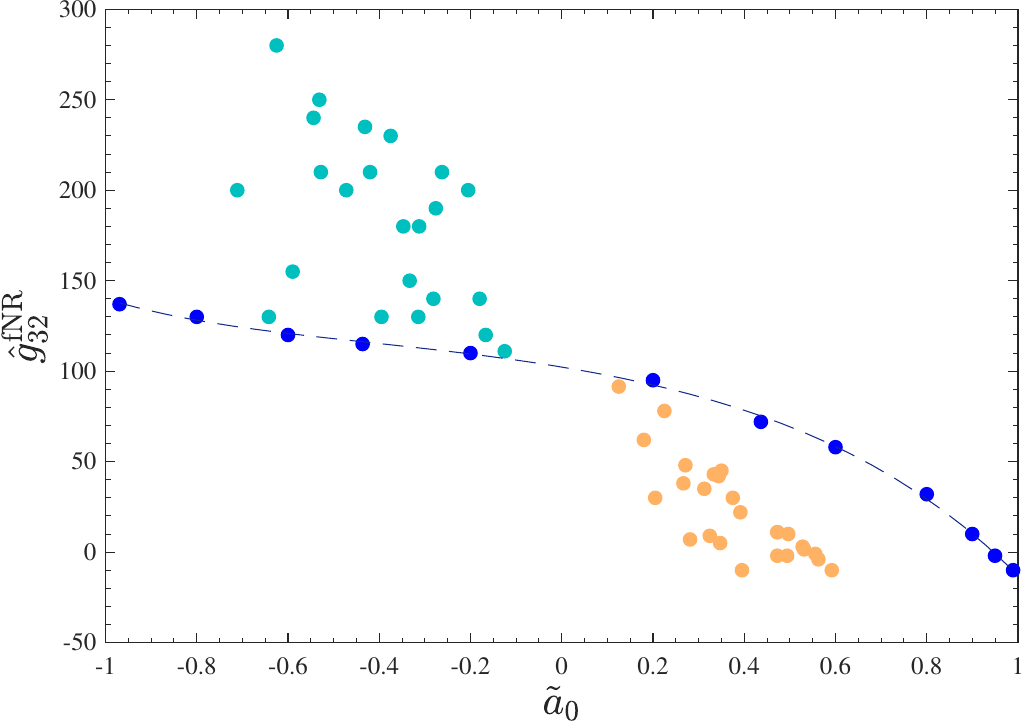}        
	\caption{\label{fig:g42_parspace}Determination of $\hat{g}_{32}^{\rm NR}$ using 3PM, 4PM and 5PM spin-spin 
	contributions in A for the \LEOBss{} model. Left panel: configuration chosen (also including the nonspinning ones used 
	to determine $a_{52}^{\rm NR}$). Right panel: the values chosen. The blue points are the 
	equal-mass, equal-spin ones and the line is the corresponding fit given by Eq.~\eqref{eq:g42fit}.}
\end{figure*}
\begin{table}[t]
	\caption{\label{tab:g42_eqmass}First-guess values for $\hat{g}_{32}^{=}$ for equal-mass, 
	equal-spin configurations. These values are then fitted with a 5-th order polynomial 
	yielding the corresponding coefficients in Table~\ref{tab:g42_coeff}. The value correspond
	to either the \LEOBa{} or \LEOBss{} models discussed in the main text.}
\begin{center}
\begin{ruledtabular}
\begin{tabular}{lllc|cc}
$\#$ & ID & $(q,\chi_1=\chi_2)$ & $\tilde{a}_0$ &$\hat{g}_{32}^{=,{\tt LEOB-PM}_{a0}}$ & $\hat{g}_{32}^{\rm =,{\tt LEOB-PM_{SS}}}$\\
\hline
1  & BBH:1137 & $(   1, -0.   9692)$ & $-0.9692$   & 440 & 137\\ 
2  & BBH:2086 & $(   1, -0.80)$        & $-0.80$     & 335 & 130 \\ 
3  & BBH:2089 & $(   1, -0.60)$        & $-0.60$     & 250  & 120 \\ 
4  & BBH: 148& $(   1, -0.4376)$        & $-0.4365$     & 200 & 115 \\ 
5  & BBH:0149 & $(   1, -0.20)$        & $-0.20$     & 130  & 110 \\
6  & BBH:0150 & $(   1, +0.   20)$     & $+0.20$     & 42  & 95\\ 
7  & BBH:0170 & $(   1, +0.   4365)$ & $+0.20$     & 8  & 72\\ 
8  & BBH:2102 & $(   1, +0.   60)$     & $+0.60$     & $-17$  & 58 \\ 
9  & BBH:2104 & $(   1, +0.   80)$     & $+0.80$     & $-32$ & 32\\  
10 & BBH:0160 & $(   1, +0.8997)$       & $+0.8997$   & $-37$  & 10\\ 
11 & BBH:0157 & $(   1, +0.9495)$   & $+0.949586$ & $-40$ & $-2$ \\ 
12 & BBH:0177 & $(   1, +0.9892)$   & $+0.989253$ & $-42$ & $-10$   
\end{tabular}
\end{ruledtabular}
\end{center}
\end{table}
Finally, Fig.~\ref{fig:g42_parspace} illustrates the full NR information used to determine $\hat{g}_{32}^{\rm NR}$ in
the case of spin-spin effects in semi-factorized form including up to residual 4PM effects and addressed 
as \LEOBss{} in the text. The left panel of the figures summarizes the part of the parameter space
covered by the simulations (including the 5 nonspinning datasets used to determine $a_{52}^{\rm NR}$,
see Table~\ref{tab:a52}). The right panel shows the values of $\hat{g}_{32}^{\rm NR}$ fitted. The dash-dotted line
is the outcome of the fit of $\hat{g}_{32}^{=}$ for the equal-mass, equal-spin configurations.

\section{LEOB gauge flexibility for nonspinning binaries}
\label{app:gauge_freedom}

In this appendix we further discuss the flexibility of the Lagrange multiplier approach, and finally examine some notable examples of gauge choice. 

For simplicity, we reduce to the case of nonspinning binaries, for which the effective metric can be taken in the general spherically symmetric form
\begin{align}
	\label{eq:PN_eff_metric}
	&g_{\mu\nu}(x, \g)dx^\mu dx^\nu=-A(r, \g)dt^2+B(r, \g)dr ^2\cr &\qquad+r^2 C(r, \g)( d\theta^2 + \sin^2 \theta d\phi^2)\,,
\end{align}
involving three EOB potentials, $(A,B,C)$. With this effective metric, the corresponding explicit mass-shell condition, Eq.~\eqref{massshell3}, reads (when
considering planar motions)
\begin{equation}
	\label{eq:constraint_gPN}
	-\frac{\gamma^2}{A\left(r,\g\right)}
	+\frac{p_r^2}{B\left(r,\g\right)}
	+\frac{p_\varphi^2}{r^2\, C\left(r,\g\right)}
	+ 1  =0\,.
\end{equation}

\subsection{Inspecting the gauge freedom}

The coordinate freedom of a general diagonal spherically
symmetric metric of the type \eqref{eq:PN_eff_metric} would be encapsulated in arbitrary changes of the 
radial coordinate: $ r \to  f(r', \g)$. The arbitrariness in the PM expansion of the function 
$f(r)= \sum_n \frac{f_n}{r^n}$ would induce essentially arbitrary changes in the PM expansion of
one of the three functions $A$, $B$ or $C$ (with correlated changes in the other two). This coordinate freedom
would be fixed by choosing, e.g., a Schwarzschild-like radial coordinate such that $C=1$. Making such a choice [i.e. choosing
$c_n(\gamma,\nu)=0$ for $n\geq1$] simplifies 
the relations between the scattering angle and the EOB potential, at the level of their PM coefficients, to
\begin{widetext}
	\begin{subequations}
		\begin{align}
			\label{eq:chipotcoefs2}
			&\chi_1 (\gamma,\nu) = \frac{p_\infty }{2}b_1(\gamma,\nu) - \frac{\gamma^2}{2p_\infty}a_1(\gamma,\nu)\,, \\
			&\chi_2 (\gamma,\nu) = \frac{\pi \gamma^2}{8}\Big\{a_1(\gamma,\nu)\left[2a_1(\gamma,\nu)-b_1(\gamma,\nu)\right] -2a_2(\gamma,\nu)\Big\}
			+\frac{\pi p_\infty^2}{32}\Big\{\left[b_1(\gamma,\nu)\right]^2 -4b_2(\gamma,\nu) \Big\}\,, \\
			&\chi_3 (\gamma,\nu) = \frac{\left(- \gamma ^2+\frac{5}{2} \gamma ^4-\frac{35 }{24}\gamma ^6\right) a_1(\gamma ,\nu
				)^3}{p_{\infty }^3}+\frac{ \left(-4\gamma ^2+5 \gamma ^4\right) a_1(\gamma ,\nu ) \left[4 a_2(\gamma ,\nu
				)+a_1(\gamma ,\nu ) b_1(\gamma ,\nu )\right]}{8 p_{\infty }}\cr&\quad-\gamma ^2 p_{\infty } \bigg\{a_3(\gamma ,\nu )+\frac{1}{2}a_2(\gamma ,\nu ) b_1(\gamma ,\nu
			)  -\frac{1}{2}a_1(\gamma ,\nu ) \left[b_1(\gamma ,\nu )^2-4 b_2(\gamma ,\nu )\right]\bigg\}+\frac{1}{24} p_{\infty }^3 \bigg\{b_1(\gamma ,\nu ) \left[b_1(\gamma ,\nu )^2\right.\cr&\left.\quad-4 b_2(\gamma
			,\nu )\phantom{^2}\right]+8 b_3(\gamma ,\nu )\bigg\}\,, \\
			&\chi_4 (\gamma,\nu) = \frac{3}{8} \pi  \gamma ^2 \left(-2+5 \gamma ^2\right) a_1(\gamma ,\nu )^4+\frac{3}{64}
			\pi  \gamma ^2 \left(-2+3 \gamma ^2\right) \bigg\{8 a_2(\gamma ,\nu )^2+8 a_1(\gamma
			,\nu ) a_2(\gamma ,\nu ) b_1(\gamma ,\nu )\cr&\quad+a_1(\gamma ,\nu ) \left[16 a_3(\gamma ,\nu
			)-a_1(\gamma ,\nu ) \left(b_1(\gamma ,\nu )^2-4 b_2(\gamma ,\nu
			)\right)\right]\bigg\}-\frac{3}{64} \pi  \pinf^2 \gamma ^2 \bigg\{16 a_4(\gamma
			,\nu )+8 a_3(\gamma ,\nu ) b_1(\gamma ,\nu )\cr&\quad-\left[2 a_2(\gamma ,\nu )-a_1(\gamma ,\nu
			) b_1(\gamma ,\nu )\right] \left[b_1(\gamma ,\nu )^2-4 b_2(\gamma ,\nu )\right]+8
			a_1(\gamma ,\nu ) b_3(\gamma ,\nu )\bigg\}\cr&\quad-\frac{3 \pi  }{1024} \bigg\{-128 \gamma ^2
			a_1(\gamma ,\nu )^2 \left[6 a_2(\gamma ,\nu )+a_1(\gamma ,\nu ) b_1(\gamma ,\nu
			)\right]+256 \gamma ^4 a_1(\gamma ,\nu )^2 \left[6 a_2(\gamma ,\nu )+a_1(\gamma ,\nu
			) b_1(\gamma ,\nu )\right]\cr&\quad+\pinf^4 \left[5 b_1(\gamma ,\nu )^4-24 b_1(\gamma
			,\nu )^2 b_2(\gamma ,\nu )+32 b_1(\gamma ,\nu ) b_3(\gamma ,\nu )+16
			\left[b_2(\gamma ,\nu )^2-4 b_4(\gamma ,\nu )\right]\right]\bigg\}\,.
		\end{align}
	\end{subequations}
\end{widetext}
The simplified Eqs. \eqref{eq:chipotcoefs2} exhibit a residual gauge freedom. Namely, after having
fixed the radial coordinate by choosing $C=1$,
one can still arbitrarily impose {\it one} functional relation among the remaining two functions $A(r, \g)$ and $B(r,\g)$. Then, the PM expansion of the scattering function $\chi(\g,j)$ will uniquely determine the expressions of both $A(r,\g)$ and $B(r,\g)$.
For instance, one can fix at will the functional form of   $B(r,\g)$ [in addition to the coordinate choice $C(r,\g)=1$] and uniquely determine the
PM expansion of $A(r,\g)$. This is what is done in Eq.~\eqref{AS*} when absorbing the post-Schwarzschild potential $Q(r,\g)$
in $A(r,\g)$, keeping Schwarzschild values for $B(r,\g)= \left( 1- \frac2{r}\right)^{-1}$ and $C(r,\g)=1$. 

For instance, denoting 
\begin{equation}
	\Delta \chi_n=\chi _n(\gamma ,\nu )-\chi _n^S(\gamma)
\end{equation}
where $\chi _n^S(\gamma)=\chi _n(\gamma ,\nu=0 )$ is the $n$PM coefficient of the Schwarzschild scattering angle, and using the fact that $\Delta \chi_1=0$, we get
\begin{widetext}
	\begin{subequations}

		\begin{align}
			&a_1(\gamma,\nu)=-2\,,\\
			&a_2(\gamma,\nu)=-\frac{8 \Delta \chi _2}{\pi  \left(3 \gamma ^2-1\right)}\,,\\
			&a_3(\gamma,\nu)=\frac{24 \left(8 \gamma ^4-8 \gamma ^2+1\right) \Delta \chi _2}{\pi  \left(3 \gamma ^2-1\right) \left(4 \gamma ^2-1\right) p_{\infty }^2}-\frac{3 \Delta \chi _3}{\left(4 \gamma ^2-1\right) p_{\infty }}\,,\\
			&a_4(\gamma,\nu)= -\frac{12 \left(140 \gamma ^8-235 \gamma ^6+123 \gamma ^4-13 \gamma ^2+1\right) \Delta \chi _2}{\pi  \left(3 \gamma ^2-1\right) \left(4 \gamma ^2-1\right) \left(5 \gamma ^2-1\right) p_{\infty }^4}+\frac{16 \left(35 \gamma ^4-30 \gamma ^2+3\right) \Delta \chi _2^2}{\pi ^2 \left(3 \gamma ^2-1\right)^2 \left(5 \gamma ^2-1\right) p_{\infty }^2}\cr
			&\quad+\frac{3 \left(35 \gamma ^4-30 \gamma ^2+3\right) \Delta \chi _3}{\left(4 \gamma ^2-1\right) \left(5 \gamma ^2-1\right) p_{\infty }^3} -\frac{32 \Delta \chi _4}{3 \pi  \left(5 \gamma ^2-1\right) p_{\infty }^2} \,.
		\end{align}
	\end{subequations}
\end{widetext}

Another way of obtaining the expressions of the $ a_n(\gamma,\nu)$ coefficients in the LJBL gauge is to relate
them to the $ q_n(\gamma,\nu)$ coefficients of the PM expansion of the $Q(r,\gamma)$ in the post-Schwarzschild
formulation of Ref.~\cite{Damour:2017zjx}. Combining the $\chi_n \leftrightarrow q_n$ map with the $\chi_n \leftrightarrow a_n$ map yields the following relations
\begin{subequations} \label{anvsqn}
	\begin{align}
		a_2(\gamma ,\nu )&=\frac{2 q_2(\gamma ,\nu )}{-1+3 \gamma ^2}\,,\\
		a_3(\gamma ,\nu )&=-\frac{3 \left(-1+10 \gamma ^2\right) q_2(\gamma ,\nu )}{\left(-1+3 \gamma ^2\right)
			\left(-1+4 \gamma ^2\right)}\cr&+\frac{3 q_3(\gamma ,\nu )}{-1+4 \gamma ^2}\,,\\
		\label{eq:a4fromQ}
		a_4(\gamma ,\nu )&=\frac{6 \gamma ^2 \left(3+25 \gamma ^2\right) q_2(\gamma ,\nu )}{\left(-1+3 \gamma
			^2\right) \left(-1+4 \gamma ^2\right) \left(-1+5 \gamma ^2\right)}\cr&+\frac{\left(-1+17 \gamma ^2\right)
			q_2(\gamma ,\nu )^2}{\left(-1+3 \gamma ^2\right)^2 \left(-1+5 \gamma ^2\right)}\cr&-\frac{\left(-5+57
			\gamma ^2\right) q_3(\gamma ,\nu )}{\left(-1+4 \gamma ^2\right) \left(-1+5 \gamma ^2\right)}\cr&+\frac{4
			q_4(\gamma ,\nu )}{-1+5 \gamma ^2}\,.
	\end{align}
\end{subequations}
It should be noted that the coefficients entering these relations depend only on $\g$, but not on $\nu$ (or $h$).
As a consequence, the (local) $ a_n(\gamma,\nu)$ coefficients inherit the special  $h$-structure of the 
(local) $ q_n(\gamma,\nu)$ coefficients discussed  in Eqs.~(2.38)-(2.41) of Ref.~\cite{Bini:2024tft}.

\subsection{Alternative gauge choices}

We list here some notable instances of gauge choices that could be made in the nonspinning case, 
showing the first few corresponding PM-coefficient relations.

\subsubsection{Post-Schwarzschild gauges}
\label{subsec:PS_gauge}

The Post-Schwarzschild (PS) gauge was introduced in Ref.~\cite{Damour:2017zjx} and then used, e.g., in Refs.~\cite{Damour:2019lcq,Antonelli:2019ytb,Khalil:2022ylj,Buonanno:2024vkx}.
It consists of imposing the Schwarzschild relations 
\begin{align}
	A\left(r,\gamma\right) = B\left(r,\gamma\right)^{-1} = 1 - 2 u \,, \qquad C\left(r,\gamma\right) =1\,,
\end{align}
with $u = (G M)/(r c^2)$, and including the full PM knowledge in the $Q\left(\gamma,r\right)$ potential. In this case, the first PM-coefficient relations read
\begin{subequations}
	\begin{align}
		&\chi_1 (\gamma,\nu) = \frac{\left[4\gamma^2-2-q_1(\gamma,\nu)\right]}{2p_\infty}\,, \\
		&\chi_2 (\gamma,\nu) = \frac{\pi}{8}\left[15\gamma^2 - 3 -2 q_1(\gamma,\nu)- 2 q_2(\gamma,\nu)\right]\,.
	\end{align}
\end{subequations}

The related PS$^*$ gauge (see \cite{Antonelli:2019ytb,Khalil:2022ylj,Buonanno:2024vkx}) can be obtained by keeping $B^{-1}= 1-2u$ and transferring all the PM information from $Q$ to $A$.
One has thus $Q=0$ and a PM-informed $A$ function that is no longer the inverse of the $B$ potential. It is easy to map $Q$ into $A$ (and viceversa), see Eqs.~\eqref{eq:QtoA} and \eqref{AS*}. The result for this map at the level of the PM coefficients can be found in Appendix B.2 of Ref.~\cite{Khalil:2022ylj}. 

\subsubsection{C = 1, fixed AB gauge}

A similar gauge is found when imposing
\begin{equation}
	A(r,\gamma) B(r,\gamma)=1\,,
\end{equation}
together with $C = 1$ and $Q = 0$.

In this case, the complete PM knowledge is incorporated into the $A$ (or, equivalently, $B$) potential.
The first two relations between its coefficients and the scattering-angle terms read as
\begin{subequations}
	\begin{align}
		&\chi_1 (\gamma,\nu) = -\frac{\left(2\gamma^2-1\right)}{2p_\infty}a_1(\gamma,\nu)\,, \\
		&\chi_2 (\gamma,\nu) = \frac{\pi}{32}\bigg[3\left(5\gamma^2-1\right)a_1^2(\gamma,\nu) \cr&\quad-4\left(3\gamma^2-1\right)a_2(\gamma,\nu)\bigg]\,.
	\end{align}
\end{subequations}

A modification of this setup consists in imposing some PN relations between the metric potentials.
For example, we could prefer to keep all-but-one PN-dependent metric functions fixed and focus on a 
particular one to include the PM terms.
For instance, we could ask 
\begin{equation}
	A(r,\gamma) B(r,\gamma)=D_{\rm 3PN}(r)\,,
\end{equation}
where 
\begin{equation}
	D_{\rm 3PN}(r)=1-\frac{6 \nu }{r^2}+\frac{2 (3 \nu -26) \nu }{r^3}
\end{equation}
is the small-velocity expansion of the $D$ potential first computed in Ref.\cite{Damour:2000we}.
Indeed this still enforces a relation between each pair of PM coefficients [$a_n(\gamma,\nu)$,$b_n(\gamma,\nu)$], leaving room for the determination of $A$ (or equivalently $B$) in terms of the scattering angle. In particular at the lowest orders we find

\begin{subequations}
	\begin{align}
		&\chi_1 (\gamma,\nu) = -\frac{\left(2\gamma^2-1\right)}{2p_\infty}a_1(\gamma,\nu) \,, \\
		&\chi_2 (\gamma,\nu) = -\frac{ \pi }{32} \bigg[ \left(3-15 \gamma ^2\right)a_1(\gamma,\nu) ^2\cr
		&\quad+4   \left(3 \gamma ^2-1\right)a_2(\gamma,\nu)+24 \nu  p_\infty^2\bigg]\,.
	\end{align}
\end{subequations}

\subsubsection{Kerr-Schild gauge}

All previous gauges have been constructed using a mass-shell condition in the form of Eq.~\eqref{eq:constraint_gPN}.
This is the quadratic-in-momenta condition which has been most commonly studied in the past.

However, this is not the only possibility. An alternative is presented by the 
``canonical Kerr-Schild gauge'', studied first within the EOB framework in 
Ref.~\cite{Ceresole:2023wxg}.
The (inverse) effective metric in the Kerr-Schild form, in Cartesian coordinates, reads
\begin{equation}
	g_{\rm eff}^{\mu\nu} = \eta^{\mu\nu} - \Phi(r, \gamma) k^\mu k^\nu\,,
\end{equation}
where $k^\mu= (1, \frac{x^i}{r})$ describes an outgoing null 4-vector centered on the origin.

For the Kerr-Schild potential $\Phi(r, \gamma)$ we can consider a PM expansion of 
the type \eqref{eq:A_pm}, with coefficients $\phi_n(\gamma,\nu)$, and determine it 
in terms of the scattering angle. For instance we find
\begin{subequations}
	\begin{align}
		&\chi_1 (\gamma,\nu) = \frac{\left(2 \gamma ^2-1\right) \phi _1(\gamma ,\nu)}{2 p_{\infty }} \,, \\
		&\chi_2 (\gamma,\nu) = \frac{\pi}{32}   \left[3 \left(5 \gamma ^2-1\right) \phi _1(\gamma ,\nu)^2\right.\cr&\quad\left.+4 \left(3 \gamma ^2-1\right) \phi _2(\gamma ,\nu)\right] \,.
	\end{align}
\end{subequations}

\section{Treatment of the 4PM dynamics of bound states in other works}
\label{app:4PM_others}

 In the PM-based HEOB models of Refs.~\cite{Khalil:2022ylj,Buonanno:2024byg} the issue 
 of deriving the 4PM contribution to the EOB bound-state dynamics was tackled by two 
 different  formal procedures of limited validity. 
 To be more specific, we start by recalling that the complete, conservative, 
 4PM-level, hyperbolic-motion coefficient $q_4^{\rm hyp,  tot}$ contains 
 the following 4PM-accurate logarithmic contribution:
   \be
   \label{eq:qhyp}
  q_4^{\rm hyp,  log}= \frac{\nu}{h^3} \mathcal E(\g) \ln\pinf^2 \,,
  \ee
where 
\begin{align}
&\mathcal E(\g)= \frac{1}{12 p_{\infty }^2} \Big( 1151-3336 \gamma +3148 \gamma ^2-912 \gamma ^3\cr 
&\quad+339 \gamma ^4-552 \gamma ^5+210 \gamma ^6\Big) \cr
&\quad + \frac{1}{2} \left(5-76 \gamma +150 \gamma ^2-60 \gamma ^3-35 \gamma ^4\right) \log
   \left(\frac{1+\gamma }{2}\right) \cr
&\quad -\frac{\gamma  \left(3-2 \gamma ^2\right) \left(11-30 \gamma ^2+35 \gamma ^4\right) \text{arcosh}(\gamma )}{4 p_\infty^3}
\end{align}
The latter term is rooted
in the specific hyperbolic-motion physics of tail-transported effects, and cannot (and should not) be 
directly analytically continued to the elliptic-motion case. A direct analytic continuation
from the hyperbolic situation, where 
$\pinf^2=\g^2-1$ is positive, to the elliptic one,
where $\pinf^2=\g^2-1$ is negative, would yield
a complex-valued Hamiltonian for elliptic motions.
Ref.~\cite{Khalil:2022ylj} partially
tackled this issue by subtracting from
the hyperbolic Hamiltonian a term
containing the leading-order, 4PN-level
contribution to $ q_4^{\rm hyp,  log}$,
namely  $q_{\rm4PM,4PN}^{\rm hyp,  log}= \frac{148 \nu  p_{\infty }^2}{15} \ln\pinf^2$.
On the one hand, such a formal log-subtraction prescription is
not based on a clear theoretical treatment
of the difference between tail effects in
the two different physical situations. On the other hand,
the approximate subtraction used in Ref.~\cite{Khalil:2022ylj} leaves some residual contributions $\propto \ln\pinf^2$ beyond the 4PN order, which are complex-valued along elliptic motions.
This explains why, in order to reproduce the results of Ref.~\cite{Khalil:2022ylj}, e.g.~Fig.~(4) therein, we found necessary to take the real part of the Hamiltonians they provide.
 
In the recent work of Buonanno et al., Ref.~\cite{Buonanno:2024byg}, an improved treatment of the continuation
of tail effects from hyperbolic to elliptic motions
has been used. There, the 
formal replacement $\ln p_\infty^2\to\ln u$ was made in a contribution canonically equivalent to $q_4^{\rm hyp,  log}$ in Eq.~\eqref{eq:qhyp}, and
an additional term, $\Delta A^{\rm 4PN}$, was added to the $A$ potential to ensure that the 4PN-level limit of the resulting
bound-state dynamics agrees, though only to second order in eccentricity, with the one derived in Ref.~\cite{Damour:2015isa}.
While the EOB model of Ref.~\cite{Buonanno:2024byg}, calibrated to Numerical Relativity (NR), 
shows reasonable agreement with simulations, we have adopted the alternative prescription 
discussed in Sec.~\ref{sec:PMstuff} due to its stronger theoretical foundation.

\section{Completing the local  nonspinning dynamics by 5PM, 6PM 
	and 7PM contributions at 6PN accuracy}
\label{app:complete_Alocal_6PN}

In the present Appendix we extend the 4PN local completion of $A$ discussed in Sec.~\ref{sec:4PN_Alocal}, using all the available results of the Tutti Frutti approach \cite{Bini:2019nra,Bini:2020wpo,Bini:2020nsb,Bini:2020hmy}, i.e.~including, up the 6PN accuracy, the 5PM, 6PM and 7PM 
contributions to the local dynamics.

At this accuracy, the completed $A$ potential reads
\begin{align}
	\label{eq:A4PM6PNcomp}
	A^{\rm loc}_{\rm 6PN completed}(r,\gamma,\nu)&=A^{\rm loc}_{\rm \leq 4PM}(r,\gamma,\nu) \nonumber\\
	&+  A^{\rm 6PN supp, loc}(r,\gamma,\nu)\,,
\end{align}
where the supplementary term has the structure
\begin{align}
	A^{\rm 6PN supp, loc}(r,\gamma,\nu)&= a_5^{\rm loc}(\pinf^2,\nu) u^5 +  a_6^{\rm loc}(\pinf^2,\nu) u^6 \nonumber\\
	&+ a_7^{\rm loc}(\pinf^2,\nu) u^7\,. 
\end{align}
Here, the velocity dependent coefficients 
$a_5(\pinf^2,\nu),  a_6(\pinf^2,\nu),  a_7(\pinf^2,\nu)$ are only known at a limited order in their expansion in powers of $\pinf^2 \equiv \g^2-1$. The 6PN accuracy corresponds to knowing 
\begin{align}
	a_5^{\rm loc}(\pinf^2,\nu) &=a_{50}^{\rm loc}(\nu) + a_{52}^{\rm loc}(\nu)\pinf^2+ a_{54}^{\rm loc}(\nu)\pinf^4 \ , \\
	a_6^{\rm loc}(\pinf^2,\nu) &= a_{60}^{\rm loc}(\nu)+ a_{62}^{\rm loc}(\nu)\pinf^2 \ , \\
	a_7^{\rm loc}(\pinf^2,\nu)&= a_{70}^{\rm loc}(\nu) \ ,
\end{align}
where $a_{50}^{\rm loc}(\nu)$ is the 4PN contribution already determined in Sec.~\ref{sec:4PN_Alocal}. Similarly as to what we did for $a_{50}^{\rm loc}(\nu)$, the values of the other five $\nu$-dependent coefficients 
$a_{52}(\nu), \cdots, a_{70}(\nu)$ can be directly obtained from the coefficients $q_n(\pinf,\nu)$ given in Table XII of \cite{Bini:2020nsb}, by means of the mapping between $A$ and $Q$ recalled in Eqs.~\eqref{anvsqn}. The resulting local contributions read
\begin{widetext}
	\begin{subequations} 
		\begin{align}
			\label{eq:4PNloc}
			&a_{52}^{\rm loc}(\nu) = \left(\frac{1549753}{4200}-\frac{840907 \pi ^2}{40960}\right) \nu
			+\left(\frac{33369}{160}+\frac{101673 \pi ^2}{20480}+\frac{\bar{d}_5^{\nu
					^2}}{5}\right) \nu ^2+\left(\frac{3607}{15}-\frac{1763 \pi ^2}{256}\right) \nu
			^3-\frac{93 \nu ^4}{16}\,, \\
			&a_{54}^{\rm loc}(\nu) = \left(-\frac{8899659907}{7056000}+\frac{89734448413 \pi ^2}{734003200}\right) \nu
			+\left(-\frac{159563213}{392000}+\frac{1700103 \pi ^2}{89600}+\frac{3 q_{45}^{\nu
					^2}}{35}-\frac{6 \bar{d}_5^{\nu ^2}}{25}\right) \nu
			^2\cr
			&\quad+\left(-\frac{954911}{3600}-\frac{2816111 \pi ^2}{245760}-\frac{2 \bar{d}_5^{\nu
					^2}}{5}\right) \nu ^3+\left(-\frac{64479}{160}+\frac{5535 \pi ^2}{512}\right) \nu
			^4+\frac{657 \nu ^5}{64}\,,
		\end{align}
	\end{subequations}
	\begin{subequations} 
		\begin{align}
			\label{eq:5PNloc}
			&a_{60}^{\rm loc}(\nu) = \left(-\frac{1724389}{4200}+\frac{339311 \pi ^2}{5120}\right) \nu
			+\left(\frac{47711}{120}-\frac{31343 \pi ^2}{5120}+a_6^{\nu
				^2}+\frac{\bar{d}_5^{\nu ^2}}{5}\right) \nu ^2+\left(\frac{2077}{15}-\frac{287 \pi
				^2}{64}\right) \nu ^3\cr
			&\quad-\frac{11 \nu ^4}{8}\,,  \\
			&a_{62}^{\rm loc}(\nu) = \left(-\frac{269524382641}{47628000}+\frac{562658476799 \pi ^2}{825753600}+\frac{45303
				\pi ^4}{524288}\right) \nu +\left(\frac{268226699}{1764000}+\frac{3835411 \pi
				^2}{1075200}+\frac{11 q_{45}^{\nu ^2}}{70}\right. \cr
			&\left.\quad-\frac{68 \bar{d}_5^{\nu
					^2}}{75}+\frac{\bar{d}_6^{\nu ^2}}{6}\right) \nu
			^2+\left(\frac{268226699}{1764000}+\frac{3835411 \pi ^2}{1075200}+\frac{11
				q_{45}^{\nu ^2}}{70}-\frac{68 \bar{d}_5^{\nu ^2}}{75}+\frac{\bar{d}_6^{\nu
					^2}}{6}\right) \nu ^2+\left(-\frac{201377}{480}+\frac{205 \pi ^2}{16}\right) \nu
			^4\cr
			&\quad+\frac{73 \nu ^5}{16} \,,
		\end{align}
	\end{subequations}
	\begin{align}
		\label{eq:6PNloc}
		&a_{70}^{\rm loc}(\nu) = \left(-\frac{31138024879}{4762800}+\frac{922426025089 \pi ^2}{990904320}-\frac{5556443
			\pi ^4}{524288}\right) \nu +\left(\frac{84714437}{88200}-\frac{6504619 \pi
			^2}{215040}+a_7^{\nu ^2}+\frac{q_{45}^{\nu ^2}}{14}\right. \cr
		&\left.\quad-\frac{13 \bar{d}_5^{\nu
				^2}}{15}+\frac{\bar{d}_6^{\nu ^2}}{6}\right) \nu
		^2+\left(-\frac{39097}{40}+\frac{91993 \pi ^2}{4096}-\frac{5 a_6^{\nu
				^2}}{2}+a_7^{\nu ^3}-\frac{\bar{d}_5^{\nu ^2}}{2}\right) \nu
		^3+\left(-\frac{14213}{96}+\frac{615 \pi ^2}{128}\right) \nu ^4+\frac{13 \nu
			^5}{16}\,.
	\end{align}
\end{widetext}
The coefficients $({\bar d}_5^{\nu^2},  {a}_6^{\nu^2};
{q}_{45}^{\nu^2}, {\bar d}_6^{\nu^2},  {a}_7^{\nu^2}, {a}_7^{\nu^3})$
entering the 5PN and 6PN orders are
numerical coefficients left undetermined by the Tutti Frutti approach, and not yet determined by other methods.
Here, we keep the notation of \cite{Bini:2020nsb}.
The coefficients ${\bar d}_n^{\nu^2}$ and ${a}_n^{\nu^2}$ belong to the $n$-PM order, while
the coefficient ${q}_{45}^{\nu^2}$ (which parametrizes
a term $\propto \nu^2 {q}_{45}^{\nu^2} p_r^4 u^5 $) belongs to the 5PM order. All these coefficients are expected to be of order unity. One could leave them as free parameters in constructing EOB waveforms, and try
to best-fit them to numerical-relativity waveforms.
Nevertheless, we have checked that they have a rather small effect on EOB waveforms.

\section{Nonlocal orbital dynamics for elliptic-like motions at 6PN accuracy.}
\label{app:nonlocal}
In this appendix we extend the computation presented in Sec.~\ref{sec:nonlocal_4PN}, i.e.~of the nonlocal PN-expanded contribution to the LJBL-gauge EOB potential $A(\gamma,\nu)$, using the full 6PN accuracy of the Tutti Frutti results \cite{Bini:2019nra,Bini:2020wpo,Bini:2020nsb,Bini:2020hmy}, up to the eight eccentricity order.

We start by recalling the structure of the 6PN-accurate Tutti Frutti results we will use.

\subsection{6PN nonlocal bound-state Hamiltonian in DJS gauge}
The DJS-gauge Tutti Frutti squared effective Hamiltonian for bound state motions,
$\hat H_{\rm eff}^2(r,p_r,p_{\varphi})$, is decomposed in a local 
and several nonlocal contributions as
\begin{align}
	\label{eq:heffTot}
	&\hat H_{\rm eff}^2(r,p_r,p_{\varphi})= \hat H_{\rm eff, loc, f}^2(r,p_r,p_{\varphi})\cr
	&\quad+ \delta \hat H_{\rm eff, nonloc, h}^2(r,p_r,p_{\varphi})+\delta \hat H_{\rm eff, f-h}^2(r,p_r,p_{\varphi})\,. \qquad
\end{align}
Here $\hat H_{\rm eff, loc, f}^2(r,p_r,p_{\varphi},\g)$ describes up to the 6PN order the local dynamics studied (in its 4PM-4PN complete version) in Sec.~\ref{sec:4PN_Alocal}. We have added a subscript $f$ to $\hat H_{\rm eff, loc, f}^2(r,p_r,p_{\varphi},\g)$ as a reminder that this local Hamiltonian is obtained by splitting the action in local and nonlocal parts with the use of a flexibility factor $f(t)$ entering the time scale $\Delta t ^f$ used as ultraviolet cutoff \cite{Bini:2020wpo}, namely
\begin{equation}
	\Delta t ^f = f(t)\Delta t ^h =  f(t) \, 2 r_{12}^h/c\,,
\end{equation}
$r_{12}^h$ denoting the two-body radial separation in harmonic coordinates. Let us discuss in turn the two other contributions,  $\delta \hat H_{\rm eff, nonloc, h}^2(r,p_r,p_{\varphi})$ and $\delta \hat H_{\rm eff, f-h}^2(r,p_r,p_{\varphi})$ to the squared effective Hamiltonian \eqref{eq:heffTot}. 

The firs contribution, $\delta \hat H_{\rm eff, nonloc, h}^2(r,p_r,p_{\varphi})$, is the nonlocal part of the squared effective Hamiltonian, obtained with $f(t)=1$ and thus with regularization time scale $\Delta t ^h=2 r_{12}^h/c$.  Its expression is precisely Eq.~\eqref{eq:Hnonloc_4PN}, that is
\begin{align}
	\label{eq:Hnonloc_h}
	&\delta \hat{H}_{\rm eff, nonloc, h}^2(r,p_r,p_{\varphi})=[1-2(1-2u)p_r^2
	\cr&\quad+p_\varphi^2 u^2] \, \delta A_{\rm nl,h}(r)
	+(1-2u)^2p_r^2\, \delta\bar{D}_{\rm nl,h}(r) \cr&\quad+ (1-2u) \, \delta Q_{\rm nl,h}(r,p_r) \,.
\end{align}
where the subscript ``h'' signals the use of a the cutoff timescale $\Delta t ^h$.\footnote{In Sec.~\ref{sec:nonlocal_4PN} there was no need for this subscript since any difference in the cutoff timescale considered has no impact before the 5PN order.}

As already mentioned in the main text, the nonlocal components $[\delta A_{\rm nl,h}(r),\delta\bar{D}_{\rm nl,h}(r),\delta Q_{\rm nl,h}(r,p_r)]$ of the EOB potentials are given at 4PN, 5PN, and 6PN in Table IV of Ref.~\cite{Bini:2020wpo} and Table VI of Ref.~\cite{Bini:2020nsb},  and reported here in Tables \ref{tab:nlc_A}, \ref{tab:nlc_D}, and \ref{tab:nlc_Q}.
\begin{table}
	\caption{\label{tab:nlc_A} Coefficients of the 4+5+6PN nonlocal potential in DJS gauge.}
	\begin{ruledtabular}
		\begin{tabular}{ll}
			Coefficient & Expression \\
			\hline
			\hline
			$a_5^{\rm nl,c}$ & $\left(\frac{128}{5}\gamma_E+\frac{256}{5}\ln2\right)\nu$\\
			$a_5^{\rm nl,\ln{}}$ & $\frac{64}{5}\nu$\\
			$a_6^{\rm nl,c}$ & $\left(-\frac{128}{5} -\frac{14008}{105} \gamma_E-\frac{31736}{105}\ln2 +\frac{243}{7}\ln3\right)\nu $\\
			& 
			$+\left( \frac{64}{5}-\frac{288}{5} \gamma_E +\frac{928}{35}\ln2-\frac{972}{7}\log3\right)\nu^2$\\
			$a_6^{\rm nl,\ln{}}$ & $-\frac{7004}{105}\nu-\frac{144}{5}\nu^2$\\
			$a_7^{\rm nl,c}$ & $
			\left(\frac{206740}{567}\ln2+\frac{12664}{105}-\frac{4617}{14}\ln3-\frac{5044}{405}\gamma_E\right)\nu$\\
			&$
			+\left(-\frac{1139672}{945}\ln2+\frac{10132}{105}+\frac{10449}{7}\ln3+\frac{101272}{315}\gamma_E\right)\nu^2$\\
			&$
			+\left(-\frac{112}{5}+32\gamma_E+\frac{1214624}{945}\ln2-\frac{4860}{7}\ln3\right)\nu^3\,$\\
			$a_7^{\rm nl,\ln}$&$
			-\frac{2522}{405}\nu+\frac{50636}{315}\nu^2+16\nu^3$\\
		\end{tabular}
	\end{ruledtabular}
\end{table}
\begin{table*}
	\caption{\label{tab:nlc_D}Coefficients of the 4+5+6PN nonlocal part of the $\bar{D}$ potential in DJS gauge.}
	\begin{ruledtabular}
		\begin{tabular}{ll}
			Coefficient & Expression \\
			\hline
			\hline
			$d_4^{\rm nl,c}$ & $\left(-\frac{992}{5}+\frac{1184}{15} \gamma_E -\frac{6496}{15}\ln2+\frac{2916}{5}\ln3 \right)\nu$\vspace{3mm}\\
			$d_4^{\rm nl,\log{}}$ & $\frac{592}{15}\nu$\vspace{3mm}\\
			$d_5^{\rm nl,c}$ & $\left(-\frac{7318}{35} -\frac{2840}{7} \gamma_E+\frac{120648}{35}\ln2 -\frac{19683}{7}\ln3\right)\nu $\\
			& $
			+\left(\frac{67736}{105} -\frac{6784}{15}  \gamma_E-\frac{326656}{21}\ln2+\frac{58320}{7}\ln3 \right)\nu^2$\vspace{3mm}\\
			$ d_5^{\rm nl,\log{}}$ & $-\frac{1420}{7}\nu-\frac{3392}{15}\nu^2$\vspace{3mm}\\
			$d_6^{\rm nl,c}$ &$
			\left(-\frac{6381680}{189}\ln2+\frac{2043541}{2835}+\frac{1765881}{140}\ln3-\frac{64096}{45}\gamma_E+\frac{9765625}{2268}\ln5\right)\nu$\\
			&$
			+\left(\frac{28429312}{189}\ln2-\frac{3576231}{70}\ln3+\frac{167906}{105}+\frac{302752}{105}\gamma_E-\frac{9765625}{378}\ln5\right)\nu^2$\\
			&$
			+\left(-\frac{9908480}{63}\ln2-\frac{744704}{945}+\frac{9765625}{252}\ln5+\frac{2944}{3}\gamma_E+\frac{1275021}{28}\ln3\right)\nu^3
			\,$\vspace{3mm}\\
			$d_6^{\rm nl,log}$&$
			-\frac{32048}{45}\nu+\frac{151376}{105}\nu^2+\frac{1472}{3}\nu^3$ \vspace{1mm}
		\end{tabular}
	\end{ruledtabular}
\end{table*}
\begin{table*}
	\caption{\label{tab:nlc_Q}Coefficients of the 4+5+6PN nonlocal part of the $Q$ potential in DJS gauge.}
	\begin{ruledtabular}
		\begin{tabular}{ll}
			Coefficient & Expression \\
			\hline
			\hline
			$ q_{43}^{\rm nl,c}$ & $\left(-\frac{5608}{15} +\frac{496256}{45}\ln2-\frac{33048}{5}\ln3 \right)\nu$\vspace{3mm}\\
			$q_{43}^{\rm nl,\ln{}}$ & 0\vspace{3mm}\\
			$q_{44}^{\rm nl,c}$ & $\left(\frac{1007633}{315} +\frac{10856}{105} \gamma_E -\frac{40979464}{315}\ln2+\frac{14203593}{280}\ln3+\frac{9765625}{504}\ln5\right)\nu$\nonumber\\
			&$+\left(\frac{74436}{35} -\frac{1184}{5}\gamma_E  +\frac{33693536}{105}\ln2  -\frac{6396489}{70}\ln3-\frac{9765625}{126}\ln5\right)\nu^2$\vspace{3mm}\\
			$q_{44}^{\rm nl,\ln{}}$ & $\frac{5428}{105}\nu-\frac{592}{5}\nu^2$\vspace{3mm}\\
			$q_{62}^{\rm nl,c}$ & $\left(-\frac{4108}{15} -\frac{2358912}{25}\ln2 +\frac{1399437}{50}\ln3+\frac{390625}{18}\ln5\right)\nu$\vspace{3mm}\\
			$q_{62}^{\rm nl,\ln{}}$ & $0$\vspace{3mm} \\
			$q_{63}^{\rm nl,c}$ & $\left(\frac{1300084}{525} +\frac{6875745536}{4725}\ln2 -\frac{23132628}{175}\ln3-\frac{101687500}{189}\ln5\right)\nu$\\
			& $ +\left(\frac{160124}{75}-\frac{4998308864}{1575}\ln2 -\frac{45409167}{350}\ln3+\frac{26171875}{18} \ln5\right)\nu^2$\vspace{3mm}\\
			$q_{63}^{\rm nl,\ln{}}$ & $0$ \vspace{3mm}\\
			$q_{45}^{\rm nl,c}$&$
			\left(\frac{70925884}{63}\ln2+\frac{13212013}{5670}-\frac{3873663}{16}\ln3-\frac{8787109375}{27216}\ln5-\frac{617716}{315}\gamma_E\right)\nu$\\
			&$+\left(\frac{92560887}{280}\ln3-\frac{12619052648}{2835}\ln2-\frac{1437979}{63}+\frac{632344}{315}\gamma_E+\frac{7755859375}{4536}\ln5\right)\nu^2$\\
			&$
			+\left(-\frac{177316}{35}+\frac{11263031264}{2835}\ln2+\frac{16544}{9}\gamma_E-\frac{4091796875}{2268}\ln5+\frac{2908467}{20}\ln3\right)\nu^3
			$\vspace{3mm}\\
			$q_{45}^{\rm nl,log}$&$
			-\frac{308858}{315}\nu+\frac{316172}{315}\nu^2+\frac{8272}{9}\nu^3
			$\vspace{3mm}\\
			$q_{64}^{\rm nl,c}$&$
			\left(-\frac{211076833264}{14175}\ln2-\frac{137711989}{28350}-\frac{9678652821}{5600}\ln3+\frac{447248}{1575}\gamma_E+\frac{153776136875}{23328}\ln5+\frac{96889010407}{116640}\ln7\right)\nu$\\
			&$
			+\left(\frac{44592947739}{2800}\ln3+\frac{2411178384736}{42525}\ln2-\frac{126070663}{4725}-\frac{26848}{175}\gamma_E-\frac{796015515625}{27216}\ln5-\frac{96889010407}{19440}\ln7\right)\nu^2$\\
			&$
			+\left(-\frac{40513708}{4725}-\frac{109566260523}{5600}\ln3+\frac{1424826953125}{54432}\ln5+\frac{96889010407}{12960}\ln7+\frac{2368}{5}\gamma_E-\frac{431564554688}{8505}\ln2\right)\nu^3
			$\vspace{3mm}\\
			$q_{64}^{\rm nl,log}$&$
			\frac{223624}{1575}\nu-\frac{13424}{175}\nu^2+\frac{1184}{5}\nu^3$\vspace{3mm}
			\\
			$q_{81}^{\rm nl,c}$&$
			\left(-\frac{35772}{175}+\frac{21668992 \ln 2}{45}+\frac{6591861 \ln
				3}{350}-\frac{27734375 \ln 5}{126}\right)\nu $\\
			$q_{82}^{\rm nl,c}$&$
			\left(\frac{5788281}{2450}-\frac{16175693888 \ln 2}{1575}-\frac{393786545409 \ln
				3}{156800}+\frac{875090984375 \ln 5}{169344}+\frac{13841287201 \ln
				7}{17280}\right)\nu$\\
			&$
			+\left(\frac{703189497728 \ln 2}{33075}+\frac{869626}{525}+\frac{332067403089 \ln
				3}{39200}-\frac{468490234375 \ln 5}{42336}-\frac{13841287201 \ln 7}{4320}\right)\nu^2$\\
			$q_{83}^{\rm nl,c}$&$
			\left(\frac{5196312336176}{35721}\ln2+\frac{17515638027261}{313600}\ln3-\frac{63886617280625}{1016064}\ln5-\frac{29247366220639}{933120}\ln7-\frac{709195549}{132300}\right)\nu$\\
			&$
			+\left(-\frac{177055674739808}{297675}\ln2-\frac{43719724468071}{156800}\ln3+\frac{366449151015625}{1524096}\ln5+\frac{26506549233199}{155520}\ln7-\frac{1746293}{70}\right)\nu^2$\\
			&$
			+\left(\frac{57604236136064}{99225}\ln2+\frac{10467583300341}{39200}\ln3-\frac{73366198046875}{381024}\ln5-\frac{7709596970957}{38880}\ln7-\frac{154862}{21}\right)\nu^3
			$\vspace{3mm}\\
			$q_{83}^{\rm nl,log}$&$0$\\
		\end{tabular}
	\end{ruledtabular}
\end{table*}

\begin{table*}
	\caption{\label{tab:nlc_A_ab1_part1} Resulting coefficients for $\delta A^{\rm 4PN}_{\rm nonlocal,h}$ and $\delta A^{\rm 5PN}_{\rm nonlocal,h}$. Here $\gamma_E$ is Euler's constant.}
	\begin{ruledtabular}
		\begin{tabular}{ll}
			Coefficient & Expression \\
			\hline
			\hline
			$a_{1,4}^{\text{nlh,c}}$ & $\left(-\frac{35772}{175}+\frac{10834496 \ln 2}{21}+\frac{6591861 \ln
				3}{350}-\frac{27734375 \ln 5}{126}\right) \nu$\vspace{3mm}\\
			$a_{2,3}^{\text{nlh,c}}$ & $\left(-\frac{7991}{15}+\frac{16486604 \ln 2}{15}+\frac{3995649 \ln
				3}{80}-\frac{68359375 \ln 5}{144}\right) \nu$\vspace{3mm}\\
			$a_{3,2}^{\text{nlh,c}}$ & $\left(-\frac{66523}{105}+\frac{1374117716 \ln 2}{1575}+\frac{145758933
				\ln 3}{2800}-\frac{387109375 \ln 5}{1008}\right)  \nu$\vspace{3mm}\\
			$a_{4,1}^{\text{nlh,c}}$ & $\left(-\frac{49682}{105}+\frac{296 \gamma_E }{15}+\frac{541043288 \ln
				2}{1575}+\frac{36333117 \ln 3}{1400}-\frac{77734375 \ln
				5}{504}\right)  \nu$\vspace{3mm}\\
			$a_{5,0}^{\text{nlh,c}}$ & $\left(-\frac{29388}{175}+\frac{136 \gamma_E }{3}+\frac{28504144 \ln
				2}{525}+\frac{876258 \ln 3}{175}-\frac{1562500 \ln
				5}{63}\right)  \nu$\vspace{3mm}\\
			$a_{4,1}^{\text{nlh,ln}}$ & $\frac{148}{15}  \nu$\vspace{3mm}\\    
			$a_{5,0}^{\text{nlh,ln}}$ & $\frac{68}{3}  \nu$\vspace{1mm}\\
			\hline
			$a_{1,5}^{\text{nlh,c}}$ & $\left(\frac{71544}{175}-\frac{21668992 \ln 2}{21}-\frac{6591861 \ln
				3}{175}+\frac{27734375 \ln 5}{63}\right)  \nu$\vspace{3mm}\\
			$a_{2,4}^{\text{nlh,c}}$ & $\left(\frac{77118197}{44800}-\frac{1623588763 \ln
				2}{315}-\frac{2256723902277 \ln 3}{2867200}+\frac{7501679921875
				\ln 5}{3096576}+\frac{96889010407 \ln 7}{442368}\right) \nu$
			\vspace{1mm}\\ & $+\left(\frac{434813}{960}+\frac{5493667951 \ln
				2}{945}+\frac{332067403089 \ln 3}{143360}-\frac{2342451171875 \ln
				5}{774144}-\frac{96889010407 \ln 7}{110592}\right) \nu ^2$\vspace{3mm}\\
			$a_{3,3}^{\text{nlh,c}}$ & $\left(\frac{9328833077}{2822400}-\frac{34463337173 \ln
				2}{3675}-\frac{5588225526843 \ln 3}{2867200}+\frac{33236879515625
				\ln 5}{7225344}+\frac{429079903231 \ln 7}{737280}\right) \nu $
			\vspace{1mm}\\ & $+\left(\frac{10137109}{6720}+\frac{165305397617 \ln
				2}{11025}+\frac{30789270513261 \ln
				3}{5017600}-\frac{14147997265625 \ln
				5}{1806336}-\frac{429079903231 \ln 7}{184320}\right) \nu ^2$\vspace{3mm}\\
			$a_{4,2}^{\text{nlh,c}}$ & $\left(\frac{5424084823}{1411200}-\frac{411 \gamma_E }{35}-\frac{260768957761
				\ln 2}{33075}-\frac{2778338476377 \ln
				3}{1433600}+\frac{43123869265625 \ln
				5}{10838016}+\frac{650540498447 \ln 7}{1105920}\right) \nu
			$
			\vspace{1mm}\\ & $+\left(\frac{7435079}{3360}-\frac{148 \gamma_E }{5}+\frac{482118519718
				\ln 2}{33075}+\frac{15468076580319 \ln
				3}{2508800}-\frac{20708641015625 \ln
				5}{2709504}-\frac{650540498447 \ln 7}{276480}\right) \nu ^2$\vspace{3mm}\\
			$a_{5,1}^{\text{nlh,c}}$ & $\left(\frac{3371152901}{1176000}-\frac{3741 \gamma_E }{25}-\frac{20705210921
				\ln 2}{6125}-\frac{3256431610833 \ln
				3}{3584000}+\frac{9395870828125 \ln 5}{5419008}+\frac{152254159211
				\ln 7}{552960}\right) \nu $
			\vspace{1mm}\\ & $+\left(\frac{2122033}{1200}-\frac{2156
				\gamma_E }{15}+\frac{215891035208 \ln 2}{33075}+\frac{3585448178811
				\ln 3}{1254400}-\frac{4649392578125 \ln
				5}{1354752}-\frac{152254159211 \ln 7}{138240}\right) \nu ^2$\vspace{3mm}\\
			$a_{6,0}^{\text{nlh,c}}$ & $\left(\frac{191444741}{196000}-\frac{7276 \gamma_E }{25}-\frac{96607902652
				\ln 2}{165375}-\frac{298360928979 \ln
				3}{1792000}+\frac{819577984375 \ln 5}{2709504}+\frac{13841287201
				\ln 7}{276480}\right) \nu $
			\vspace{1mm}\\ & $+\left(\frac{873471}{1400}-\frac{2576
				\gamma_E }{15}+\frac{37411207016 \ln 2}{33075}+\frac{321481778769 \ln
				3}{627200}-\frac{403390234375 \ln 5}{677376}-\frac{13841287201
				\ln 7}{69120}\right) \nu ^2$\vspace{3mm}\\
			$a_{4,2}^{\text{nlh,ln}}$ & $-\frac{411 \nu }{70}-\frac{74 \nu ^2}{5}$\vspace{3mm}\\
			$a_{5,1}^{\text{nlh,ln}}$ & $-\frac{3741 \nu }{50}-\frac{1078 \nu ^2}{15}$\vspace{3mm}\\
			$a_{6,0}^{\text{nlh,ln}}$ & $-\frac{3638 \nu }{25}-\frac{1288 \nu ^2}{15}$
			\vspace{1mm}
		\end{tabular}
	\end{ruledtabular}
\end{table*}

\begin{table*}
	\caption{\label{tab:nlc_A_ab1_part2} Resulting coefficients for $\delta A^{\rm 5.5PN}_{\rm nonlocal,h}$ and $\delta A^{\rm 6PN}_{\rm nonlocal,h}$. Here $\gamma_E$ is Euler's constant}
	\begin{ruledtabular}
		\begin{tabular}{ll}
			Coefficient & Expression \\
			\hline
			\hline
			$a_{2,4}^{\text{nlh,}\sqrt{u}}$ & $\frac{5994461 \pi  \nu }{75184200}$\vspace{3mm}\\    
			$a_{3,3}^{\text{nlh,}\sqrt{u}}$ & $-\frac{242219431 \pi  \nu }{1608106500}$\vspace{3mm}\\   
			$a_{4,2}^{\text{nlh,}\sqrt{u}}$ & $\frac{2995946714 \pi  \nu }{723647925}$\vspace{3mm}\\    
			$a_{5,1}^{\text{nlh,}\sqrt{u}}$ & $\frac{7276481179 \pi  \nu }{187960500}$\vspace{3mm}\\    
			$a_{6,0}^{\text{nlh,}\sqrt{u}}$ & $\frac{83295355783 \pi  \nu }{1378377000}$\vspace{1mm}\\
			\hline
			$a_{1,6}^{\text{nlh,c}}$ & $\nu  \left(-\frac{143088}{175}+\frac{43337984 \ln 2}{21}+\frac{13183722
				\ln 3}{175}-\frac{55468750 \ln 5}{63}\right)$\vspace{3mm}\\ 
			$a_{2,5}^{\text{nlh,c}}$ & $\left(-\frac{313773279}{89600}+\frac{2111141083 \ln
				2}{210}+\frac{7256176434639 \ln 3}{5734400}-\frac{9546479921875
				\ln 5}{2064384}-\frac{96889010407 \ln 7}{294912}\right) \nu
			$
			\vspace{1mm}\\ & $+\left(-\frac{434813}{640}-\frac{5493667951 \ln
				2}{630}-\frac{996202209267 \ln 3}{286720}+\frac{2342451171875 \ln
				5}{516096}+\frac{96889010407 \ln 7}{73728}\right) \nu ^2$\vspace{3mm}\\ 
			$a_{3,4}^{\text{nlh,c}}$ & $\left(-\frac{179112164569}{30481920}+\frac{94783454497241 \ln
				2}{3214890}+\frac{67243020174387 \ln
				3}{8028160}-\frac{15571051587341875 \ln
				5}{1170505728}-\frac{877215707820361 \ln 7}{214990848}\right) \nu
			$
			\vspace{1mm}\\ & $+\left(-\frac{153263413}{40320}-\frac{88831649433773 \ln
				2}{1071630}-\frac{10703991524229 \ln 3}{286720}+\frac{31037329281015625 \ln
				5}{877879296}+\frac{1910557705211407 \ln 7}{89579520}\right) \nu
			^2$
			\vspace{1mm}\\ & $+\left(-\frac{154862}{189}+\frac{57604236136064 \ln
				2}{893025}+\frac{1163064811149 \ln 3}{39200}-\frac{73366198046875
				\ln 5}{3429216}-\frac{7709596970957 \ln 7}{349920}\right) \nu ^3$\vspace{3mm}\\ 
			$a_{4,3}^{\text{nlh,c}}$ & $\left(-\frac{30492532967}{6096384}+\frac{11612 \gamma_E 
			}{315}+\frac{45246337124803 \ln 2}{918540}+\frac{153054539029089
				\ln 3}{8028160}-\frac{2812054420105625 \ln
				5}{130056192}-\frac{2243037691275955 \ln 7}{214990848}\right) \nu
			$
			\vspace{1mm}\\ & $+\left(-\frac{2988885587}{302400}+\frac{1751 \gamma _E}{70}-\frac{42852920634415
				\ln 2}{214326}-\frac{232856210028969 \ln 3}{2508800}+\frac{2551367966640625
				\ln 5}{31352832}+\frac{1259330630494909 \ln 7}{22394880}\right) \nu
			^2$
			\vspace{1mm}\\ & $+\left(-\frac{92957869}{30240}+37 \gamma_E  +\frac{164772199472387 \ln
				2}{893025}+\frac{106829150283279 \ln
				3}{1254400}-\frac{6635329272265625 \ln
				5}{109734912}-\frac{714307308582007 \ln 7}{11197440}\right) \nu ^3$\vspace{3mm}\\ 
			$a_{5,2}^{\text{nlh,c}}$ & $\left(-\frac{1343296717549}{592704000}+\frac{492514 \gamma_E 
			}{6125}+\frac{4011985837440181 \ln
				2}{89302500}+\frac{19587014970857253 \ln
				3}{1003520000}-\frac{154259538621875 \ln
				5}{7962624}-\frac{6631680016091291 \ln 7}{597196800}\right) \nu
			$
			\vspace{1mm}\\ & $+\left(-\frac{50618075581}{3528000}+\frac{331157
				\gamma _E}{1050}-\frac{613647346154779 \ln 2}{2976750}-\frac{305083228792563
				\ln 3}{3136000}+\frac{501076363859375 \ln 5}{6096384}+\frac{5860271869361
				\ln 7}{97200}\right) \nu ^2$
			\vspace{1mm}\\ & $+\left(-\frac{1693354921}{352800}+\frac{3743 \gamma_E 
			}{15}+\frac{790301920439 \ln 2}{3969}+\frac{115626711539763 \ln
				3}{1254400}-\frac{783616805078125 \ln
				5}{12192768}-\frac{17370815437255 \ln 7}{248832}\right) \nu ^3$\vspace{3mm}\\ 
			$a_{6,1}^{\text{nlh,c}}$ & $\left(-\frac{109102461857}{666792000}-\frac{3119672 \gamma_E 
			}{165375}+\frac{4210267059801826 \ln
				2}{200930625}+\frac{1197825049766241 \ln
				3}{125440000}-\frac{1311429915923125 \ln
				5}{146313216}-\frac{3704618057777623 \ln 7}{671846400}\right) \nu
			$
			\vspace{1mm}\\ & $+\left(-\frac{619513589}{55125}+\frac{595496 \gamma
				_E}{525}-\frac{1337797760209928 \ln 2}{13395375}-\frac{149608543096413 \ln
				3}{3136000}+\frac{2159121776609375 \ln 5}{54867456}+\frac{837731963172853
				\ln 7}{27993600}\right) \nu ^2$
			\vspace{1mm}\\ & $+\left(-\frac{1474770263}{396900}+\frac{7928 \gamma_E 
			}{15}+\frac{86858343124312 \ln 2}{893025}+\frac{7071646351059 \ln 3}{156800}-\frac{422975045703125 \ln 5}{13716864}-\frac{48430663916299 \ln 7}{1399680}\right) \nu ^3$\vspace{3mm}\\ 
			$a_{7,0}^{\text{nlh,c}}$ & $\left(\frac{63540120697}{166698000}-\frac{784991 \gamma_E 
			}{99225}+\frac{157306259361746 \ln 2}{40186125}+\frac{11443865567679
				\ln 3}{6272000}-\frac{61083824984375 \ln
				5}{36578304}-\frac{35593777993529 \ln 7}{33592320}\right) \nu
			$
			\vspace{1mm}\\ & $+\left(-\frac{3809736181}{1058400}+\frac{375754 \gamma
				_E}{315}-\frac{50267711665576 \ln 2}{2679075}-\frac{22712673038121 \ln
				3}{2508800}+\frac{1608509992109375 \ln 5}{219469824}+\frac{128297312282489
				\ln 7}{22394880}\right) \nu ^2$
			\vspace{1mm}\\ & $+\left(-\frac{189399347}{158760}+348 \gamma_E  +\frac{3241904436308 \ln
				2}{178605}+\frac{527833234161 \ln 3}{62720}-\frac{154715013671875
				\ln 5}{27433728}-\frac{3667941108265 \ln 7}{559872}\right) \nu ^3$\vspace{3mm}\\ 
			$a_{4,3}^{\text{nlh,ln}}$ & $\frac{5806 \nu }{315}+\frac{1751 \nu ^2}{140}+\frac{37 \nu ^3}{2}$\vspace{3mm}\\
			$a_{5,2}^{\text{nlh,ln}}$ & $\frac{246257 \nu }{6125}+\frac{331157 \nu ^2}{2100}+\frac{3743 \nu ^3}{30}$\vspace{3mm}\\ 
			$a_{6,1}^{\text{nlh,ln}}$ & $-\frac{1559836 \nu }{165375}+\frac{297748 \nu ^2}{525}+\frac{3964 \nu
				^3}{15}$\vspace{3mm}\\ 
			$a_{7,0}^{\text{nlh,ln}}$ & $-\frac{784991 \nu }{198450}+\frac{187877 \nu ^2}{315}+174 \nu ^3$
			\vspace{1mm}
		\end{tabular}
	\end{ruledtabular}
\end{table*}

Each of the nonlocal components in Eq.~\eqref{eq:Hnonloc_h} has been completed by a 5.5PN contribution, obtained in 
Sec.~XII of Ref.~\cite{Bini:2020wpo}) [see Eq.~(12.9) there].
These additional 5.5PN contributions explicitly read
\begin{subequations}
	\begin{align}
		&\delta A_{\rm nl,h}^{\rm 5.5PN}= \dfrac{13696\pi}{525} \nu \, u^{13/2} \, , \\
		&\delta \bar{D}_{\rm nl,h}^{\rm 5.5PN}= \dfrac{264932\pi}{1575}\nu \, u^{11/2} \, ,
		\\
		&\delta Q_{\rm nl,h}^{\rm 5.5PN} = \dfrac{88703\pi}{1890}\nu \, p_r^4 \, u^{9/2} -\dfrac{2723471\pi}{756000}\nu \, p_r^6 \, u^{7/2}\cr
		&\quad+\dfrac{5994461\pi}{12700800}\nu \, p_r^8 \, u^{5/2} +O(p_r^{10}) \,.
	\end{align}
\end{subequations}
Note that, generalizing Eq.~\eqref{eq:Q4PNdjs}, the quantity
\begin{align}
	&\delta Q_{\rm nl,h}(r,p_r)\equiv\delta Q_{\rm nl,h}(r,p_r)^{\rm 4PN}+\delta Q_{\rm nl,h}(r,p_r)^{\rm 5PN}\cr&\quad+\delta Q_{\rm nl,h}(r,p_r)^{\rm 5.5PN}+\delta Q_{\rm nl,h}(r,p_r)^{\rm 6PN}\,,
\end{align}
has the structure
\begin{align}
	\label{eq:Qtotdjs}
	&\delta Q_{\rm nl,h}(r,p_r) =  c_4 p_r^4 ( u^3 q_{43}+ u^4 \bar q_{44} + u^{9/2} q_{4,4.5} \cr&\quad+ u^5 \bar q_{45})+ c_6 p_r^6 ( u^2 q_{62} + u^3 \bar q_{63}+u^{7/2}q_{6,3.5}+ u^4 \bar q_{64})\cr&\quad+ c_8 p_r^8 ( u q_{81}+u^2 \bar q_{82}+u^{5/2} \bar q_{8,2.5} + u^3 q_{83}) \cr&\quad+\mathcal{O}(p_r^{10})\,.
\end{align}
Here the undecorated coefficients $q_{nm}$ depend only on $\nu$, whereas the barred coefficients $\bar q_{nm}$ have a linear dependence on $\ln{u}$ as well. As in Eq.~\eqref{eq:Q4PNdjs}, the parameters $c_4,c_6,c_8$ appearing in Eq.~\eqref{eq:Qtotdjs} are eccentricity-keying parameters.

The last term, $\delta \hat H_{\rm eff, f-h}^2(r,p_r,p_{\varphi})$, in Eq.~\eqref{eq:heffTot} is an additional contribution coming from the flexibility factor $f$. Its expression in EOB coordinates has the same structure of Eq.~\eqref{eq:Hnonloc_h}, with the corresponding coefficients of the EOB potentials, say $(\delta^{\rm f-h}A,\delta^{\rm f-h}\bar{D},\delta^{\rm f-h}Q)$, that are given in Eqs.~(7.32)-(7.34) of Ref.~\cite{Bini:2020hmy} in terms of six flexibility parameters, $(C_2,C_3,D^0_2,D^0_3,D^0_4,D^0_1)$; see Sec.VII of Ref.~\cite{Bini:2020hmy} for more details.

\subsection{Transforming the 6PN nonlocal bound-state dynamics to the LJBL gauge}
Let us indicate how it is possible to map the results recalled in the previous subsection into corresponding additional contributions [respectively $\delta A_{\rm nonlocal,h}(\gamma, u)$ and $\delta A_{\rm f-h}(\gamma, u)$] to the basic $A(\gamma,u)$ potential of our LJBL gauge, so that
\begin{align}
	\label{eq:Atot}
	&A(\gamma,u)=A^{\rm loc}_{\rm 6PN completed}(\gamma,u)+\delta A_{\rm nonlocal,h}(\gamma,u)\cr&\quad+\delta A_{\rm f-h}(\gamma,u) \,.
\end{align}
Here $A^{\rm loc}_{\rm 6PN completed}(\gamma,u)$ is the local 6PN-completed component, determined in Appendix \ref{app:complete_Alocal_6PN}.

The mapping between the DJS gauge dynamics, described by the Hamiltonian $H_{\rm eff}^2(r,p_r,p_{\varphi})$ in Eq.~\eqref{eq:heffTot}, and the (PN expansion of the) LJBL-gauge dynamics described by the $A(\gamma,u)$ potential of Eq.~\eqref{eq:Atot} is obtained, as we did at 4PN accuracy in Sec.~\ref{sec:nonlocal_4PN}, by looking for a canonical transformation such that the DJS gauge restriction 
is equivalent to the corresponding LJBL condition.

The resulting $\delta A_{\rm nonlocal,h}(\gamma,u)$ and $\delta A_{\rm f-h}(\gamma, u)$ contributions to $A(\gamma,u)$, and the canonical transformation itself,  are obtained order by order in the PN expansion. This canonical transformation involves logarithms of $u$. See Refs.~\cite{Bini:2019nra,Bini:2020wpo,Bini:2020nsb,Bini:2020hmy} for similar computations.

Focusing first on the nonlocal part, it has the structure 
\begin{align}
	\label{eq:Anonloc}
	&\delta A_{\rm nonlocal,h}(\gamma,u) = \delta A^{\rm 4PN}_{\rm nonlocal,h}(p_\infty,u) \cr&\quad+ \delta A^{\rm 5PN}_{\rm nonlocal,h}(p_\infty,u) +\delta A^{\rm 5.5PN}_{\rm nonlocal,h}(p_\infty,u) \cr&\quad+ \delta A^{\rm 6PN}_{\rm nonlocal,h}(p_\infty,u) \,
\end{align}
where each PN contribution is a polynomial in $\pinf^2$:
\begin{align}
	&\delta A^{ n \rm PN}_{\rm nonlocal,h}= \sum _{k=1}^{n+1} a_{k,n+1-k}^{\text{nlh,c}} \, u^k p_\infty ^{2(n+1-k)} \cr&\quad+ \ln u \sum_{k=4}^{n+1} a_{k,n+1-k}^{\text{nlh,ln}} \,
	u^k p_\infty ^{2(n+1-k)} \,  ,
\end{align}
for $n=4,5,6$ and
\begin{equation}
	\delta A^{\rm 5.5PN}_{\rm nonlocal,h}= a_{k,n+1-k}^{\text{nlh,}\sqrt{u}} \, \sqrt{u} \sum _{k=2}^{n+1} u^k p_\infty ^{2(n+1-k)} \, .
\end{equation}
Note the presence of square roots and logarithms of $u$. 
All the $\nu$-dependent parameters $(a_{m,n}^{\text{nlh,c}},a_{m,n}^{\text{nlh,ln}},a_{m,n}^{\text{nlh,}\sqrt{u}})$ are collected in Tables \ref{tab:nlc_A_ab1_part1} and \ref{tab:nlc_A_ab1_part2}. 
The 6PN-accurate nonlocal contribution to the $A$ potential, $\delta A_{\rm nonlocal,h}(\gamma,u)$ in Eq.~\eqref{eq:Anonloc}, has the same eccentricity structure of its 4PN-accurate analogue, i.e.
\begin{align}
	\label{eq:Anonloc_struct}
	&\delta A_{\rm nonlocal,h}(\gamma,u) = \delta A_{\rm nonlocal,h}^{\leq e^2}(\gamma,u)\cr&\quad+c_4\delta A_{\rm nonlocal,h}^{e^4}(\gamma,u)+c_6\delta A_{\rm nonlocal,h}^{e^6}(\gamma,u)\cr&\quad+c_8\delta A_{\rm nonlocal,h}^{e^8}(\gamma,u)\,.
\end{align}

In the ancillary file we explicitly provide the full expression (6PN-accurate and up to $e^8$) of $\delta A_{\rm nonlocal,h}(\gamma,u)$, according to the separation shown in Eq.~\eqref{eq:Anonloc_struct}.

Coming now to the f-h component, our result has the structure
\begin{align}
	\label{eq:Afmh_ansatz}
	&\delta A_{\rm f-h} = (a^{\rm f-h}_{5,2} p_\infty^2 + a^{\rm f-h}_{5,4} p_\infty^4)u^5 \cr&\quad+ (a^{\rm f-h}_{6,0} + a^{\rm f-h}_{6,2} p_\infty^2)u^6 + a^{\rm f-h}_{7,0}u^7 \,.
\end{align}
This contribution starts at 5PN and 5PM, and has no logarithms or square roots.
The canonical transformation yields the result
\begin{subequations}
	\begin{align}
		&a^{\rm f-h}_{5,2} = \frac{2 C_2}{5} \nu ^2 \,, 
		\\
		&a^{\rm f-h}_{5,4}=\left(\frac{6 D_2^0}{35}-\frac{7 C_2}{25}\right)\nu ^2 -\frac{4 C_2}{5}
		\nu ^3\,,
		\\
		&a^{\rm f-h}_{6,0} = \left(\frac{2 C_2}{5}+2 C_3\right) \nu ^2\,, 
		\\
		&a^{\rm f-h}_{6,2} = \left(\frac{11 D_2^0}{35}+\frac{D_3^0}{3}+\frac{104 C_2}{75}-\frac{4
			C_3}{3}\right) \nu ^2\cr
		&\quad+\left(-\frac{9 C_2}{5} -5
		C_3\right) \nu ^3\,,
		\\
		&a^{\rm f-h}_{7,0} = \left(\frac{D_2^0}{7}+\frac{D_3^0}{3}+2 D_4^0+\frac{19 C_2}{15}+\frac{23
			C_3}{3}\right) \nu ^2\cr
		&\quad+\left(2 D_4^1-C_2-6 C_3\right) \nu ^3 \,.
	\end{align}
\end{subequations}
When adding $\delta A_{\rm f-h}$ to $A^{\rm loc}_{\rm 6PN completed}(\gamma,u)$ we checked
that the six flexibility parameters $(C_2,C_3,D^0_2,D^0_3,D^0_4,D^0_1)$ have only
the effect of shifting the values of the six
undetermined coefficients $({\bar d}_5^{\nu^2},  {a}_6^{\nu^2};
{q}_{45}^{\nu^2}, {\bar d}_6^{\nu^2},  {a}_7^{\nu^2}, {a}_7^{\nu^3})$ [entering Eqs.~\eqref{eq:4PNloc}-\eqref{eq:6PNloc}], in keeping with
Eqs. (7.35) of Ref.~\cite{Bini:2020hmy}.

We highlight that we explicitly checked that the results of canonical transformation detailed above can be equivalently derived from a scattering angle matching analogous to the one used for the 6PN-accurate local completion of Appendix \ref{app:complete_Alocal_6PN}.

\bibliographystyle{apsrev4-1}
\bibliography{refs20260130.bib, local.bib}
	
\end{document}